\documentclass[aps,pra,reprint,superscriptaddress,nofootinbib]{revtex4-2}
\usepackage{graphicx}
\usepackage{amsmath}
\usepackage{amssymb}
\usepackage{todonotes}
\usepackage{nicefrac}
\usepackage{booktabs}
\usepackage{csquotes}
\usepackage{slashed}
\usepackage{hyperref}
\usepackage{tikz,tikz-3dplot}
\usepackage{subcaption}
\usepackage[export]{adjustbox}
\usepackage{float}
\usepackage[newfloat]{minted}
\usepackage{rotating}
\usepackage{xcolor}

\SetupFloatingEnvironment{listing}{name=Listing}

\begin{document}

\title{Hypothesis Tests for Observing Quantum Entanglement in \texorpdfstring{$\mathbf{ H \rightarrow WW^*}$}{HWW} at the LHC}

\author{Vincent Croft}
\affiliation{LIACS, Universiteit Leiden, Niels Bohrweg 1, 2333 CA Leiden, Netherlands}
\affiliation{Nikhef, Dutch National Institute for Subatomic Physics, Science Park 105, 1098 XG Amsterdam}
\author{Lennart Voelz}
\affiliation{Department of Physics, TU Dortmund University, Dortmund, Germany}
\author{Andrii Vak}
\affiliation{Czech Technical University in Prague, Institute of Experimental and Applied Physics, Husova 240/5, CZ-11000, Prague1, Czechia}
\author{André Sopczak}
\affiliation{Czech Technical University in Prague, Institute of Experimental and Applied Physics, Husova 240/5, CZ-11000, Prague1, Czechia}
\author{Carsten Burgard}
\affiliation{Department of Physics, TU Dortmund University, Dortmund, Germany}
\affiliation{Institut für Experimentalphysik, Universität Hamburg, 22761 Hamburg, Germany}

\begin{abstract}
We present a novel experimental strategy for testing quantum entanglement in Higgs boson decays to $W$ boson pairs at the Large Hadron Collider. Unlike theoretical approaches that rely on expectation values of Bell operators, which are highly sensitive to outliers and detector effects, we introduce a continuous formulation of the CGLMP inequality that enables standard hypothesis testing between entangled and separable states. To overcome the fundamental challenge of reconstructing invisible neutrino momenta in the $H \rightarrow WW^* \rightarrow \ell\nu\ell\nu$ channel, we employ conditional denoising diffusion probabilistic models (cDDPM), which provide unbiased, multidimensional unfolding applicable to the full measured dataset, including backgrounds. We evaluate the diffusion-based reconstruction against analytical methods through profile likelihood hypothesis tests implemented in RooFit, with systematic uncertainties from background normalisation and unfolding shape fully propagated. Our results demonstrate that the diffusion-based approach enables robust hypothesis testing of quantum entanglement in a realistic collider environment, with 3$\sigma$ evidence of quantum entanglement projected at approximately 555~fb$^{-1}$ and exceeding 5$\sigma$ at 1600~fb$^{-1}$ to be well within the expected limits of the HL-LHC luminosity targets.
\end{abstract}

\maketitle

\section{Introduction}

Quantum entanglement represents one of the most profound departures of quantum mechanics from classical physics. Since the seminal work of Einstein, Podolsky, and Rosen~\cite{Einstein:1935rr} questioning the completeness of quantum mechanics, and Bell's subsequent formulation of testable inequalities~\cite{Bell:1964kc}, experiments have consistently confirmed the predictions of quantum theory over local hidden variable models. Modern high-energy colliders offer a unique environment to probe entanglement in relativistic quantum field processes, where particles are produced at energies far exceeding those accessible in traditional quantum optics experiments.

Recent observations of quantum entanglement in top quark pair production by the ATLAS~\cite{ATLASttbar2024} and CMS~\cite{CMS:2024pts} collaborations have demonstrated that collider experiments can serve as precision laboratories for quantum information science. Given that quantum tomography provides the full reconstruction of the density matrix of a state, couplings beyond the Standard Model can be constrained by means of observables easily defined in terms of the polarization density matrix, giving limits competitive with the best available bounds with the potential for tomography based measurements to lead to more stringent global fit limits\cite{Grzadkowski:2010es,Aoude:2022imd,Fabbrichesi:2023cev,Sullivan:2024wzl,Severi:2022qjy,Fabbrichesi:2023cev,Maltoni:2024tul}.

The Higgs boson, as a scalar particle, decays to pairs of vector bosons in a quantum state that is predicted to be maximally entangled within the Standard Model. The decay channel $H \to WW^* \to \ell\nu\ell\nu$ is particularly attractive for entanglement studies because the parity-violating nature of $W$ boson decays provides a direct mapping from spin orientation to measurable lepton kinematics, enabling quantum state tomography through angular distributions~\cite{barr1,barr2,barr3,Aguilar-Saavedra:2022mpg}. A theoretical framework for extracting the spin density matrix and testing Bell inequalities has been developed~\cite{barr2}, but no experimental measurement strategy accounting for the practical challenges of collider environments has yet been demonstrated.

\subsection{Motivation: From Discrete Inequalities to Continuous Hypothesis Tests}

The standard approach to testing Bell inequalities at colliders involves computing expectation values of Bell operators constructed from spin correlation coefficients.
For two-qutrit systems such as $WW^*$ pairs, the Collins-Gisin-Linden-Massar-Popescu (CGLMP) inequality~\cite{CGLMP} provides the optimal framework, with the Bell parameter $\mathfrak{I}_3$ computed as a linear combination of Gell-Mann correlation coefficients extracted through quantum state tomography.

However, this approach faces fundamental limitations in collider environments. The expectation value $\langle \mathfrak{I}_3 \rangle$ is computed over the ensemble of events in such a way that events with extreme kinematic configurations arising from detector mismeasurements, background contamination, or rare kinematic fluctuations can significantly bias the extracted Bell parameter. Furthermore, the Gell-Mann coefficients $c_{ij}$ are extracted event-by-event through projections onto angular basis functions. The resulting distributions are typically non-Gaussian, with heavy tails that complicate the interpretation of central values and uncertainties. The assumption of a well-defined mean, central to the standard tomography procedure, breaks down when these distributions are skewed or multimodal.

Selection cuts required to suppress backgrounds and satisfy detector acceptance inevitably bias the angular distributions as well.
Cuts on lepton $p_T$ and $\eta$ preferentially select certain event topologies, distorting the extracted correlation coefficients even at truth level.
In a realistic analysis, the measured sample contains contributions from multiple physics processes, and computing a single expectation value over a mixed sample conflates the quantum properties of different states.
These issues are explored in detail in Appendix~\ref{app:ggm}.

To address these challenges, we propose a paradigm shift. Rather than computing discrete expectation values, we construct continuous distributions of the CGLMP observable and perform standard hypothesis tests to distinguish entangled from separable states. 
Hypothesis tests based on the full distribution shape are less sensitive to individual extreme events than expectation values, and naturally incorporate background models through simultaneous fitting of signal and background components.
Profile likelihood methods implemented in tools such as RooFit~\cite{RooFit} provide a mature, well-validated framework for hypothesis testing with systematic uncertainties, and the output is a $p$-value or significance for rejecting the null hypothesis (separable state), rather than an expectation value whose statistical interpretation requires additional assumptions.
The fit model and treatment of systematic uncertainties are detailed in Appendix~\ref{app:fit}.

\subsection{The Neutrino Reconstruction Challenge}

A fundamental obstacle to quantum state tomography in $H \to WW^* \to \ell\nu\ell\nu$ is the presence of two invisible neutrinos in the final state. Complete reconstruction of the $W$ boson rest frames, essential for projecting onto the spin basis, requires knowledge of all four neutrino momentum components ($p_x$, $p_y$, $p_z$, and $E$) for each neutrino.

The experimental signature provides only partial information. The missing transverse momentum $\vec{p}_T^{\text{miss}}$ constrains the sum of the neutrino transverse momenta, but the individual neutrino momenta and their longitudinal components remain undetermined. Traditional analytical approaches~\cite{aben} exploit kinematic constraints such as mass shell conditions and missing momentum balance but introduce model-dependent assumptions and fail for a significant fraction of events.

Machine learning approaches have recently shown promise for neutrino reconstruction in similar channels. The study of entanglement in $\tau^+\tau^-$ production~\cite{Zhang:2025mmm} demonstrated that diffusion-based models can significantly outperform traditional methods for reconstructing neutrino kinematics, achieving mass resolutions of approximately 6\% compared to approximately 20\% for the Missing Mass Calculator.

In this work, we employ conditional denoising diffusion probabilistic models (cDDPM)~\cite{Pazos:2024} for neutrino reconstruction. This approach offers a crucial advantage over alternative methods such as OmniFold~\cite{OmniFold}. The diffusion model operates on individual events and can be applied directly to the measured dataset, including background contributions, without requiring truth-level labels during inference. This enables the unfolding to be incorporated naturally into the analysis chain, with the hypothesis test performed on fully reconstructed events. The methodology and performance comparisons are presented in Section~\ref{sec:reconstruction}. Appendix~\ref{app:unfolding} provides a comprehensive assessment of alternative approaches, demonstrating that classical unfolding of the analytically-reconstructed $\mathfrak{I}_3$ has zero coverage due to the near-total loss of per-event information, that OmniFold cannot handle the multi-background environment, and documenting the cDDPM architecture and hyperparameter optimisation in detail.

\subsection{Paper Organization}

The remainder of this paper is organized as follows. Section~\ref{sec:theory} reviews the theoretical framework for quantum state tomography in $WW^*$ systems and introduces our continuous formulation of the CGLMP observable. Section~\ref{sec:simulation} describes the Monte Carlo simulation and detector modeling. Section~\ref{sec:reconstruction} presents the neutrino reconstruction methods, comparing analytical and diffusion-based approaches. Section~\ref{sec:hypothesis} develops the hypothesis testing framework using profile likelihood methods. Section~\ref{sec:results} presents the reconstruction performance, Boosted Decision Tree (BDT)-based event selection, systematic hierarchy, and projected sensitivity. Section~\ref{sec:discussion} discusses implications and future directions.

\section{Theoretical Framework}\label{sec:theory}
\subsection{Spin Density Matrix and Polarisation Projections in \texorpdfstring{$WW^*$}{WW*} Decays}

The Higgs boson, being a scalar particle, decays isotropically in its rest frame. However, the spin correlations of its decay products, in this case two spin-1 $W$ bosons, carry information about the quantum structure of the underlying interaction. These bosons serve as polarimeters for the Higgs decay, encoding angular momentum features that may be sensitive to physics beyond the Standard Model. The decay channel $H \to WW^* \to \ell\nu \ell\nu$ (with $\ell = e, \mu$) is particularly suited for such studies, as the parity-violating nature of the $W$ boson decay provides a direct mapping from spin orientation to measurable lepton angular distributions.

In the narrow-width and non-relativistic approximations, the $WW^*$ system from Higgs decay can be represented in the spin basis as~\cite{barr1}
\begin{equation}
|\psi_s\rangle = \frac{1}{\sqrt{3}}\left(|{+}\rangle|{-}\rangle + |{0}\rangle|{0}\rangle + |{-}\rangle|{+}\rangle\right),
\label{eq:singlet_state}
\end{equation}
where $|{+}\rangle$, $|{0}\rangle$, and $|{-}\rangle$ denote the helicity eigenstates in the W rest frames (back-to-back convention, cf. Ref.~[barr1]) with eigenvalues $+1$, $0$, and $-1$ respectively. This maximally entangled state arises from angular momentum conservation in the decay of the spin-0 Higgs boson.

To describe the spin correlations, the quantum state of the $WW^*$ system is expressed through a spin density matrix $\rho$, which captures the probabilities and correlations associated with the spin configurations. Since the $W$ bosons are spin-1 particles, the system resides in a $3 \otimes 3$-dimensional Hilbert space. The density matrix can be expanded in the basis of Gell-Mann matrices $\lambda_i$, which generate the $\mathrm{SU}(3)$ algebra~\cite{barr2,Bertlmann:2008}:
\begin{adjustbox}{minipage=1.2\columnwidth,width=\columnwidth}
    \begin{align}
\rho = \frac{1}{9} \mathbb{I}_3 \otimes \mathbb{I}_3 
+ \sum_{i=1}^8 a_i \lambda_i \otimes \mathbb{I}_3 
+ \sum_{j=1}^8 b_j \mathbb{I}_3 \otimes \lambda_j 
+ \sum_{i,j=1}^8 c_{ij} \lambda_i \otimes \lambda_j \,.
\label{eq:density_matrix}
\end{align}
\end{adjustbox}

Here, the vectors $a_i$ and $b_j$ represent the single-particle polarisation of each $W$ boson, while the matrix $c_{ij}$ encodes their spin-spin correlations. The eight Gell-Mann matrices $\lambda_i$ satisfy the trace orthogonality relation $\mathrm{Tr}(\lambda_i \lambda_j) = 2\delta_{ij}$, which enables extraction of individual coefficients from experimental data. The coefficients $c_{ij}$ are used as the basis for entanglement-sensitive observables discussed in later sections.

\begin{figure}[ht]
    \centering
    \tikzset{RPY/.code args={#1,#2,#3}{
    \pgfmathsetmacro{\rollangle}{#1}%
    \pgfmathsetmacro{\pitchangle}{#2}%
    \pgfmathsetmacro{\yawangle}{#3}%
    \pgfmathsetmacro{\newxx}{cos(\yawangle)*cos(\pitchangle)}
    \pgfmathsetmacro{\newxy}{sin(\yawangle)*cos(\pitchangle)}
    \pgfmathsetmacro{\newxz}{-sin(\pitchangle)}
    \path (\newxx,\newxy,\newxz);
    \pgfgetlastxy{\nxx}{\nxy};
    \pgfmathsetmacro{\newyx}{cos(\yawangle)*sin(\pitchangle)*sin(\rollangle)-sin(\yawangle)*cos(\rollangle)}
    \pgfmathsetmacro{\newyy}{sin(\yawangle)*sin(\pitchangle)*sin(\rollangle)+ cos(\yawangle)*cos(\rollangle)}
    \pgfmathsetmacro{\newyz}{cos(\pitchangle)*sin(\rollangle)}
    \path (\newyx,\newyy,\newyz);
    \pgfgetlastxy{\nyx}{\nyy};
    \pgfmathsetmacro{\newzx}{cos(\yawangle)*sin(\pitchangle)*cos(\rollangle)+ sin(\yawangle)*sin(\rollangle)}
    \pgfmathsetmacro{\newzy}{sin(\yawangle)*sin(\pitchangle)*cos(\rollangle)-cos(\yawangle)*sin(\rollangle)}
    \pgfmathsetmacro{\newzz}{cos(\pitchangle)*cos(\rollangle)}
    \path (\newzx,\newzy,\newzz);
    \pgfgetlastxy{\nzx}{\nzy};
    \pgfkeysalso{%
      /tikz/x={(\nxx,\nxy)},
      /tikz/y={(\nyx,\nyy)},
      /tikz/z={(\nzx,\nzy)}
    }
  }
}

\tdplotsetmaincoords{50}{120} 
\begin{tikzpicture}[tdplot_main_coords, scale=1.3]
    \begin{scope}[canvas is xy plane at z=0]
      \fill[gray,opacity=0.3] (-1.5,-1.5) rectangle (1.5,1.5);
      \draw[thick,->,shorten >=5pt] (-3,0) -- (0,0) node[pos=0.4, below]{$\hat{p}$};
      \draw[thick,->,shorten >=5pt] ( 3,0) -- (0,0) node[pos=0.4, below]{$\hat{p}$};
    \end{scope}
    
    \begin{scope}[RPY={0,50,-30}]
      \fill[gray,opacity=0.3] (0,1) coordinate (V1) -- ++ (1,1) -- ++ (-2,0) -- cycle;
      \draw[->] (V1) -- ++ ( 0.5,.8) node[pos=1.2] {$\ell$};
      \draw[->] (V1) -- ++ (-0.5,.8) node[pos=1.1] {$\nu$};
      \draw[thick,red,->] (0,1) -- ++ (.75,0,0) node [pos=1.2]{$\hat{r}$};
      \draw[thick,red,->] (0,1) -- ++ (0,.75,0) node [pos=1.2]{$\hat{k}$};
      \draw[thick,red,->] (0,1) -- ++ (0,0,.75) node [pos=1.2]{$\hat{n}$};                      
    \end{scope}
    \begin{scope}[RPY={0,30,125}]
      \fill[gray,opacity=0.3] (0,1) coordinate (V2) -- ++ (1,1) -- ++ (-2,0) -- cycle;
      \draw[->] (V2) -- ++ ( 0.5,.8) node[pos=1.1] {$\ell$};
      \draw[->] (V2) -- ++ (-0.5,.8) node[pos=1.1] {$\nu$};      
      \draw[thick,red,->] (0,1) -- ++ (.75,0,0) node [pos=1.2]{$\hat{r}$};
      \draw[thick,red,->] (0,1) -- ++ (0,.75,0) node [pos=1.2]{$\hat{k}$};
      \draw[thick,red,->] (0,1) -- ++ (0,0,.75) node [pos=1.2]{$\hat{n}$};                
    \end{scope}

    \draw[dashed] (0,0) -- (V1) node[midway,left] {$V_1$};
    \draw[dashed] (0,0) -- (V2) node[midway,left] {$V_2$};
    
\end{tikzpicture}
    \caption{Visualization of coordinate systems used for spin analysis. The basis vectors $\{\mathbf{\hat{n}}, \mathbf{\hat{r}}, \mathbf{\hat{k}'}\}$ define the reference frame in each $W$ boson rest frame.}
    \label{fig:coordinate-systems}
\end{figure}
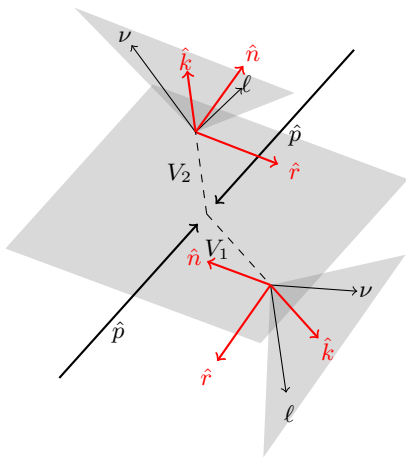

To connect this formalism to experimentally accessible observables, the spin degrees of freedom must be expressed in terms of angular distributions of the decay leptons. Each $W$ boson decay defines a coordinate system in its rest frame. For this purpose, the direction of flight of each $W$ boson in the $H$ rest frame is denoted as $\mathbf{\hat{k}}$, while $\mathbf{\hat{p}}$ represents the direction of the beam in the same frame. This is supplemented with the definition of $\mathbf{\hat n}$ as the vector orthogonal to the plane in which the leptons decay, and $\mathbf{\hat r}$ as the vector orthogonal to both $\mathbf{\hat{k}}$ and $\mathbf{\hat n}$ as shown in Figure~\ref{fig:coordinate-systems}.

After applying a Lorentz boost into the rest frame of each $W$ boson, this defines a new orthonormal basis, given by the set of vectors $\{\mathbf{\hat{x}}, \mathbf{\hat{y}}, \mathbf{\hat{z}}\} = \{\mathbf{\hat{n}}, \mathbf{\hat{r}}, \mathbf{\hat{k}'}\}$.
In this frame, the lepton decay directions can be described by their polar and azimuthal angles $(\theta, \phi)$, measured with respect to the spin basis. These angular distributions reflect the polarization states of the $W$ bosons and thus carry information about the spin correlations of the $WW^*$ system.

Previous studies of angular observables in $H \to WW^*$ decays, including those focused on CP and spin-parity structure~\cite{aben,ATLAS:2021pkb}, have employed similar kinematic reconstructions. However, those analyses targeted specific coupling hypotheses. In contrast, the present approach aims to reconstruct the full spin density matrix and test entanglement through the angular structure of the final-state leptons.

\subsection{Study of Entanglement}

Since the $W$ bosons are spin-1 particles, their spin states are appropriately described by the $\mathrm{SU}(3)$ group, whose generators are the Gell-Mann matrices $\lambda_i$. To extract the components of the spin density matrix associated with these generators from the angular decay distributions, the Wigner-Weyl transformation is employed~\cite{barr2}. This transformation establishes a correspondence between quantum mechanical operators and phase-space functions. In particular, each Gell-Mann matrix is mapped to a Wigner $P$ symbol $\Phi_i^{P\pm}$, where the superscripts $+$ and $-$ refer to the decay products of the positively and negatively charged $W$ bosons, respectively.

The $W$ boson decays maximally violate chirality: a $W^+$ boson decay preferentially emits a charged lepton along the $W^+$ spin direction, while a $W^-$ boson decay preferentially emits a charged lepton moving against its spin direction~\cite{barr1}. This means that decaying $W$ bosons are ``their own polarimeters'', with each decay causing a spin measurement to be made along the axis of the emitted lepton.

The key feature of the Wigner-Weyl transformation is that it preserves expectation values: the quantum mechanical trace $\mathrm{Tr}(\rho \, \lambda_i \otimes \lambda_j)$ corresponds to the average of the product of the associated Wigner $P$ symbols over the angular phase space. Working within the orthonormal basis established earlier ensures that the emission angles are defined consistently across events, allowing for a meaningful projection onto the spin operators.

The Wigner $P$ symbols are functions of the lepton emission direction cosines in the $W$ boson rest frames:
\begin{equation}
\xi_x = \sin\theta\cos\phi, \quad \xi_y = \sin\theta\sin\phi, \quad \xi_z = \cos\theta,
\label{eq:direction_cosines}
\end{equation}
where $(\theta, \phi)$ are the polar and azimuthal angles of the charged lepton in the corresponding $W^\pm$ rest frame with respect to the basis $\{\mathbf{\hat{x}}, \mathbf{\hat{y}}, \mathbf{\hat{z}}\}$.

The spin-spin correlation coefficients $c_{ij}$ in the parameterization of the spin density matrix can thus be extracted as
\[
    c_{ij} = \frac{1}{4}\left\langle \Phi_i^{P+} \Phi_j^{P-} \right\rangle_{\text{avg}}.
\]
Here, the average $\langle \cdots \rangle_{\text{avg}}$ is taken over the ensemble of decay events. This averaging, however, requires the events to follow a distribution with a well-defined mean, an assumption that will later be shown not to be obvious in all cases.

\begin{figure}[h]
    \centering
    \includegraphics[width=0.5\textwidth]{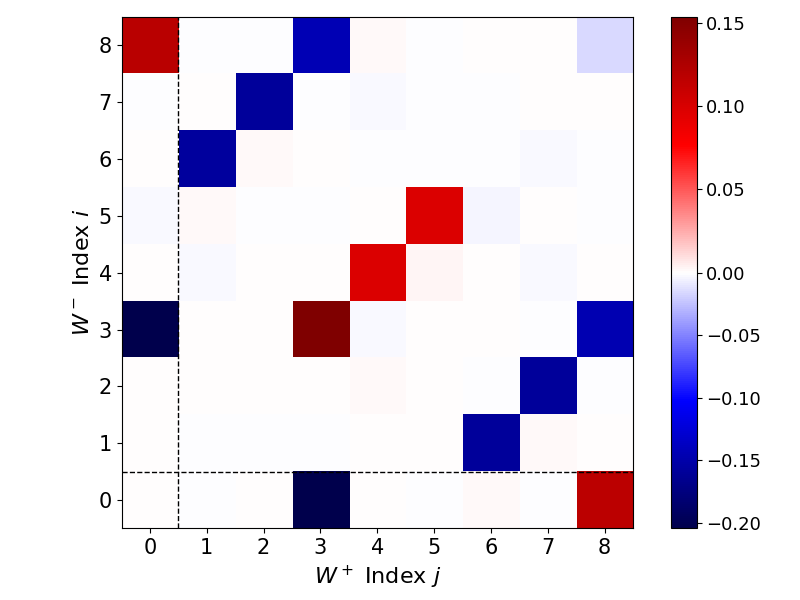}
    \caption{Gell-Mann correlation parameters estimated from the MC simulated $W^\pm$ bosons. The marked first column and row (excluding the meaningless (0, 0) element) represent the Bloch vectors of the individual $W^\pm$ bosons.}
    \label{fig:GMmatrix_truth}
\end{figure}

To conveniently visualize the coefficients, these ensemble averages are presented in a two-dimensional heatmap. In \autoref{fig:GMmatrix_truth}, the resulting histogram of the MC generated data (described in section~\ref{sec:simulation}) is shown. This agrees with the findings of Barr et al.~\cite{barr1,barr2,barr3}.

\begin{figure}[h]
    \centering
    \includegraphics[width=0.5\textwidth]{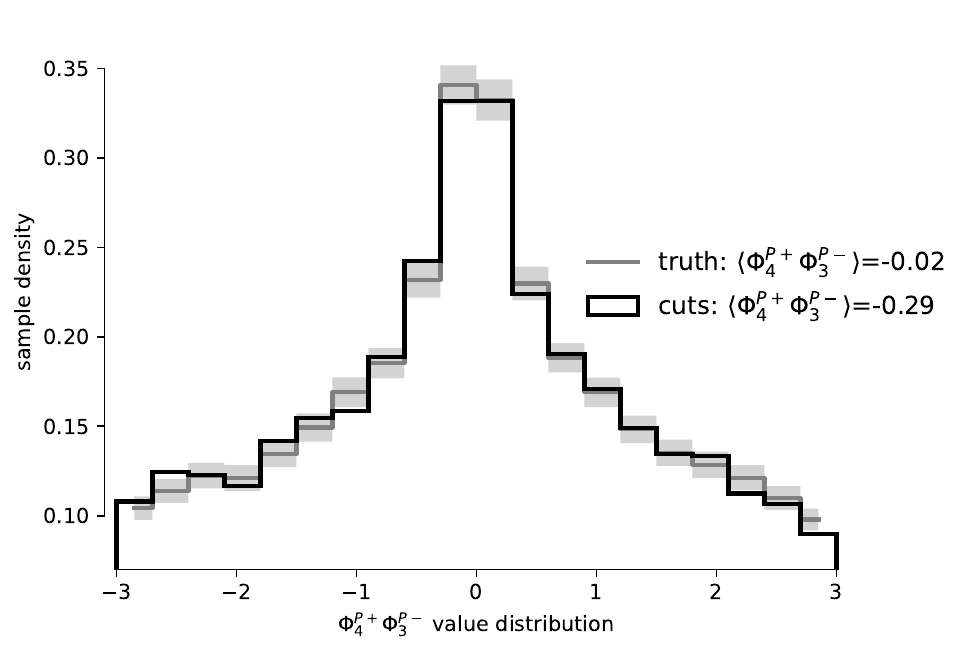}
    \caption{Distribution of the per-event product $\Phi^{P+}_4\,\Phi^{P-}_3$ whose mean defines the Gell-Mann coefficient via $c_{43} = \tfrac{1}{4}\langle\Phi^{P+}_4\,\Phi^{P-}_3\rangle$. Grey: truth-level (no cuts); black: with a minimal lepton fiducial selection applied (one lepton $p_T>22$~GeV, the other $p_T>10$~GeV, both within the detector acceptance). Despite the distributions being functionally identical, the per-event mean shifts from $\langle\Phi^{P+}_4\Phi^{P-}_3\rangle = -0.02$ to $-0.29$, illustrating the heavy-tail bias that operator-style expectation values pick up from selection cuts (see Appendix~\ref{app:ggm}). Among all $c_{ab}$ correlation entries this is the largest single-element shift; the same effect appears in every other off-diagonal entry to varying degree. The light-grey envelope is the $\pm 1\sigma$ bootstrap band on the truth histogram (1000 resamples of the expected number of Higgs events at 139~fb$^{-1}$.}
    \label{fig:operator_shift}
\end{figure}

 It should be noted, however, as described in appendix \ref{app:ggm}, that the reconstruction of this matrix from the ensemble averages is highly dependent on the measured distribution. In practice the distributions described in ~\cite{barr1,barr2,barr3} tend towards a cauchy distribution rather than a normal distribution assumed in the literature, meaning that the expectation value is highly unstable. This is demonstrated in figure \ref{fig:operator_shift} as even though the distributions look very similar under selection criteria the expectation value varies dramatically.
 
Though outside of the scope of this study, it should be noted that the Cauchy-like form of these distributions is reminiscent of the Breit-Wigner lineshape that governs resonant scattering. Recent work has shown that maximal entanglement can serve as a foundational principle capable of reproducing gauge theory structures~\cite{CerveraLierta:2017} and motivates further study into the possibility that the statistical properties of spin correlation observables may be deeply connected to the quantum information content of the underlying interaction.

\subsection{Tests of Quantum Entanglement with the CGLMP Inequality}

The coefficients $c_{ij}$ can be combined into observables sensitive to quantum entanglement. In particular, the Bell operator $\hat{B}_{\text{CGLMP}}^{xy}$ is constructed as a specific linear combination of tensor products of Gell-Mann matrices, following the CGLMP  framework for testing Bell inequalities in three-level (qutrit) systems~\cite{CGLMP,barr1}. 

For pairs of spin-1 particles like the $WW^*$ system, the CGLMP inequality is optimal for detecting Bell-nonlocality.
While a generalised CHSH inequality for spin-1 systems assigns measurement outcomes as the eigenvalues $\{+1, 0, -1\}$ of the spin operators, the additional zero outcome dilutes the expectation values and tends to decrease violation~\cite{barr1}.
By contrast, the CGLMP inequality is designed specifically for qutrit systems and is near-maximally violated by the $H \to WW^*$ state.

The CGLMP inequality states that for any local hidden variable theory~\cite{CGLMP}:
\begin{equation}
\mathfrak{I}_3 \leq 2.
\label{eq:cglmp_bound}
\end{equation}
Quantum mechanics allows values up to $\mathfrak{I}_3 \approx 2.87$ for maximally entangled qutrits, with the $H \to WW^*$ system predicted to achieve $\mathfrak{I}_3 \approx 2.6$--$2.8$ depending on the kinematic selection~\cite{barr1}.

The expectation value of the Bell operator can be directly expressed in terms of the reconstructed spin correlation coefficients:
\begin{equation*}
    \begin{aligned}
        \mathfrak{I}_3 &= \langle \hat{B}_{\text{CGLMP}}^{xy} \rangle = 4(c_{44} + c_{55}) \\
                    &- 4\sqrt{\frac{2}{3}}(c_{11} + c_{16} + c_{66} + c_{61} + c_{22} + c_{27} + c_{77} + c_{72}).
    \end{aligned}
\end{equation*}

\begin{figure}[h]
    \centering
    \includegraphics[width=0.5\textwidth]{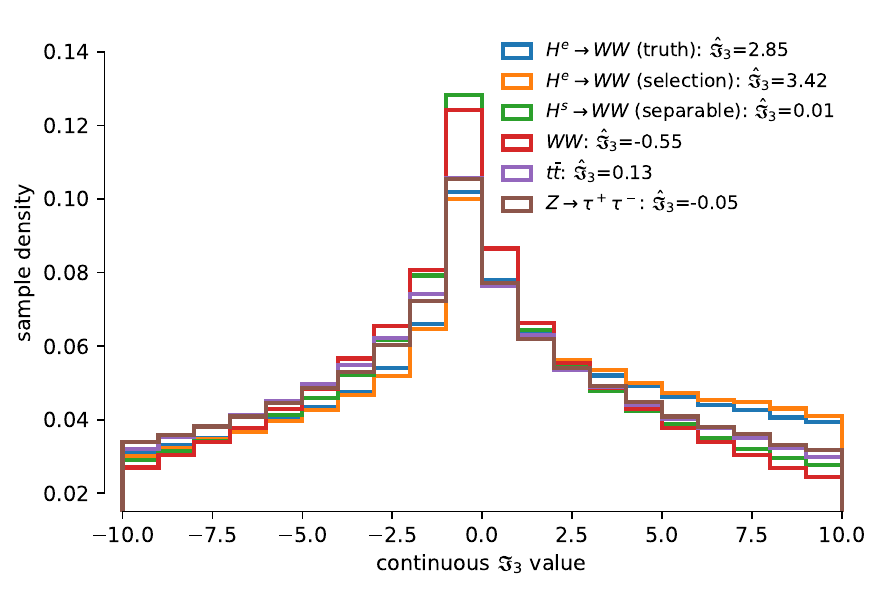}
    \caption{Continuous CGLMP distribution for various processes. Where $H^e$ denotes the fully entangled state and $H^s$ a fully separable state constructed from separating and randomising the on-shell W bosons in simulated Higgs events. The expectation value of this distribution (denoted $\hat{\mathfrak{I}}_3$ ) is identical to that derived from the components of Generalised Gell-Mann (GGM) matrix. The selection-dependent shift from 2.85 to 3.42 illustrates the heavy-tail sensitivity of the operator expectation value discussed in Sec. II.B and Appendix C, and motivates the shape-based hypothesis test we adopt below.}
    \label{fig:continous_cglmp}
\end{figure}

Following the effect demonstrated in figure \ref{fig:operator_shift} this formula can be constructed from the continuous distributions directly rather than as a product of expectation values.
This departure from the traditional constructions as in equation \ref{eq:cglmp_bound}, allows us to instead construct a pseudo continuous Bell\emph{-like} test statistic on which traditional hypothesis testing of quantum entanglement can be applied.
This pseudo continuous distribution can be seen in figure \ref{fig:continous_cglmp}. It should be noted the distinct shape difference between entangled processes and non-entangled is clearly visible regardless of various reconstruction effects especially when the tails of the distribution are included.

The methodology for extracting correlation coefficients and testing Bell inequalities from the spin density matrix extends naturally to other bipartite systems involving spin-1 particles, including $H \to ZZ^*$, $t\bar{t}$ production, and diboson processes~\cite{barr2,barr3}. 
The Gell-Mann parametrisation provides a universal framework for quantum state tomography at colliders including test of new physics as explored in appendix \ref{app:smeft}.

Within the measurement context of the LHC the $\mathfrak{I}_3 > 2$ inequality serves as a separating quantity between entangled and separable density-matrix hypotheses; loopholes that affect a strict interpretation as a model-independent test of local realism (detection efficiency, locality, freedom-of-choice) are not closed by collider measurements, and we consequently frame the result as evidence of quantum entanglement rather than as a loophole-free Bell test.
\section{Simulation}\label{sec:simulation}

\subsection{Signal Generation}

We generate $pp \to H \to WW^* \to \ell\nu\ell\nu$ events at $\sqrt{s} = 13$~TeV using \texttt{MadGraph5\_aMC@NLO} version 3.4.2~\cite{MadGraph}. The Higgs boson is produced via gluon fusion, which dominates the production cross section at the LHC. Parton showering and hadronization are performed with Pythia8~\cite{Pythia8}. The signal sample consists of $5 \times 10^5$ events prior to selection.

Detector effects are simulated using DELPHES 3.5.0~\cite{DELPHES} with the default ATLAS card. This provides a realistic treatment of lepton reconstruction efficiency, momentum resolution, and missing transverse energy reconstruction. We validate the DELPHES simulation against a smaller sample generated with Sherpa~\cite{Sherpa}.

\subsection{Event Selection}

The event selection is designed to minimize bias on the angular distributions while maintaining adequate background rejection. Events are required to contain exactly two opposite-sign leptons (electrons or muons) with leading lepton $p_T > 22$~GeV, subleading lepton $p_T > 15$~GeV, and both leptons within $|\eta| < 2.5$. We deliberately avoid selection criteria based on angular correlations between decay products, as these would directly bias the observables of interest.

After selection, the signal sample contains 286,026 events, corresponding to an efficiency of approximately 29\%. The impact of selection cuts on the Gell-Mann coefficient extraction is studied in Appendix~\ref{app:ggm}.

\subsection{Background Processes}

In the dilepton final state, several background processes contribute significantly. Non-resonant $WW$ production constitutes an irreducible background with the same final state topology. Top quark pair production $t\bar{t}$ and single top production in the $Wt$ channel produce $WW$ pairs from top decays, accompanied by $b$-jets that can fail identification. The $Z/\gamma^* \to \tau\tau$ process contributes when both $\tau$ leptons decay leptonically.

The relative composition and kinematic properties of these backgrounds are modified by the selection requirements. Following the treatment in ATLAS measurements of $H \to WW^* \to e\nu\mu\nu$~\cite{ATLASHWWdiff}, we normalise each background component to the theoretical cross section at the appropriate order in perturbation theory ($WW$ at NNLO, $t\bar{t}$ at NNLO+NNLL, $Z{\to}\tau\tau$ at NLO) and validate data-simulation agreement in dedicated control regions.

\begin{figure}[h]
    \centering
    \includegraphics[width=0.48\textwidth]{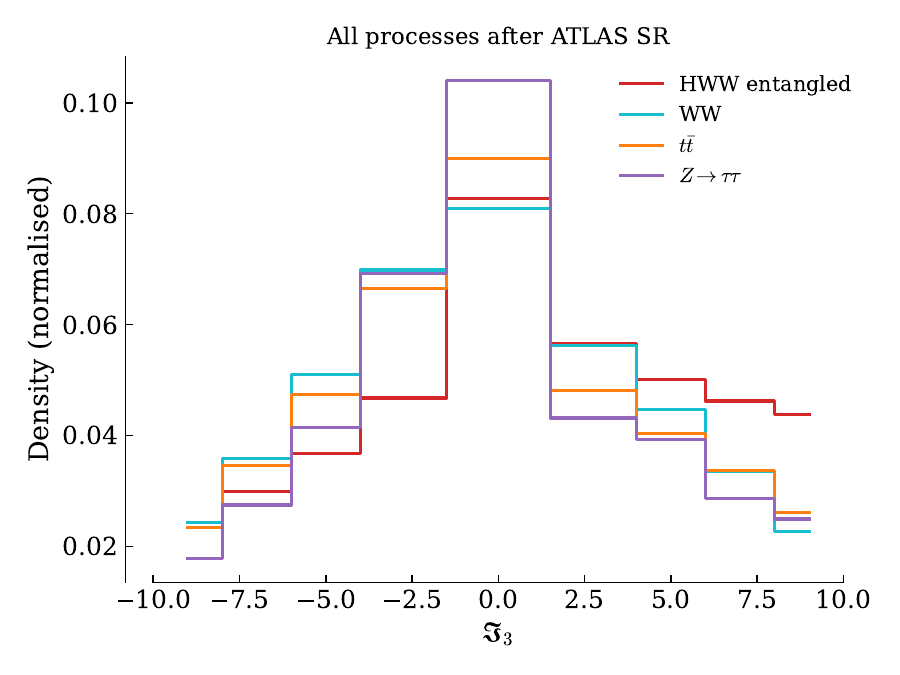}
    \caption{Normalised $\mathfrak{I}_3$ distributions for signal and each background process after the ATLAS signal region selection. The HWW entangled signal has broader tails than the background processes, providing the discrimination power for the hypothesis test.}
    \label{fig:backgrounds}
\end{figure}

The background composition after the ATLAS signal region selection is summarised in Table~\ref{tab:cutflow}. The $\mathfrak{I}_3$ distributions for each background process are shown in Figure~\ref{fig:backgrounds}. Non-resonant WW production dominates the background, with a shape similar to the entangled signal, constituting the irreducible component. The $t\bar{t}$ and $Z{\to}\tau\tau$ contributions have distinct shapes that are more centrally peaked.

\begin{table}[h]
    \centering
    \caption{Expected event yields at 139~fb$^{-1}$ before and after the ATLAS signal region selection.}
    \label{tab:cutflow}
    \begin{tabular}{l r r r}
        \toprule
        Process & Pre-selection & After SR & Efficiency \\
        \midrule
        $H{\to}WW^*$ & 4,000 & 768 & 19.2\% \\
        WW & 95,000 & 8,265 & 8.7\% \\
        $t\bar{t}$ & 150,000 & 7,800 & 5.2\% \\
        $Z{\to}\tau\tau$ & 185,000 & 1,110 & 0.6\% \\
        \midrule
        Total background & 430,000 & 17,175 & 4.0\% \\
        \midrule
        S/B & 0.9\% & 4.5\% & \\
        \bottomrule
    \end{tabular}
\end{table}

\section{Neutrino Reconstruction}\label{sec:reconstruction}

Complete reconstruction of the $W$ boson rest frames requires determining the four-momenta of both neutrinos. With only the missing transverse momentum measured, this is an underconstrained problem. We compare two approaches: analytical reconstruction exploiting kinematic constraints, and conditional denoising diffusion for unfolding detector effects.

\subsection{Analytical Reconstruction}

The analytical method exploits the constraint that the visible decay products and neutrinos must reconstruct to the Higgs boson mass. Assuming the $W$ boson masses and imposing the missing transverse momentum constraint yields a system of polynomial equations that can be solved for the neutrino longitudinal momenta. Following Ref.~\cite{aben}, we implement this procedure and select the solution minimizing the sum of reconstructed $W$ masses.

This method fails for a substantial fraction of events, approximately 35\%, where no real solution exists due to resolution effects and the off-shell $W$ boson. For these events, we adopt the procedure of minimizing the constraint violation. The algorithm is explained in more detail in Appendix \ref{app:ana}. The analytical method provides a baseline with no training requirements but suffers from significant inefficiency and model dependence.

A Deep feed-forward Neural Network (DNN) was optimized for the given dataset and found to be a possible alternative to the analytical approach~\cite{vak_2025}.

\subsection{Conditional Denoising Diffusion Probabilistic Models}

Correcting for detector effects in experimental data through unfolding is critical for enabling precision measurements. Traditional unfolding methods face challenges in scalability, flexibility, and dependence on simulations. Following Pazos et al.~\cite{Pazos:2024}, we employ conditional denoising diffusion probabilistic models (cDDPM) for multidimensional, object-wise unfolding.

The cDDPM approach utilizes a non-iterative, flexible posterior sampling method. The forward diffusion process gradually corrupts truth-level quantities $\vec{x}$ with Gaussian noise over $T$ timesteps, yielding a sequence of increasingly noisy versions $\vec{x}_1, \vec{x}_2, \ldots, \vec{x}_T$. The reverse process, parametrized by a neural network $\epsilon_\theta$, learns to denoise conditioned on the detector-level measurements $\vec{y}$:
\begin{equation}
p_\theta(\vec{x}_{t-1}|\vec{x}_t, \vec{y}) = \mathcal{N}(\vec{x}_{t-1}; \mu_\theta(\vec{x}_t, t, \vec{y}), \sigma_t^2 \mathbb{1}).
\end{equation}

A key innovation of the cDDPM approach is the incorporation of distribution moments as conditioning information, which exhibits a strong inductive bias that allows generalization to unseen physics processes without explicitly assuming the underlying distribution~\cite{Pazos:2024}. 
We condition on the detector-level lepton 4-vectors and missing transverse momentum together with an explicit one-hot process tag; an earlier variant that used $p_\mathrm{T}$ moments as a soft tag is documented in Appendix \ref{app:cddpm_systematics} along with its failure mode.

The critical advantage of the cDDPM over alternative unfolding approaches, such as OmniFold, is that it operates on individual events. During inference, the model requires only detector-level measurements without truth-level labels, enabling its application to the full measured dataset, including background contributions. This property is essential for our application, where backgrounds must be included in the hypothesis test. By contrast, OmniFold and related iterative reweighting methods require truth-level information for all events to be unfolded, which is incompatible with the multi-background environment of Higgs analyses. Furthermore, as shown in Appendix~\ref{app:unfolding}, the analytical neutrino reconstruction produces $\mathfrak{I}_3$ values that are essentially uncorrelated with truth ($r = 0.08$), rendering classical binned unfolding of the 1D $\mathfrak{I}_3$ distribution ineffective for hypothesis testing despite formally recovering the marginal distribution. The cDDPM resolves this by targeting the helicity angles directly, initialising from the analytical solution and learning to correct the residual distortions.

\section{Hypothesis Testing Framework}\label{sec:hypothesis}

The central question of this analysis is whether the measured continuous CGLMP distribution is more consistent with an entangled or a separable quantum state. We address this through a binned likelihood ratio test, treating the fully separable configuration as the null hypothesis.

The expected number of events in bin $i$ is modelled as
\begin{equation}
\nu_i(f, \theta) = f \cdot s_i^{\text{ent}}(\theta) + (1{-}f) \cdot s_i^{\text{sep}}(\theta) + \sum_j b_{ij}(\theta),
\end{equation}
where $s_i^{\text{ent}}$ and $s_i^{\text{sep}}$ are the entangled and separable signal templates, $b_{ij}$ represents the background contributions from process $j$, and $\theta$ denotes the nuisance parameters. The parameter of interest $f \in [0,1]$ interpolates between the separable ($f{=}0$) and entangled ($f{=}1$) hypotheses, while the total signal yield is held fixed.

The separable template is constructed by randomly reassigning the on-shell and off-shell $W$ bosons between events, destroying quantum correlations while preserving the marginal angular distributions. This ``mangled'' sample is validated by confirming that its Gell-Mann correlation matrix $c_{ij}$ is consistent with zero.

The templates for signal and each background process are constructed as bootstrapped histograms of the continuous $\mathfrak{I}_3$ observable in 9 variable-width bins over $[-10, 10]$, with a wide central bin $[-2, 2]$ that absorbs the cDDPM zero-peak artefact (Section~\ref{sec:results}). Background normalisation uncertainties are treated as independent nuisance parameters with Gaussian constraints: 20\% for WW, 10\% for $t\bar{t}$, and 15\% for $Z{\to}\tau\tau$, following the treatment in Ref.~\cite{ATLAS:2025hki}. Per-bin statistical uncertainties are incorporated through Barlow--Beeston lite~\cite{histfactory} Poisson constraints on gamma parameters.

The model is implemented using the HistFactory~\cite{histfactory} formalism within RooFit~\cite{RooFit}, using the \texttt{HistoToWorkspaceFactoryFast} interface. The simple likelihood ratio test statistic is employed for frequentist hypothesis testing via the \texttt{FrequentistCalculator}, with binned toy generation. Further details of the likelihood construction and systematic treatment are given in Appendix~\ref{app:fit}, with exhaustive robustness studies in Appendix~\ref{app:fit_study}.

\section{Results}\label{sec:results}

\subsection{Diffusion-Based Reconstruction}

The cDDPM v3 model is trained on a balanced mixture of HWW signal and the dominant backgrounds (WW, $t\bar{t}$, $Z\to\tau\tau$), using an explicit one-hot process tag and 6-dimensional boosted-frame helicity-angle targets (Appendix~\ref{app:unfolding}). The design evolved through three iterations: a signal-only model (v1) that hallucinated entangled neutrinos for every background event, a mixed-training model with soft $p_T$-moment conditioning (v2) that the network learned to ignore, and the production v3 with an explicit process tag that the network cannot ignore. Table~\ref{tab:cddpm_variants} in Appendix~\ref{app:unfolding} documents the bias reduction across variants.

\begin{figure}[h]
    \centering
    \includegraphics[width=0.48\textwidth]{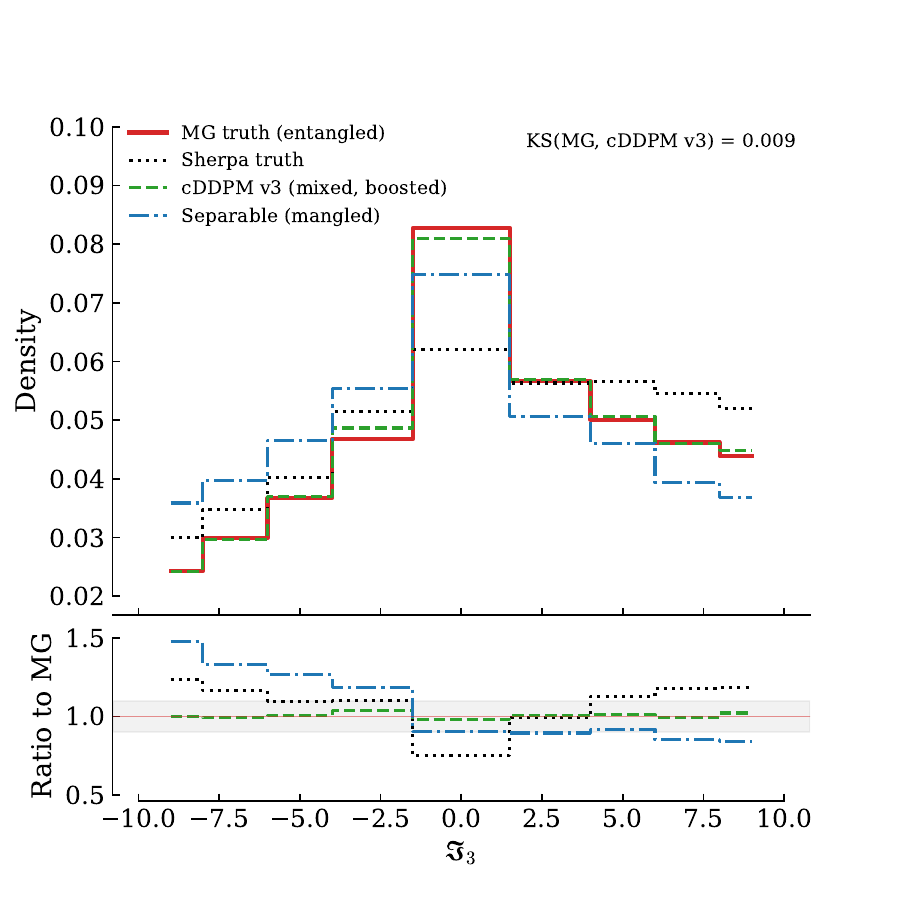}
    \caption{Top: normalised $\mathfrak{I}_3$ density for MadGraph truth, Sherpa truth, cDDPM reconstruction, and the separable (mangled) template. Bottom: ratio to MadGraph truth, with a $\pm 10\%$ band.}
    \label{fig:truth_vs_reco}
\end{figure}

Figure~\ref{fig:truth_vs_reco} compares the reconstructed $\mathfrak{I}_3$ distribution to truth for the signal. The MadGraph and Sherpa generators agree at the sub-percent level, validating generator independence. The cDDPM introduces a residual shape distortion concentrated near $\mathfrak{I}_3 = 0$; the tail asymmetry that drives the entangled--separable discrimination is preserved. We adopt a 9-bin variable-width binning with a wide central bin $[-2, 2]$ to absorb this artefact.

\begin{figure*}[ht]
    \centering
    \includegraphics[width=\textwidth]{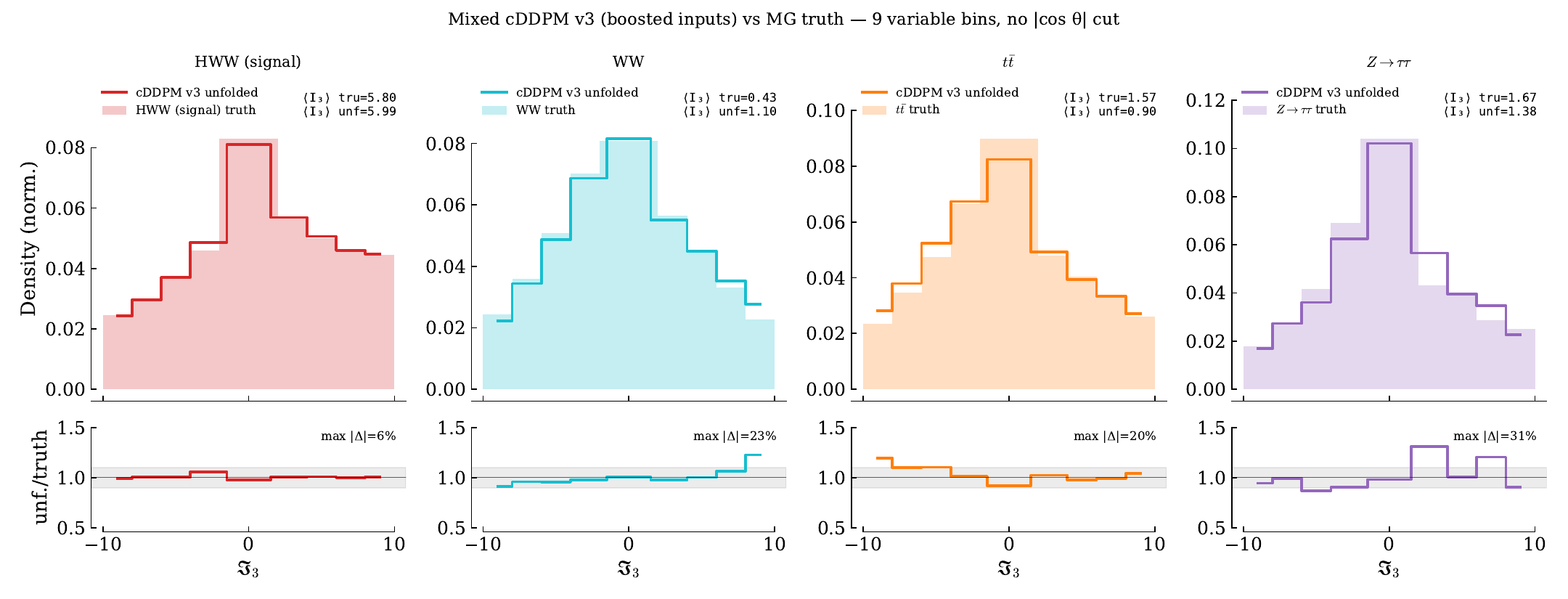}
    \caption{Per-process closure of the cDDPM v3 reconstruction. Top: density histograms (truth shaded, unfolded stepped). Bottom: per-bin unfolded/truth ratio with $\pm 10\%$ band; the maximum residual is annotated. These residuals define the unfolding shape systematic in the fit.}
    \label{fig:unfolded_vs_truth}
\end{figure*}

Figure~\ref{fig:unfolded_vs_truth} shows the per-process closure: the unfolded $\mathfrak{I}_3$ shape is compared to truth for each of the four processes individually. Under the production cDDPM (standard sampling; see Appendix~\ref{app:cddpm_warmstart}) the per-bin residuals are rms 2\% on the HWW signal template, with per-bin residuals rms $\sim$10--15\% on the backgrounds dominated by their low-statistics tail bins. These residuals enter the fit as HistoSys nuisance parameters.

\subsection{BDT-Based Event Selection}

The ATLAS signal region selection yields $S/B \approx 4.5\%$, dominated by the irreducible WW background. A BDT-based selection using only non-angular kinematic variables (Appendix~\ref{app:selection}) improves this to 13.9\% while preserving the $\mathfrak{I}_3$ shape (KS~$= 0.016$, max bin residual 4\%). The key finding from the BDT study is that six different classifier architectures (single combined, three split variants, and multiclass) all converge to $S/\sqrt{B} \approx 8.5$--$9.4$ (Table~\ref{tab:bdt_arch} in Appendix~\ref{app:selection}). This ceiling is physics-limited: every feature that discriminates HWW from the irreducible WW background also biases $\mathfrak{I}_3$, because the kinematic differences between Higgs-mediated and non-resonant $WW$ production encode the same spin correlations that define the Bell observable. The BDT improvement comes almost entirely from rejecting the reducible $t\bar{t}$ and $Z{\to}\tau\tau$ backgrounds.

Table~\ref{tab:bdt_cutflow_main} summarises the event yields before and after the BDT selection.

\begin{table}[ht]
    \centering
    \caption{Expected event yields at 139~fb$^{-1}$ after the ATLAS signal region selection and after the additional BDT selection.}
    \label{tab:bdt_cutflow_main}
    \begin{tabular}{l r r r}
    \toprule
    Process & ATLAS SR & SR + BDT & BDT eff. \\
    \midrule
    $H{\to}WW^*$     & 768  & 526  & 68.5\% \\
    WW                & 8{,}265 & 3{,}331 & 40.3\% \\
    $t\bar{t}$        & 7{,}800 & 55   & 0.7\%  \\
    $Z{\to}\tau\tau$   & 1{,}110 & 398  & 35.9\% \\
    \midrule
    Total background   & 17{,}175 & 3{,}784 & 22.0\% \\
    \midrule
    S/B               & 4.5\% & 13.9\% & --- \\
    \bottomrule
    \end{tabular}
\end{table}

\subsection{Systematic Hierarchy and Sensitivity}

The expected sensitivity is computed using the full HistFactory profile likelihood with three classes of nuisance parameter: background normalisations, cDDPM unfolding shape, and a bootstrap-PCA template stat-error model that respects the multinomial bin-to-bin correlations of the cDDPM-unfolded templates (Appendix~\ref{app:fit}, paragraph~\ref{par:bootstrap_pca}). Table~\ref{tab:syst_waterfall} shows how each systematic source degrades the significance from the idealised stat-only case. The dominant degradation arises from background normalisation uncertainties, reflecting the background-dominated regime; the unfolding shape uncertainty is the second-largest contributor; the template stat error is sub-dominant at the production yield once the bin-to-bin correlations are properly modelled.
Additional systematic concerns relevant for a full analysis such as the lepton efficiency and PDF uncertainty are addressed in appendix \ref{app:fit_syst}, and are typically of the order of 2-5\%.
These should have minimal effect on the sensitivity with the exception of the signal cross-section uncertainty in which quantum entanglement might have a large effect on higher order contributions\cite{CerveraLierta:2017} and as such are not investigated here. 

\begin{table}[h]
    \centering
    \caption{Expected Asimov significance for the inclusive selection at 139~fb$^{-1}$ under different systematic configurations. Each row enables only the indicated source; the final row includes all sources simultaneously. 
    The template stat-error model is the production bootstrap-PCA construction in which the leading modes are shape-orthogonal to the signal direction meaning that they have minimal pull on the parameter of interest when profiled. 
    The legacy, per-bin Poisson value is reported for comparison.}
    \label{tab:syst_waterfall}
    \begin{tabular}{l c}
        \toprule
        Configuration & $Z_{\text{Asimov}}$ ($\sigma$) \\
        \midrule
        Stat-only (no NPs)                                   & 5.97 \\
        Template stat (bootstrap-PCA) only                   & 5.97 \\
        Template stat (Barlow-Beeston - for comparison only)              & 4.01 \\
        Unfolding shape only                                 & 2.94 \\
        Background normalisation only                        & 1.33 \\
        All systematics (production)                         & 1.13 \\
        \bottomrule
    \end{tabular}
\end{table}

\subsection{Reducing Systematic Uncertainties}
\label{sec:reduce_systematics}

Table~\ref{tab:syst_waterfall} demonstrates that the dominant limitation
on the expected significance is the background normalisation uncertainty,
which degrades the Asimov significance from the idealised stat-only value
of $5.97\,\sigma$ to $1.33\,\sigma$ at $139\,\text{fb}^{-1}$.
The unfolding shape uncertainty is the second-largest contributor
($2.94\,\sigma$ in isolation). Several concrete strategies can reduce
both sources.

The background normalisation uncertainties (20\% for $WW$, 10\% for
$t\bar{t}$, 15\% for $Z\to\tau\tau$) follow the treatment of
Ref.~\cite{ATLAS:2025hki} and constitute the dominant systematic source.
Dedicated control regions in the same dataset could provide data-driven
constraints on these normalisations simultaneously with the signal fit,
replacing the Gaussian constraints with data-driven ones.
The $t\bar{t}$ contribution could be further constrained through a
$b$-tagged control region orthogonal to the signal region, and the
$Z\to\tau\tau$ contribution through a dedicated control region
exploiting the distinctive $\tau$ decay kinematics.

The cDDPM v3 model achieves per-bin residuals of rms~2\% on the HWW
signal template, with maximum per-bin residuals of 5\% for HWW,
25\% for $WW$, 41\% for $t\bar{t}$, and 33\% for $Z\to\tau\tau$,
as annotated in Figure~\ref{fig:unfolded_vs_truth}. Two observations from
Appendix~\ref{app:unfolding} are relevant. First, the 60k and 120k
training configurations give comparable performance, so that increasing
the training sample size provides only marginal improvement. Second,
the residual $+5\%$ HWW bias reflects an intrinsic limitation of the
reconstruction: even the analytical neutrino solution achieves only
modest angular resolution for the helicity angles that determine $I_3$
(Table~\ref{tab:cddpm_variants}).

If the background normalisation uncertainties can be reduced through
data-driven control regions and the unfolding bias further suppressed,
the sensitivity could improve beyond the projections presented here,
potentially bringing $3\,\sigma$ evidence within reach of the full
Run-3 dataset and $5\,\sigma$ observation at the HL-LHC.

\begin{figure}[h]
    \centering
    \includegraphics[width=0.48\textwidth]{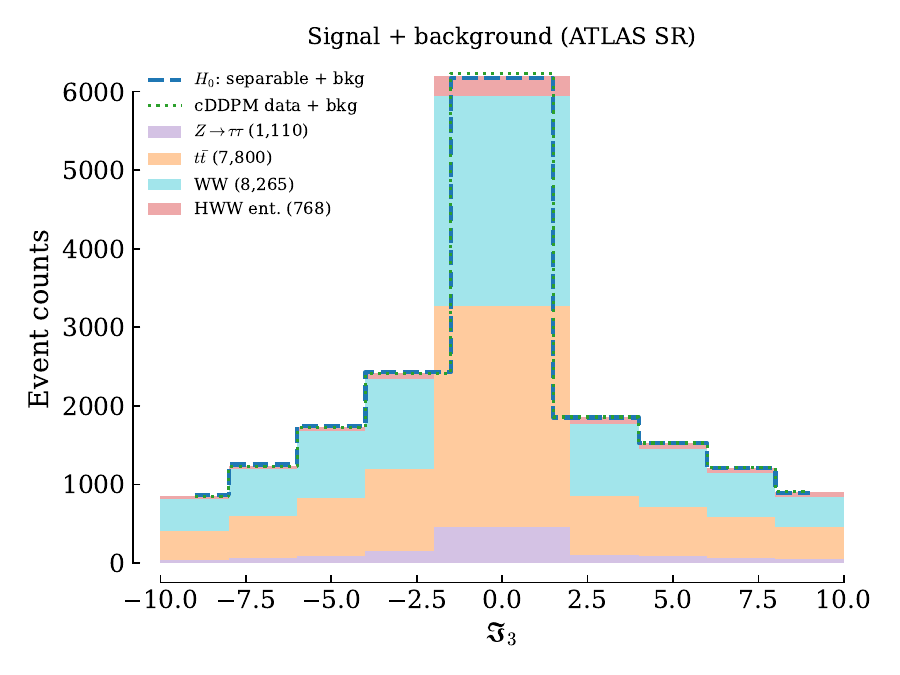}
    \caption{Stacked signal-plus-background $\mathfrak{I}_3$ distribution after ATLAS SR selection at 139~fb$^{-1}$. The entangled signal (red) sits atop $Z{\to}\tau\tau$ (purple), $t\bar{t}$ (orange), and WW (cyan). The separable hypothesis total (blue dashed) is overlaid.}
    \label{fig:stacked}
\end{figure}

Figure~\ref{fig:stacked} shows the signal-plus-background $\mathfrak{I}_3$ composition. The entangled signal sits beneath a factor-of-20 background, underscoring the importance of the profile likelihood approach over simple background subtraction.

\subsection{Expected Sensitivity}

Table~\ref{tab:sensitivity} summarises the expected Asimov significance at several luminosity benchmarks for both the inclusive and BDT-enhanced selections. The BDT improves sensitivity by the equivalent of approximately a factor of three in integrated luminosity. At high luminosities the growth slows below $\sqrt{L}$ scaling, reflecting the systematic floor.

\begin{table}[h]
    \centering
    \caption{Expected Asimov significance $Z$ (in $\sigma$) for rejecting the separable hypothesis at several integrated-luminosity benchmarks, including all systematic uncertainties described in table \ref{tab:syst_waterfall}.}
    \label{tab:sensitivity}
    \begin{tabular}{l c c c c c c}
        \toprule
        Selection & S/B & \multicolumn{5}{c}{$L$ [fb$^{-1}$]} \\
        \cmidrule(lr){3-7}
        & & 139 & 500 & 1000 & 2000 & 3000 \\
        \midrule
        ATLAS SR only & 4.5\%  & 1.6 & 2.0 & 2.6 & 3.4 & 4.0 \\
        SR + BDT      & 13.9\% & 2.3 & 2.9 & 4.0 & 5.5 & 6.7 \\
        \bottomrule
    \end{tabular}
\end{table}

\begin{figure}[h]
    \centering
    \includegraphics[width=0.48\textwidth]{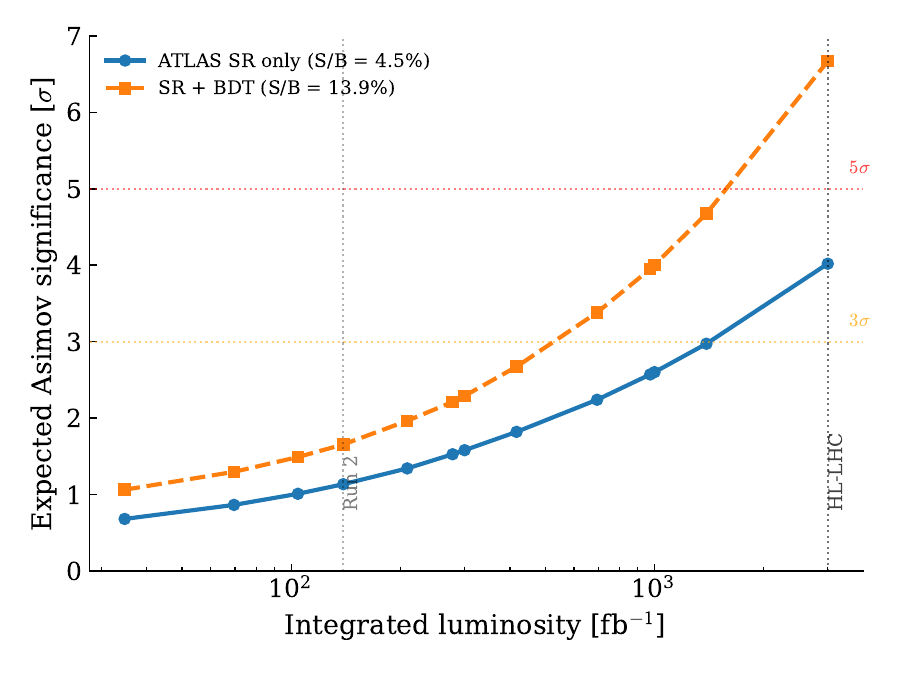}
    \caption{Expected Asimov significance for rejecting the separable hypothesis as a function of integrated luminosity, for the ATLAS SR selection alone (blue) and with BDT-based background rejection (orange). Horizontal lines indicate the $3\sigma$ and $5\sigma$ thresholds.}
    \label{fig:lumi_scan}
\end{figure}

Figure \ref{fig:lumi_scan} demonstrates the core finding of this paper, that with careful calibration of a continuous hypothesis based fit test and through the use of machine learning for observable reconstruction including neutrino and Higgs rest frame reconstruction, it will be possible to observe evidence for quantum entanglement in Higgs decays to W boson pairs with 3$\sigma$ at roughly 550~fb$^{-1}$ and 5$\sigma$ at 1600~fb$^{-1}$ using a BDT enhanced signal region.

\section{Discussion}\label{sec:discussion}

\subsection{Summary of Findings}

This work demonstrates that the continuous CGLMP formulation enables standard hypothesis testing for quantum entanglement at colliders, avoiding the sensitivity issues inherent in expectation-value-based approaches. The per-event $\mathfrak{I}_3$ distribution provides robust shape discrimination between entangled and separable hypotheses even in the presence of significant background contamination and reconstruction effects.

The conditional denoising diffusion model preserves the $\mathfrak{I}_3$ shape with KS distances below 0.05 relative to truth, while operating on individual events without requiring truth-level labels during inference. This enables direct application to the full measured dataset including backgrounds, a key advantage over reweighting-based approaches such as OmniFold.

With a BDT-based selection using non-angular variables and the full systematic model, the analysis achieves an expected Asimov significance of 1.7$\sigma$ at 139~fb$^{-1}$, 3.4$\sigma$ at 700~fb$^{-1}$, and 6.7$\sigma$ at the HL-LHC target of 3000~fb$^{-1}$. The sensitivity is limited by the systematic floor arising from background normalisation and unfolding shape uncertainties rather than statistical precision alone.

\subsection{Comparison to Previous Work}

Our approach differs from previous theoretical studies~\cite{barr1,barr2,barr3} in several key aspects. We explicitly address detector effects and reconstruction challenges, provide a measurement strategy compatible with standard LHC analysis frameworks, and quantify sensitivity including systematic uncertainties.

The recent study of $\tau^+\tau^-$ entanglement~\cite{Zhang:2025mmm} similarly employs machine learning for neutrino reconstruction. Our work extends this to the $WW^*$ system, which has richer spin structure (qutrits vs qubits) but poses greater reconstruction challenges in that the qutrit system uses the Wigner P symbols (as described in \ref{app:wigner}) where the per-event integrand is unbounded.
The spin 1/2 tau system therefore is only limited by the issue of neutrino imputation rather than the moment estimator itself and therefore focuses on a very different pathology. 

\subsection{Future Directions}

Several extensions of this work are envisioned. Implementation with full detector simulation and data-driven background estimation would enable a complete ATLAS or CMS analysis and is expected to significantly reduce the background normalisation uncertainties that currently limit the sensitivity. Recent work~\cite{Grossi:2024jae} has highlighted that NLO corrections can significantly modify angular coefficients, and incorporating these effects will be essential for precision measurements. 
%
A further avenue is the regression of the neutrino four-momenta via a deep feed-forward neural network, taking as input the dilepton kinematics $(p^\mu_{\ell_1}, p^\mu_{\ell_2})$ and the missing transverse momentum components $(p^{\mathrm{miss}}_x, p^{\mathrm{miss}}_y)$. The reconstructed neutrino four-momenta can then be used for quantum state tomography, and initial studies show promising performance~\cite{vak_2025}.
This neural representation provides an excellent proof of concept that suggest that modern advances in Neural Simulation Based Inference (NSBI) would allow future analyses to leverage the ability of neural processes to approximate the likelihood ratio.
Represented here in the form of a single continuous dimension, the network itself would learn the high dimensional representation with the promise of increased sensitivity, while also lending extensive experience in statistical rigour from years of experiments on LHC data to the growing field  of research into machine learning of quantum systems.

Beyond entanglement, the per-event correlation matrix $c_{ij}$ provides sensitivity to new physics through alternative linear combinations of its elements. As shown in Appendix~\ref{app:smeft}, observables targeting the CP-odd and CP-even sectors of the Gell-Mann matrix can probe SMEFT operators affecting the $HWW$ vertex, with the continuous-distribution hypothesis testing framework developed here applying directly to such searches. These observables are statistically independent of $\mathfrak{I}_3$ (Appendix~\ref{app:smeft}, Figure~\ref{fig:smeft_2d}), meaning a single tomographic dataset simultaneously tests for entanglement, CP violation, and anomalous couplings. Finally, the $H \to ZZ^* \to 4\ell$ and $H \to \tau^+\tau^-$ channels may offer complementary sensitivity through similar constructions, though differences in helicity structure may make combinations prohibitive.

The datasets used to produce this study are available for open science based re-analysis\cite{croft_2026_20285937} and future studies and published through zenodo \href{10.5281/zenodo.20285937}{https://doi.org/10.5281/zenodo.20285937}.

\subsection{Conclusions}

We have presented a comprehensive strategy for testing quantum entanglement in Higgs boson decays to $W$ boson pairs at the LHC. By reformulating the Bell inequality test as a continuous hypothesis test and employing state-of-the-art diffusion-based unfolding for neutrino reconstruction, we enable a measurement that is robust, compatible with standard analysis tools, and sensitive to quantum entanglement with existing LHC datasets.

Our principal result is that with a BDT-enhanced selection and the full systematic model, 3$\sigma$ evidence of quantum entanglement in $H \to WW^*$ decays is projected at approximately 550~fb$^{-1}$, 5$\sigma$ observation at approximately 1600~fb$^{-1}$, with an expected significance of 6.7$\sigma$ at the HL-LHC target of 3000~fb$^{-1}$. The dominant systematic limitation is the background normalisation uncertainty, followed by the cDDPM unfolding shape bias. Further reductions, through data-driven background constraints and improved diffusion architectures, would tighten the projection. The continuous formulation of the CGLMP observable, combined with diffusion-based neutrino reconstruction, provides a practical and robust path to measuring quantum entanglement in Higgs boson decays.

\section{Acknowledgements}

We acknowledge support by the DFG under Germany’s Excellence Strategy 390833306 – EXC 2121: Quantum Universe, and by the Ministry of
Education, Youth and Sports of the Czech
Republic under the project number LM 2023040.


\onecolumngrid
\bibliography{references}

\newpage

\appendix

\section{Wigner \texorpdfstring{$P$}{P}-Symbols}\label{app:wigner}

The Wigner $P$-symbol provides a mapping from spin operators to functions on the unit sphere that is central to quantum state tomography~\cite{barr2,Li_2013}. For a spin-$s$ system, the $P$-symbol of an operator $A$ is defined as
\begin{equation}
\Phi_A^P(\Omega) = \sum_{m=-s}^{s} \langle s,m|A|s,m\rangle\, P_m(\Omega),
\end{equation}
where $P_m(\Omega)$ is the probability distribution for measuring spin projection $m$ when the spin is aligned along direction $\Omega = (\theta,\phi)$.

For spin-1, the probabilities are $P_{+1} = \cos^4(\theta/2)$, $P_0 = \frac{1}{2}\sin^2\theta$, $P_{-1} = \sin^4(\theta/2)$. For a projective decay such as $W \to \ell\nu$, the $P$-symbols of the Gell-Mann matrices $\lambda_i$ are expressed in terms of emission cosines $\xi_x$, $\xi_y$, $\xi_z$ and $\theta = \arccos(\xi_z)$:
\begin{align}
\Phi_1^{P\pm} &= \sqrt{2}(5\xi_z \pm 1)\xi_x \nonumber\\
\Phi_2^{P\pm} &= \sqrt{2}(5\xi_z \pm 1)\xi_y \nonumber\\
\Phi_3^{P\pm} &= \frac{1}{4}(\pm 4\xi_z + 15\cos(2\theta) + 5) \nonumber\\
\Phi_4^{P\pm} &= 5(\xi_x^2 - \xi_y^2) \label{eq:Wigner_P}\\
\Phi_5^{P\pm} &= 10\xi_x\xi_y \nonumber\\
\Phi_6^{P\pm} &= \sqrt{2}(\pm 1 - 5\xi_z)\xi_x \nonumber\\
\Phi_7^{P\pm} &= \sqrt{2}(\pm 1 - 5\xi_z)\xi_y \nonumber\\
\Phi_8^{P\pm} &= \frac{1}{4\sqrt{3}}(\pm 12\xi_z - 15\cos(2\theta) - 5) \nonumber
\end{align}
The $\pm$ superscripts arise from the parity-violating nature of $W$ boson decay: $W^+$ couples only to left-handed fermions while $W^-$ couples only to right-handed antifermions, producing opposite angular distributions and entering the $P$-symbols through sign differences in terms linear in $\xi_z$.

The correlation coefficients are obtained by averaging products over the angular distribution:
\begin{align}
a_i^+ &= \frac{1}{2}\langle \Phi_i^{P+} \rangle_{\text{avg}} \nonumber\\
b_j^- &= \frac{1}{2}\langle \Phi_j^{P-} \rangle_{\text{avg}} \label{eq:tomography}\\
c_{ij} &= \frac{1}{4}\langle \Phi_i^{P+} \Phi_j^{P-} \rangle_{\text{avg}} \nonumber
\end{align}
The $P$-symbols satisfy the orthogonality relation $\int (d\Omega/4\pi)\, \Phi_i^P\, \Phi_j^P = \delta_{ij}$ when properly normalised, ensuring unique recovery of the density matrix elements.

\section{Analytical Reconstruction}\label{app:ana}

Reconstructing the separate neutrino four-momenta is required for quantum state tomography. Together with the lepton four-momenta, this provides the full final state, enabling the W bosons and Higgs boson to be reconstructed. We found that the algorithm presented in Ref.~\cite{aben} for the separate neutrino solution is based on strong assumptions, which limits the expressiveness of the results. Therefore, we use the presented algorithm up to the point of the combined missing 4-vector solution, after which we deploy a modified algorithm.
Applying momentum conservation leaves us with an equation for the first neutrino, i.e. the one corresponding to the leading lepton:
\begin{equation}\label{eq:momentum_conservation}
    p_{\nu,1} = \frac{1}{2}(p_{\nu\nu} + p_{l,2} - p_{l,1} - p_{W,2} + p_{W,1}).
\end{equation}
$p_{\nu\nu}$ is known from the previously mentioned algorithm, which leaves $p_{W,1}$ and $p_{W,2}$ as the remaining unknowns.
To find a solution for the underconstrained system, we estimate the invariant mass of the $W$ bosons by applying the on-shell constraint.
In the massless lepton approximation, the on-shell constraint $(p_{\ell} + p_{\nu})^2 = M_W^2$, yields $M_W^2 = 2(E_{\ell}E_{\nu} - \vec{p}_{\ell} \cdot \vec{p}_{\nu})$. If the result is greater than zero and within $3\sigma$ of the Breit-Wigner width of 2.1~GEV~\cite{pdg_w_width}, the solution is accepted. Otherwise, the algorithm falls back to the PDF nominal value of 80.4~GEV. The direction of each W boson is estimated by combining the direction of the lepton momentum with the equal split of the missing transverse momentum.
The individual neutrino four-vectors are then decomposed from the combined system via the linear decomposition in \autoref{eq:momentum_conservation} and $p_{\nu,\text{off}} = p^{VV} - p_{\nu,\text{on}}$.


\section{Gell-Mann Matrix Extraction and Stability}\label{app:ggm}

This appendix investigates the extraction of Gell-Mann correlation coefficients and the stability of the procedure under realistic experimental conditions.

\begin{figure}[ht]
  \centering
  \begin{subfigure}{0.46\textwidth}
    \centering
    \includegraphics[width=\linewidth]{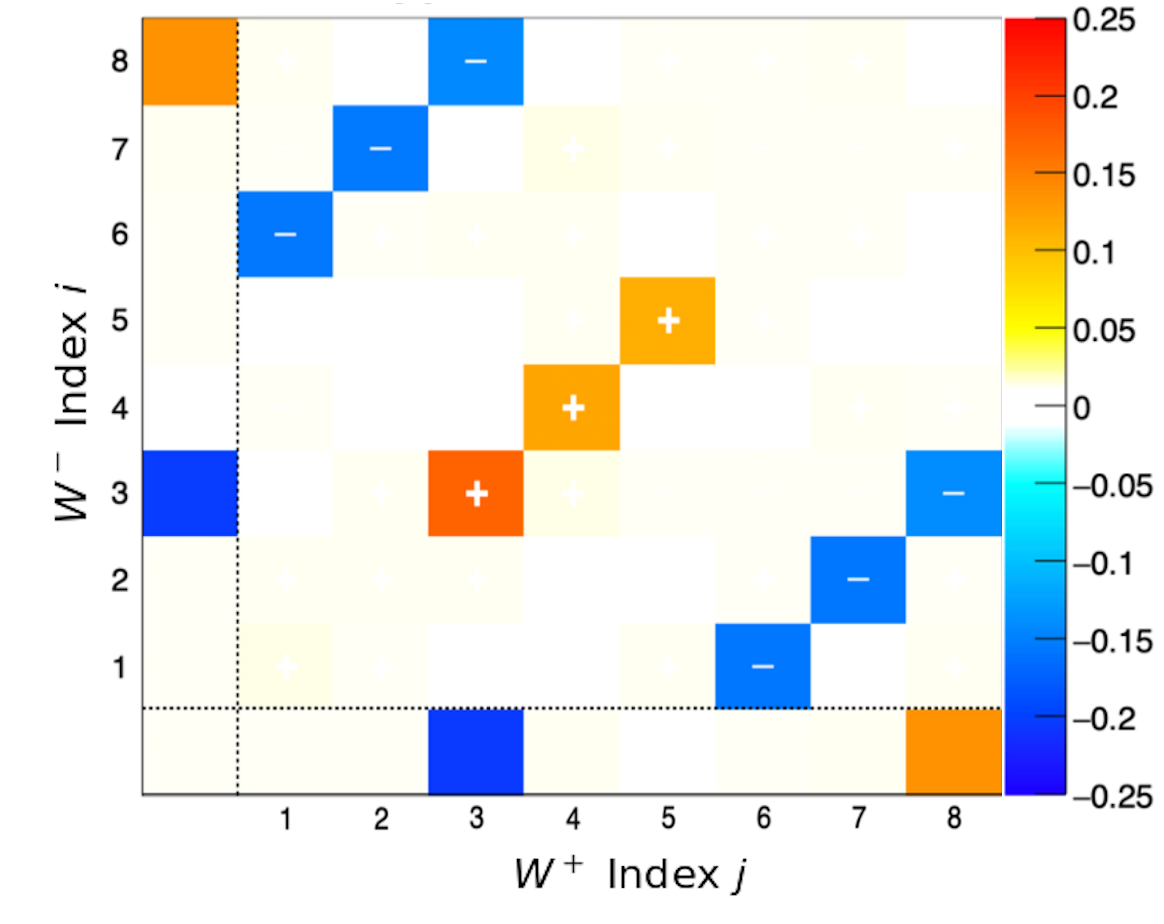}
  \end{subfigure}
  \hfill
  \begin{subfigure}{0.50\textwidth}
    \centering
    \includegraphics[width=\linewidth]{figures/madgraph_gellmann_result.png}
  \end{subfigure}
  \caption{Comparison of the theoretical prediction of Ref.~\cite{barr2} (left) against the calculated Gell-Mann coefficients from the generated MadGraph sample (right).}
  \label{fig:gellmann_truth_validation}
\end{figure}

At truth level, without selection cuts, the extracted Gell-Mann matrix agrees with the theoretical prediction of Ref.~\cite{barr2} within statistical uncertainties (Figure~\ref{fig:gellmann_truth_validation}). The diagonal elements $C_{44}$ and $C_{55}$ that dominate the CGLMP parameter are recovered with sub-percent precision.

\subsection{Sensitivity to Selection Cuts}

\begin{figure}[ht]
    \centering
    \includegraphics[width=0.65\linewidth]{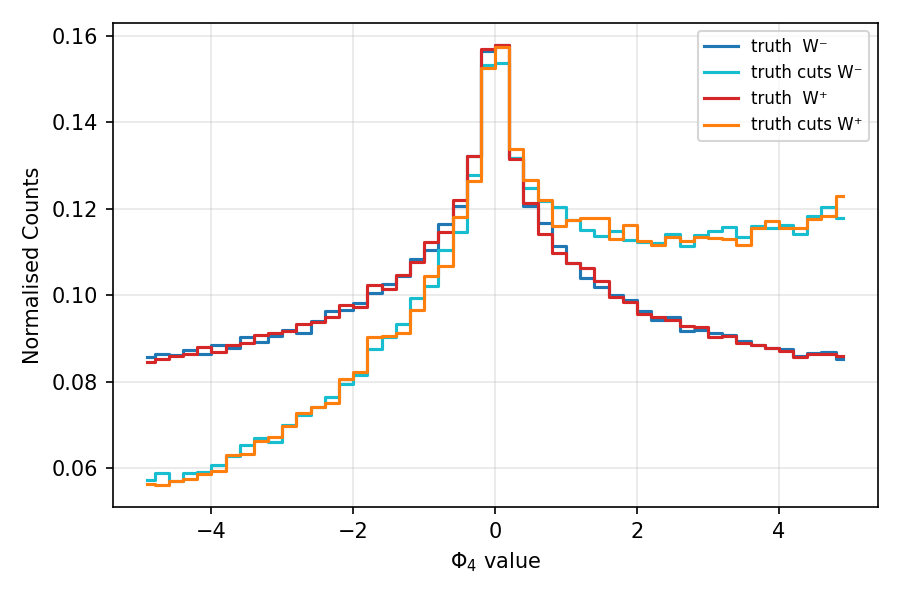}
    \caption{Distribution of the $C_{44}$ Gell-Mann coefficient before and after selection cuts, showing how the mean shifts despite an apparently stable shape.}
    \label{fig:c44_distribution_shift}
\end{figure}

Selection cuts on lepton kinematics introduce significant distortions in the extracted coefficients. Each element of the correlation matrix is derived from an expectation value over a distribution that is sharply peaked but possesses no intrinsic normalisation, so small biases produce disproportionately large effects. Figure~\ref{fig:c44_distribution_shift} illustrates this for the coefficient $C_{44}$, whose underlying distribution $\Phi_4^{P+}\Phi_4^{P-} = 25(\xi_x^{+2} - \xi_y^{+2})(\xi_x^{-2} - \xi_y^{-2})$ is peaked near zero with tails to $\pm 25$. Selection cuts shift the mean even though the overall shape appears nearly unchanged.

This sensitivity propagates directly to the CGLMP Bell parameter. An expectation-value-based test would require either acceptance correction (model-dependent) or restriction to uniform-acceptance regions (reduced statistics). The template-based hypothesis testing framework developed in this paper avoids these issues: both signal templates are subjected to the same selection, so acceptance effects cancel in the comparison.

\section{Unfolding Studies}\label{app:unfolding}

This appendix presents a comprehensive study of unfolding methodologies for the Bell test analysis. We begin by assessing the viability of classical binned unfolding applied to the $\mathfrak{I}_3$ distribution computed from the analytical neutrino reconstruction, and show that the analytical reconstruction destroys essentially all discriminating power between the entangled and separable hypotheses. We then discuss the OmniFold reweighting approach, before presenting the conditional denoising diffusion (cDDPM) method adopted in this analysis, which operates at the level of the helicity angles where per-event information is preserved.

\subsection{Overview of Approaches}

The continuous CGLMP observable $\mathfrak{I}_3$ is defined in terms of the helicity angles $(\theta^{\pm}, \phi^{\pm})$ of the $W$ boson decay products, which in turn require knowledge of the neutrino momenta. Since the neutrinos are not directly measured, any strategy for extracting $\mathfrak{I}_3$ from data must reconstruct or infer the missing degrees of freedom. Three broad classes of approach are available: analytical reconstruction followed by classical unfolding, in which the neutrino momenta are estimated event-by-event using the Higgs and $W$ mass constraints~\cite{Sonnenschein:2006ud} and the resulting $\mathfrak{I}_3$ distribution is corrected for reconstruction effects using standard binned unfolding methods~\cite{RooUnfold}; ensemble reweighting (OmniFold)~\cite{OmniFold}, in which a machine-learning classifier iteratively reweights simulated events to match data without explicit neutrino reconstruction; and generative unfolding (cDDPM)~\cite{Pazos:2024}, in which a conditional diffusion model samples per-event truth-level quantities directly from detector-level inputs, bypassing both the analytical reconstruction step and the need for binned unfolding.

\subsection{Analytical Reconstruction and the \texorpdfstring{$\mathfrak{I}_3$}{I3} Response}\label{app:inverse_problem}

We first assess the baseline approach: computing $\mathfrak{I}_3$ from the analytical neutrino reconstruction and applying classical unfolding to correct the resulting distribution. The analytical method~\cite{Sonnenschein:2006ud} uses the Higgs and $W$ mass constraints to solve for the neutrino momenta from the measured leptons and $E_T^{\mathrm{miss}}$; a valid solution is obtained for 82\% of HWW events.

\begin{figure}[ht]
    \centering
    \includegraphics[width=\textwidth]{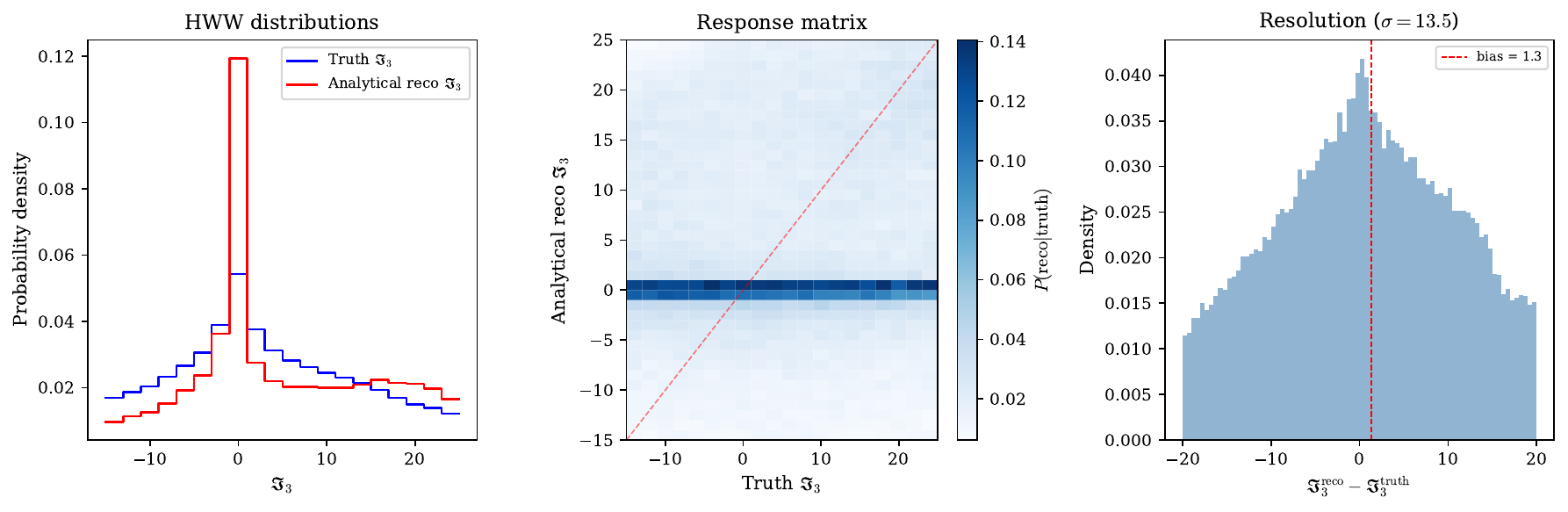}
    \caption{The $\mathfrak{I}_3$ response from analytical neutrino reconstruction. Left: truth (blue) and analytically-reconstructed (red) $\mathfrak{I}_3$ distributions; the reco distribution shows a characteristic spike near zero from poorly reconstructed events. Centre: normalised 2D response matrix $P(\mathrm{reco}|\mathrm{truth})$; the near-uniform columns confirm that the reconstruction has lost the truth-level information. Right: resolution distribution with $\sigma \approx 17$, comparable to the distribution width itself.}
    \label{fig:i3_response}
\end{figure}

\begin{figure}[ht]
    \centering
    \includegraphics[width=0.45\textwidth]{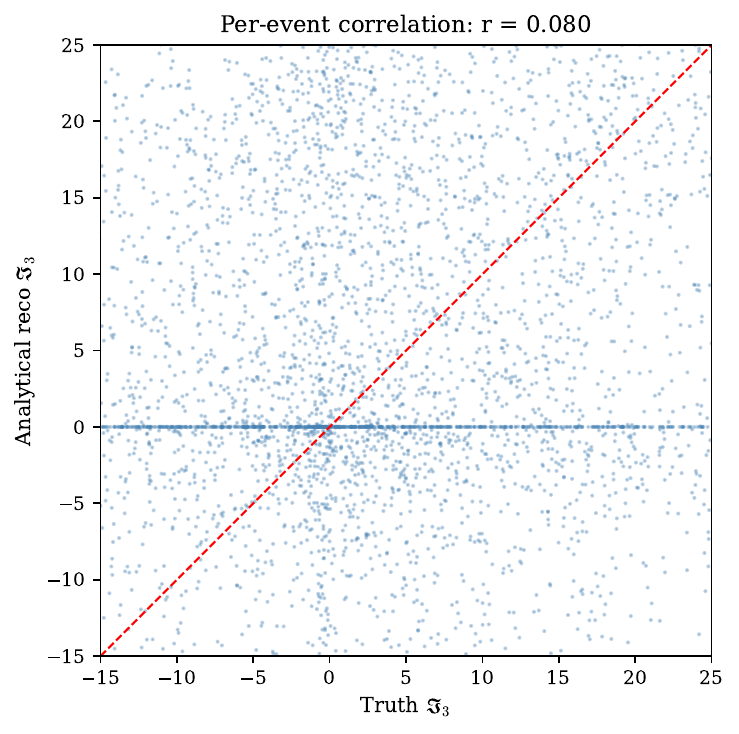}
    \caption{Per-event scatter of truth vs.\ analytically-reconstructed $\mathfrak{I}_3$ for 5\,000 HWW events. The correlation coefficient $r = 0.08$ confirms that the analytical reconstruction carries negligible per-event information about the true $\mathfrak{I}_3$.}
    \label{fig:i3_scatter}
\end{figure}

From 165\,000 events with valid analytical solutions, we construct paired (truth $\mathfrak{I}_3$, analytical-reco $\mathfrak{I}_3$) values and build a 1D response matrix with 20 truth bins and 40 measured bins spanning $\mathfrak{I}_3 \in [-15, 25]$. Figure~\ref{fig:i3_response} shows the truth and reconstructed distributions, the 2D response matrix, and the reconstruction resolution. Two features are immediately apparent. First, the per-event correlation between truth and reconstructed $\mathfrak{I}_3$ is $r = 0.08$, effectively zero; and the scatter plot (Fig.~\ref{fig:i3_scatter}) confirms that the analytical reconstruction carries almost no information about the true $\mathfrak{I}_3$ on an event-by-event basis. Second, the response matrix rows are nearly identical (mean inter-row correlation $> 0.99$), so that $P(\mathfrak{I}_3^{\mathrm{reco}}\mid\mathfrak{I}_3^{\mathrm{truth}})$ is approximately independent of the truth value: the reconstruction effectively maps all truth distributions to the same measured distribution.

\subsection{Classical Unfolding of \texorpdfstring{$\mathfrak{I}_3$}{I3}}\label{app:classical_methods}

Despite the poor per-event correlation, classical binned unfolding can in principle recover the true $\mathfrak{I}_3$ \emph{distribution} by inverting the response matrix. We test this using four standard methods implemented in the RooFitUnfold framework~\cite{RooUnfold}: iterative Bayesian (D'Agostini, 4 iterations)~\cite{DAgostini:1994fjx}, SVD ($k_{\mathrm{reg}} = 10$)~\cite{Hocker:1995kb}, Tikhonov (TUnfold, L-curve $\tau$)~\cite{Schmitt:2012kp}, and matrix inversion. We also test BRU~\cite{Croft:2026bru} with 25 spline knots.

The RooUnfold bias and coverage diagnostics~\cite{RooUnfold} reveal the fundamental problem. Using 200 internal toy pseudo-experiments, the built-in \texttt{CalculateBias} routine reports a bias-to-variance ratio of $\mathrm{bias}^2/\mathrm{var} \approx 5 \times 10^5$ for the Bayesian method and \texttt{ScanCoverage} returns \emph{zero} coverage at all iteration counts from 1 to 8. The unfolding is completely bias-dominated: the variance from statistical fluctuations is negligible compared to the systematic bias introduced by the near-degenerate response matrix.

The physical reason is straightforward: because the response matrix rows are nearly identical, the unfolding has no data-driven information about which truth distribution produced the observed data. The ``unfolded'' result is determined almost entirely by the regularisation prior (for Bayesian and BRU) or by the pseudo-inverse of a nearly singular matrix (for TUnfold and inversion), not by the measured distribution.

\subsection{Discrimination Loss}\label{app:discrimination_loss}

\begin{figure}[ht]
    \centering
    \includegraphics[width=0.9\textwidth]{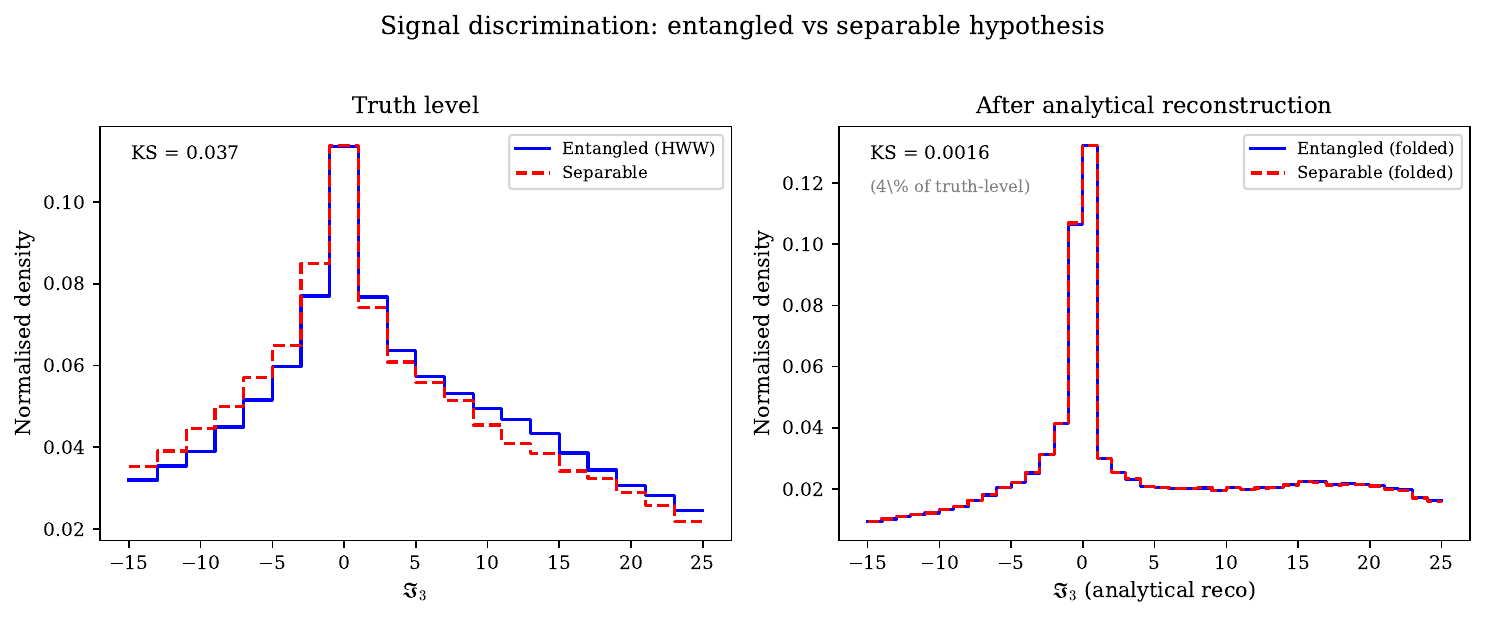}
    \caption{Discrimination between entangled and separable hypotheses. Left: at truth level, the $\mathfrak{I}_3$ distributions are clearly distinguishable ($\mathrm{KS} = 0.037$). Right: after folding through the analytical reconstruction response, the distributions become indistinguishable ($\mathrm{KS} = 0.002$, 4\% of the truth-level discrimination).}
    \label{fig:i3_discrimination_loss}
\end{figure}

The practical consequence is that the analytical reconstruction destroys the signal that the Bell test is designed to measure. Figure~\ref{fig:i3_discrimination_loss} quantifies this directly: the entangled and separable truth-level $\mathfrak{I}_3$ distributions differ by $\mathrm{KS} = 0.037$, but after folding through the analytical reconstruction response, the reco-level distributions are indistinguishable ($\mathrm{KS} = 0.002$). Only 4\% of the truth-level discrimination survives the analytical reconstruction. No amount of unfolding can recover information that has been destroyed at the reconstruction stage.

This result establishes that the analytical neutrino reconstruction, while providing a valid per-event estimate of the neutrino momenta, produces $\mathfrak{I}_3$ values that are too poorly correlated with truth to support a hypothesis test. The template fit would see effectively identical $\mathfrak{I}_3$ shapes for the entangled and separable hypotheses and would have no sensitivity.

\subsection{Implications for the Analysis Strategy}\label{app:classical_limitations}

The classical unfolding study motivates two requirements for the reconstruction method. First, the reconstruction must preserve per-event angular information: the failure of the analytical-reconstruction-plus-unfolding pipeline is not a failure of the unfolding methods themselves but of the reconstruction step that precedes them. The $\mathfrak{I}_3$ observable is a highly nonlinear function of the helicity angles, and the analytical reconstruction, despite providing reasonable neutrino momentum estimates, does not resolve the angular correlations that determine $\mathfrak{I}_3$; a reconstruction that operates directly on the helicity angles, rather than on the full neutrino 4-vectors, can concentrate its capacity on the degrees of freedom that matter. Second, background treatment must be native: even if the reconstruction quality were improved, classical unfolding of the 1D $\mathfrak{I}_3$ requires process-specific response matrices and prior background subtraction, which in the multi-background Higgs environment introduces model dependence.

The cDDPM approach (Section~\ref{app:cddpm_method}) addresses both requirements. It targets the 6-dimensional helicity angles directly (the minimal representation needed for $\mathfrak{I}_3$) and initialises the reverse diffusion from the analytical reconstruction, allowing the network to correct the residual distortions rather than reconstructing from scratch. The one-hot process tag enables native background handling without prior subtraction.

\subsection{OmniFold: Methodology}\label{app:omnifold_method}

OmniFold~\cite{OmniFold} provides a method to simultaneously unfold all observables through iterative reweighting with machine learning classifiers. Unlike binned approaches, OmniFold operates on the full multidimensional feature space and requires no explicit binning of the observables, making it in principle an unbinned, high-dimensional unfolding method.

\subsubsection{Algorithm}

The OmniFold procedure alternates between two reweighting steps in each iteration $n$:

\paragraph{Step 1 (detector level).} A binary classifier $f_1^{(n)}$ is trained to distinguish simulation (label 0) from data (label 1) in detector-level feature space. The simulation events carry weights $w_{\mathrm{push}}^{(n)}$ from the previous generator-level step (initially uniform). The classifier output is converted to a likelihood ratio:
\begin{equation}
    w_{\mathrm{pull}}^{(n)}(x) = w_{\mathrm{push}}^{(n)}(x) \cdot \frac{f_1^{(n)}(x)}{1 - f_1^{(n)}(x)}\,,
\end{equation}
which reweights the simulated detector-level events to match the data distribution.

\paragraph{Step 2 (generator level).} A second classifier $f_2^{(n)}$ is trained to distinguish unweighted generator-level MC (label 0) from the same MC events weighted by $w_{\mathrm{pull}}^{(n)}$ (label 1). The resulting reweighting factors are applied at generator level:
\begin{equation}
    w_{\mathrm{push}}^{(n+1)}(x) = \frac{f_2^{(n)}(x)}{1 - f_2^{(n)}(x)}\,.
\end{equation}

These weights are propagated back to Step~1 for the next iteration. After convergence, $w_{\mathrm{push}}^{(n)}$ applied to the generator-level MC yields the unfolded distribution for any observable simultaneously.

The classifiers are implemented as multi-layer perceptrons (3 hidden layers of dimensions $\{64, 128, 64\}$ with GELU activations) trained with the weighted binary cross-entropy loss. We use the implementation of Ref.~\cite{OmniFoldNeutrino} with Adam optimisation ($\mathrm{lr} = 5\times 10^{-5}$), batch size 512, and early stopping with patience 10 on the validation loss.

\subsubsection{Signal-Only Validation}\label{app:omnifold_signal}

To validate the OmniFold implementation, we apply it to the signal-only case: 100\,000 MadGraph5 HWW events serve as simulation and 50\,000 Sherpa HWW events serve as pseudo-data, both passing the ATLAS signal region selection. The detector-level features ($e$ and $\mu$ transverse momenta, pseudorapidities, azimuthal angles, and $E_T^{\mathrm{miss}}$ components; 9 features) and the truth-level features (lepton and neutrino 4-momenta; 12 features) are used for Steps~1 and~2 respectively.

\begin{figure}[ht]
    \centering
    \includegraphics[width=0.48\textwidth]{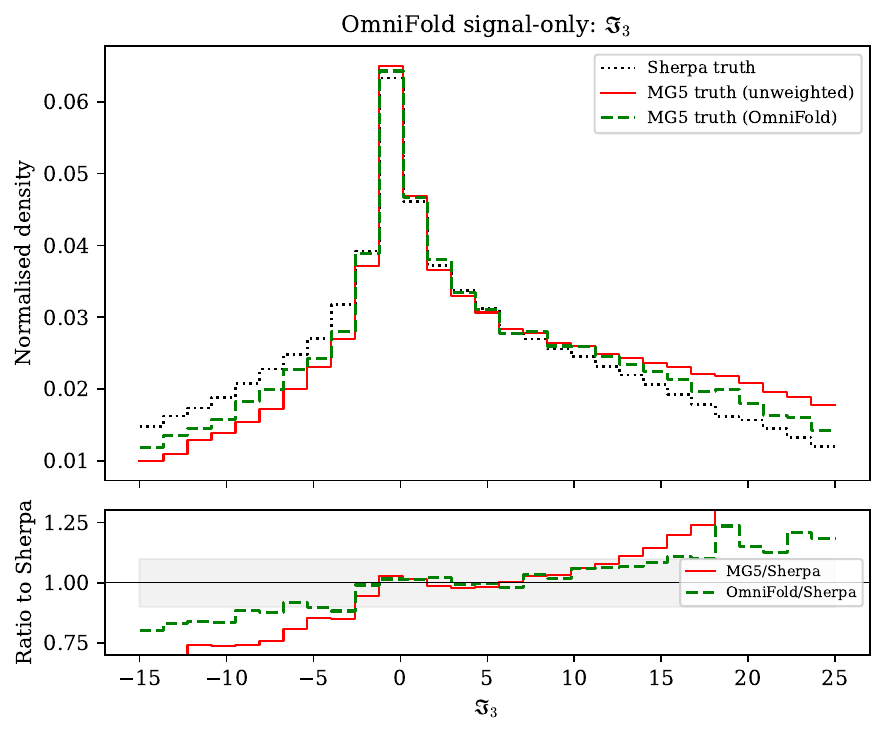}
    \caption{OmniFold signal-only validation on $\mathfrak{I}_3$. Top: MG5 truth-level $\mathfrak{I}_3$ before (red) and after (green dashed) OmniFold reweighting, compared to the Sherpa truth (black dotted). The reweighting successfully corrects the generator-level difference. Bottom: ratio to Sherpa truth.}
    \label{fig:omnifold}
\end{figure}

\begin{figure}[ht]
    \centering
    \includegraphics[width=\textwidth]{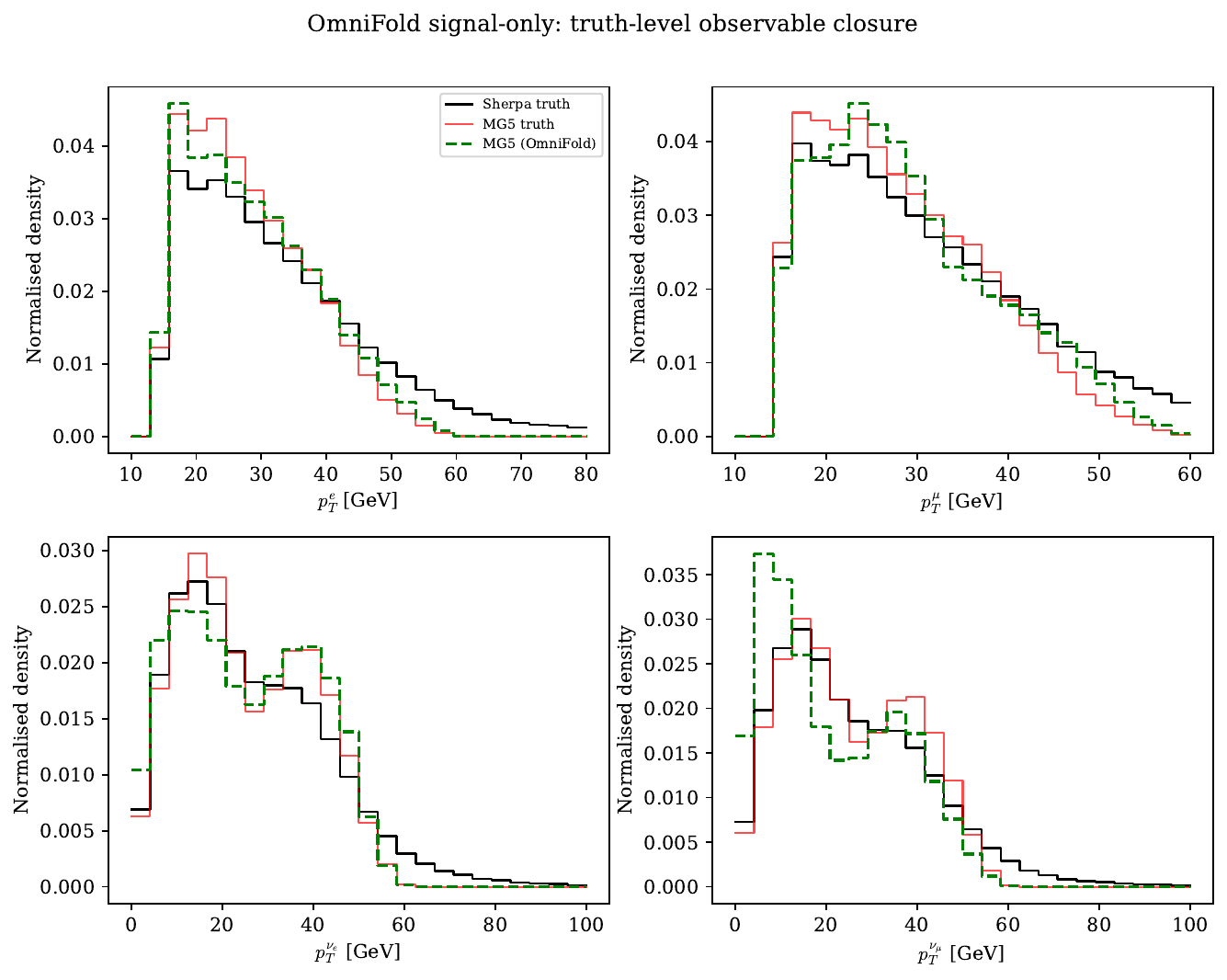}
    \caption{OmniFold signal-only closure on four truth-level observables: electron $p_T$ (top left), muon $p_T$ (top right), electron neutrino $p_T$ (bottom left), and muon neutrino $p_T$ (bottom right). The reweighted MG5 distributions (green dashed) agree with Sherpa truth (black) across the full kinematic range.}
    \label{fig:omnifold_observables}
\end{figure}

Figure~\ref{fig:omnifold} shows the $\mathfrak{I}_3$ distribution before and after OmniFold reweighting, compared to the independent Sherpa truth. The reweighted distribution agrees well with truth, confirming that OmniFold provides an excellent unfolding of the signal shape in isolation. Figure~\ref{fig:omnifold_observables} demonstrates that this closure extends beyond $\mathfrak{I}_3$ to individual truth-level observables: the reweighted MG5 distributions for the lepton and neutrino transverse momenta agree with the Sherpa reference across the full kinematic range.

\subsubsection{Convergence}\label{app:omnifold_convergence}

The OmniFold procedure is run for up to 10 iterations. Convergence is monitored via the two-sample Kolmogorov--Smirnov (KS) statistic and binned $\chi^2$ between the reweighted MG5 truth-level distributions and the Sherpa reference, evaluated on the leading lepton $p_T$. Both metrics decrease rapidly over the first 3--4 iterations and plateau thereafter, confirming that the default $n_{\mathrm{iter}} = 5$ used in the signal-only study is sufficient. Increasing to 10 iterations provides no measurable improvement and risks overtraining the classifiers.

\subsubsection{Limitations in the Mixed-Background Environment}\label{app:omnifold_limitations}

The success of OmniFold in the signal-only case does not extend to the mixed-background environment of the Higgs analysis. OmniFold is fundamentally a reweighting procedure that operates on the full ensemble of simulated events, with the reweighting factors learned by training classifiers to distinguish simulation from data at detector level, then propagating these weights to generator level. This requires that truth-level information be available for all events being unfolded.

\begin{figure}[ht]
    \centering
    \includegraphics[width=0.85\textwidth]{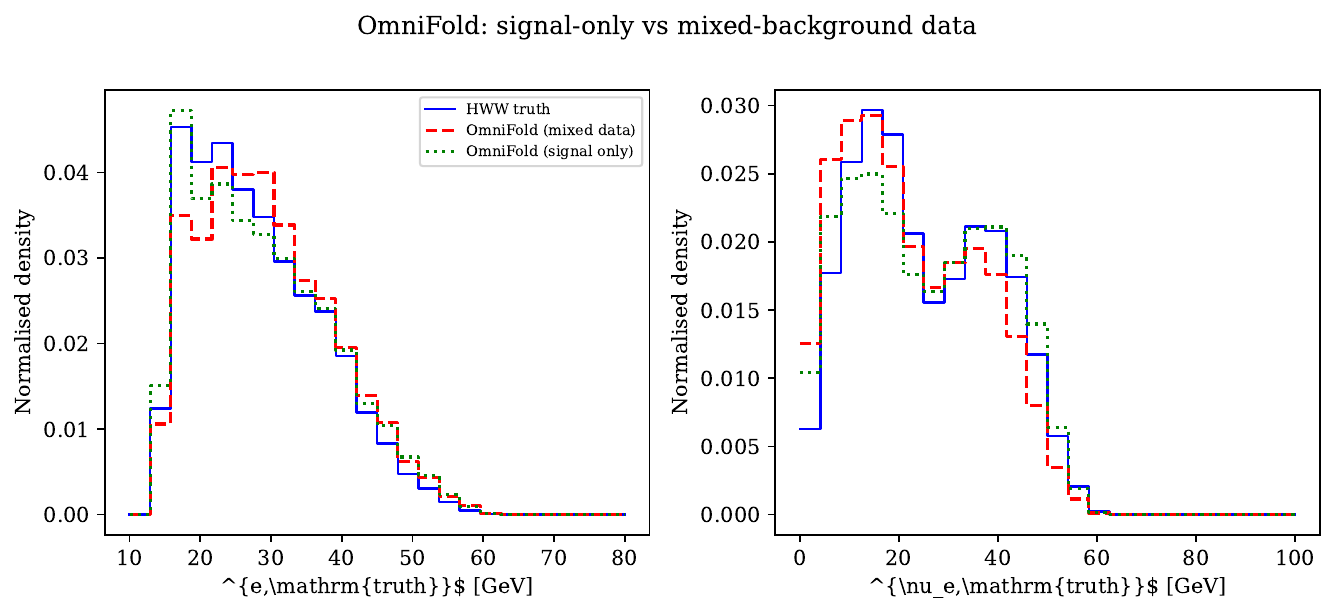}
    \caption{OmniFold failure in the mixed-background environment. Left: truth-level electron $p_T$. Right: truth-level electron neutrino $p_T$. The HWW truth (blue solid) is compared to OmniFold trained on signal-only data (green dotted, good agreement) and on mixed signal+background data (red dashed, systematic distortion). The mixed-data reweighting shifts the distributions toward the harder background spectra.}
    \label{fig:omnifold_mixed_failure}
\end{figure}

In the context of $H\to WW^*$ analyses with multiple dominant backgrounds, this requirement poses a fundamental incompatibility. The measured dataset contains signal events, for which we wish to extract truth-level neutrino momenta, interleaved with background events ($WW$, $t\bar{t}$, $Z\to\tau\tau$) that have entirely different truth-level origins. Three possible strategies for applying OmniFold in this environment all fail:

\begin{enumerate}
    \item \textbf{Prior background subtraction.} Subtracting backgrounds from the measured distribution before unfolding reintroduces model dependence: the subtracted sample depends on the assumed background cross sections, kinematic shapes, and detector response. This undermines the data-driven philosophy of OmniFold and can introduce negative bin contents in background-dominated regions.

    \item \textbf{Including backgrounds in the unfolding.} Training OmniFold with the combined signal-plus-background ``data'' against a signal-only MC simulation forces the Step~1 classifier to learn the data/MC discrepancy that arises from the missing backgrounds, not from the physics of interest. The resulting generator-level weights are distorted by the attempt to make signal MC mimic a signal-plus-background mixture, producing a biased truth-level distribution.

    To verify this, we train OmniFold with ``data'' consisting of HWW signal ($n = 10{,}000$), WW ($n = 40{,}000$), and $t\bar{t}$ ($n = 20{,}000$) backgrounds in realistic proportions, against HWW-only MC. Figure~\ref{fig:omnifold_mixed_failure} shows the resulting reweighted truth-level distributions compared to the actual HWW truth and the signal-only OmniFold result. The mixed-data reweighting systematically distorts the lepton and neutrino $p_T$ distributions, as the classifier pushes the HWW MC toward the harder $t\bar{t}$ and broader WW spectra that dominate the data sample.

    \item \textbf{Multi-process MC.} Including all processes in the MC with their associated truth-level information is conceptually inconsistent with the goal of measuring signal properties. The reweighted truth-level MC would be a mixture of signal and background truths, with no mechanism to isolate the signal component without additional model-dependent decomposition.
\end{enumerate}

\subsubsection{OmniFold vs cDDPM: StructuralComparison}\label{app:omnifold_vs_cddpm}

Table~\ref{tab:omnifold_vs_cddpm} summarises the key structural differences between OmniFold and the cDDPM approach. The fundamental distinction is that OmniFold operates on the ensemble level through reweighting, while cDDPM operates on the event level through generative sampling. This makes cDDPM naturally compatible with mixed-background environments: each event is processed independently, conditioned on its detector-level measurements and an explicit process tag, without requiring ensemble-level consistency between data and simulation.

\begin{table}[H]
    \centering
    \caption{Structural comparison of OmniFold and cDDPM unfolding approaches.}
    \label{tab:omnifold_vs_cddpm}
    \begin{tabular}{l l l}
        \toprule
        Property & OmniFold & cDDPM \\
        \midrule
        Operating level          & Ensemble (reweighting) & Event (generative) \\
        Output                   & Event weights          & Per-event predictions \\
        Truth labels at inference & Required (MC)         & Not required \\
        Background treatment     & Requires subtraction   & Process tag conditioning \\
        Dimensionality           & Unbinned, arbitrary    & Unbinned, arbitrary \\
        Generator dependence     & Corrects MC $\to$ data & Learned posterior \\
        Per-event uncertainty    & Not available          & Via multi-sample \\
        \bottomrule
    \end{tabular}
\end{table}

A complementary advantage of OmniFold is its model-agnostic nature: it corrects arbitrary mismodelling between any two distributions without assuming a specific detector response model, and simultaneously unfolds all observables. In the signal-only case, this makes OmniFold an excellent choice for correcting generator-level differences (e.g.\ MadGraph vs.\ Sherpa). For the Bell test, however, the mixed-background environment and the need for event-level neutrino reconstruction make the cDDPM approach the appropriate tool.

\subsection{Conditional Denoising Diffusion}\label{app:cddpm_method}

The conditional denoising diffusion probabilistic model (cDDPM) approach of Pazos et al.~\cite{Pazos:2024} overcomes the limitations of both classical and reweighting methods by operating on individual events. The model learns to sample from the posterior distribution $P(\vec{x}|\vec{y})$ of truth-level quantities $\vec{x}$ given detector-level measurements $\vec{y}$, without requiring explicit evaluation of the prior distribution.

The cDDPM is trained on paired simulation data $(\vec{x}, \vec{y})$ but during inference requires only the detector-level quantities $\vec{y}$. This enables application to measured events regardless of their true origin, making the method compatible with analyses containing multiple backgrounds.

\subsubsection{Architecture}

The denoising network $\epsilon_\theta$ is a fully-connected architecture with 4 hidden layers of 512 units each, using SiLU activations. Sinusoidal positional embeddings encode the diffusion timestep $t$, which is concatenated with the conditioning vector before each residual block.

\subsubsection{Conditioning Information}

The conditioning vector $\vec{y}$ consists of the detector-level lepton four-momenta packaged as $(E, p_x, p_y, p_z)$ for each lepton (8 components), a pseudo-four-vector for the missing transverse momentum $(p_T^{\mathrm{miss}}, p_x^{\mathrm{miss}}, p_y^{\mathrm{miss}}, 0)$ (4 components), and an explicit 4-dimensional one-hot process tag (1 for the originating process, 0 elsewhere). This yields a total conditioning dimension of 16 features. The one-hot tag provides an unambiguous process identity so that the network can learn process-specific neutrino kinematics; an earlier variant (``v2'') that used per-process $p_T$ distribution moments as a soft tag failed to distinguish processes in practice (see Section~\ref{app:cddpm_design}).

The target vector $\vec{x}$ comprises the helicity angles of the charged leptons in the $W^{\pm}$ rest frames, encoded in a diffusion-safe representation: $\vec{x} = [\operatorname{arctanh}(\cos\theta^-), \sin\phi^-, \cos\phi^-, \operatorname{arctanh}(\cos\theta^+), \sin\phi^+, \cos\phi^+]$, yielding a target dimension of~6. This is a substantially lower-dimensional target than the neutrino four-momenta (8 components) used in earlier variants, and directly encodes the angular information that determines the CGLMP observable. The $\operatorname{arctanh}$ encoding maps $\cos\theta\in[-1,1]$ to an unbounded domain, avoiding boundary artefacts; the $(\sin\phi, \cos\phi)$ encoding eliminates the $\pm\pi$ wrap-around discontinuity.

\subsubsection{Training Configuration}

The model is trained using the simplified loss function
\begin{equation}
L = \mathbb{E}_{t, \vec{x}_0, \epsilon}\left[\|\epsilon - \epsilon_\theta(\vec{x}_t, t, \vec{y})\|^2\right],
\end{equation}
where $\epsilon \sim \mathcal{N}(0, \mathbb{1})$ is the noise added in the forward process and $\vec{x}_t$ is the noised truth-level quantity at timestep $t$.

We use the ``mixed v3'' (boosted-angle targets, one-hot process tag, analytical warm start) variant as the baseline reconstruction throughout the rest of this note. The training set is a balanced mixture of $6\times 10^4$ events from each of the four processes (HWW, WW, $t\bar{t}$, $Z\to\tau\tau$), with all events passing the dilepton fiducial preselection but \emph{not} the full ATLAS SR cuts, so that the network sees the broad kinematic envelope of each process. A single normaliser is fit on the combined feature and target spaces and shared by all processes. Training uses the Adam optimiser with a learning rate of $3\times 10^{-4}$, batch size $2048$, a linear noise schedule over $T=1000$ diffusion steps, and early stopping with a patience of 80 epochs on the validation loss. The network converges after approximately 520 epochs with a validation loss of 0.19.

\subsection{Inference}

During inference, each event's detector-level measurements are first passed through the analytical neutrino reconstruction~\cite{Sonnenschein:2006ud} to obtain a starting guess for the helicity angles. These analytical angles, encoded in the same $(\operatorname{arctanh}\cos\theta, \sin\phi, \cos\phi)$ representation as the training targets, serve as the initialisation for a \emph{warm-start} reverse diffusion: the analytical estimate is forward-diffused to an intermediate timestep $t_{\mathrm{start}}=300$ (of $T=1000$) and then denoised back to $t=0$. This procedure seeds the reverse process near a physically plausible configuration, reducing the effective number of denoising steps and concentrating model capacity on correcting detector-level distortions rather than reconstructing kinematics from scratch. The warm-start inference produces one sample per event with no post-processing or physical constraints applied.

\subsection{Closure and Bias}

Closure is validated process-by-process by comparing the unfolded $\mathfrak{I}_3$ distribution to the truth distribution on the same events (Figure~\ref{fig:unfolded_vs_truth}). Under the production cDDPM (standard sampling, no warm start; Sec.~\ref{app:cddpm_warmstart}), the per-bin HWW residuals have rms 2\% with all bins below 5\%; for $t\bar{t}$, $WW$, and $Z\to\tau\tau$ the per-bin residuals have rms 15\%, 10\%, and 15\% respectively, dominated by the tail bins with lower statistics. With the chosen 9-bin variable-width binning these residuals enter the fit as the \texttt{unf\_<proc>} HistoSys nuisances of Appendix~\ref{app:fit}.

Two earlier variants informed the design of v3. The \emph{signal-only} ``v1'' variant, trained exclusively on HWW, gave $\langle\mathfrak{I}_3\rangle$ shifts of order $+5$ for every background, comparable to the full dynamic range of the observable, because the network had no information that distinguished WW or $t\bar{t}$ events from HWW events with similar detector-level kinematics. The ``v2'' variant retrained on the balanced four-process mixture and used per-process $p_T$ moments as a soft process tag (following Ref.~\cite{Pazos:2024}) but predicted lab-frame neutrino 4-vectors (8-dimensional targets); the network learned to ignore the moment features and still produced HWW-like neutrinos for backgrounds. Switching to an explicit one-hot process tag and lower-dimensional boosted-angle targets as per the v3 design resolved both pathologies.

Generator independence is confirmed by the sub-percent agreement between MadGraph and Sherpa truth-level distributions on the HWW signal sample.

\subsection{cDDPM Systematic Studies}\label{app:cddpm_systematics}

The remainder of this appendix documents the systematic exploration of the cDDPM design space. We vary the network architecture, noise schedule, inference parameters, training data volume, and regularisation, totalling over 40 configurations. The goal is to demonstrate that the chosen v3 design (512-unit, 4-layer network with linear noise schedule, warm-start at $t_{\mathrm{start}}=300$, and single-sample inference) represents a near-optimal configuration for the $\mathfrak{I}_3$ observable.

\subsubsection{Conditioning and Target Design}\label{app:cddpm_design}

Three conditioning and target designs were evaluated. Table~\ref{tab:cddpm_variants} summarises the key properties and performance of each variant. The ``v1'' signal-only model, trained exclusively on HWW, produces $\langle\mathfrak{I}_3\rangle$ shifts comparable to the full dynamic range of the observable for all background processes, as the network has no mechanism to distinguish WW or $t\bar{t}$ events from HWW events with similar detector-level kinematics. The ``v2'' variant introduced a balanced four-process training mixture with per-process $p_T$ moments as soft conditioning tags, but the network learned to ignore these features and continued to produce HWW-like neutrinos for backgrounds. The ``v3'' design resolved both pathologies by switching to (i) an explicit 4-dimensional one-hot process tag that the network cannot ignore, and (ii) 6-dimensional boosted-frame helicity-angle targets that directly encode the physics-relevant observables with reduced dimensionality.

\begin{table}[H]
    \centering
    \caption{Summary of cDDPM conditioning and target design variants. The $\mathfrak{I}_3$ bias is defined as the fractional shift $(\langle\mathfrak{I}_3\rangle_{\mathrm{unfolded}} - \langle\mathfrak{I}_3\rangle_{\mathrm{truth}}) / |\langle\mathfrak{I}_3\rangle_{\mathrm{truth}}|$.
    }
    \label{tab:cddpm_variants}
    \begin{tabular}{lcccccc}
        \toprule
        Variant & Cond.\ dim & Target dim & Process tag & HWW bias & WW bias \\
        \midrule
        v1 (signal-only)     & 18 & 8 (lab $\nu$)    & None     & $+$78\%  & $>$1000\% \\
        v2 (mixed, moments)  & 18 & 8 (lab $\nu$)    & $p_T$ moments & $+$40\%  & $>$1000\% \\
        v3 (mixed, one-hot)  & 16 & 6 (angles)        & One-hot  & $+$14\%  & $+$316\% \\
        \bottomrule
    \end{tabular}
\end{table}

The residual $+$14\% HWW bias in v3 reflects an intrinsic limitation of the reconstruction: even the analytical neutrino solution, which exploits the Higgs and $W$ mass constraints, achieves only modest angular resolution for the helicity angles that determine $\mathfrak{I}_3$. 
The $+$316\% WW fractional bias is a fractional shift on a process whose truth $\langle\mathfrak{I}_3\rangle = +0.43$ sits near zero.
The corresponding absolute shift ($+$1.4 units) is comparable to the HWW absolute shift ($+$0.8 units), so the two processes are in fact reconstructed with similar absolute fidelity. 
The \emph{per-bin} WW closure residual on the chosen 9 variable-width bins is what enters the fit (rms 10\%, maximum 25\%; Fig.~\ref{fig:unfolded_vs_truth}), profiled as a HistoSys nuisance (\texttt{unf\_WW}) jointly with the 20\% WW normalisation Gaussian prior.
The Asimov fit converges across all 87 robustness configurations in Appendix~\ref{app:fit_study} with no boundary fits or sign-flipped pulls; the leave-one-out NP ranking (Appendix~\ref{app:fit_ranking}) places the unfolding-shape NP at $\Delta Z = +0.199\,\sigma$ and the WW-normalisation NP at $\Delta Z = +0.354\,\sigma$, both well-behaved.

\subsubsection{Architecture Study}\label{app:cddpm_arch}

We systematically vary the hidden dimension $\{128, 256, 512, 1024\}$ and depth $\{2, 3, 4, 6\}$ of the denoising network while holding all other hyperparameters at the v3 baseline values ($T{=}1000$, $\mathrm{lr}{=}3{\times}10^{-4}$, batch size 2048, dropout 0.1, linear schedule). All 13 configurations are trained on the same balanced 240k-event dataset with identical normalisation.

\begin{figure}[ht]
    \centering
    \includegraphics[width=\textwidth]{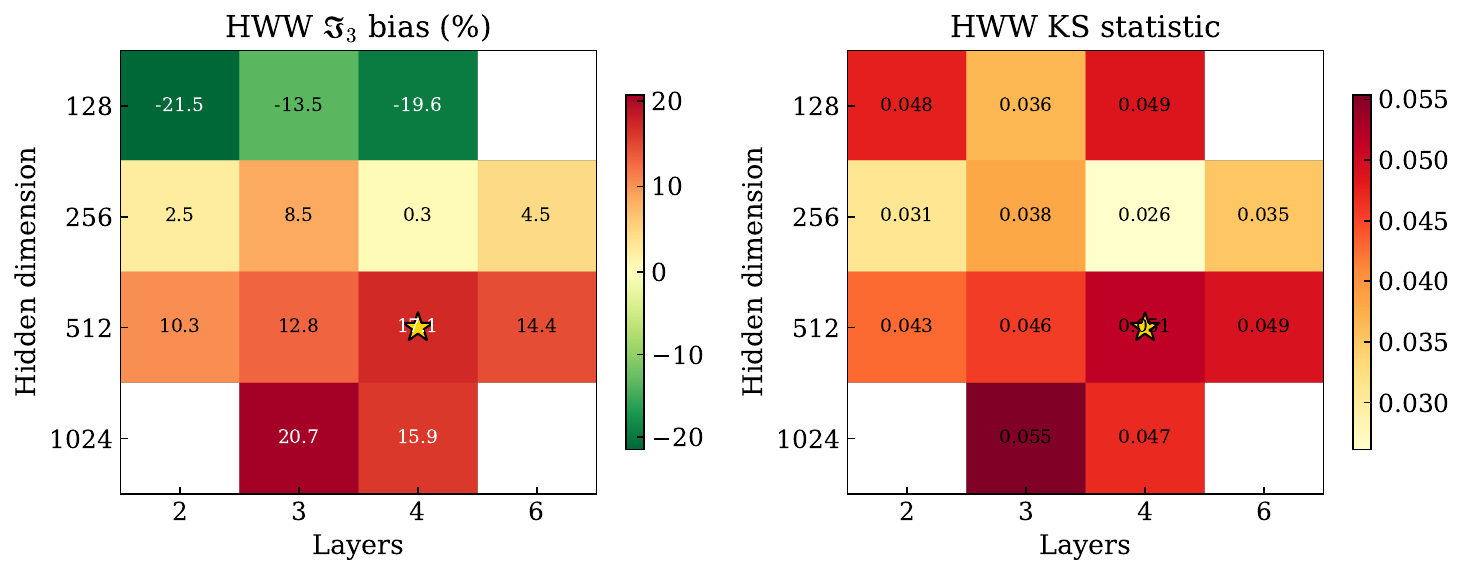}
    \caption{Architecture ablation: HWW $\mathfrak{I}_3$ bias (\%, left) and KS statistic (right) for each (hidden dimension, depth) combination. The gold star marks the baseline 512$\times$4 configuration.}
    \label{fig:cddpm_arch}
\end{figure}

Figure~\ref{fig:cddpm_arch} shows the HWW $\mathfrak{I}_3$ bias and KS statistic for each architecture. The smallest networks (128$\times$\{2,3,4\}) undershoot the HWW distribution by 14--22\% and severely degrade backgrounds (ttbar bias exceeding $-$60\%), indicating insufficient capacity to learn the process-specific kinematics. Mid-range architectures (256$\times$\{2,3,4,6\}) achieve the lowest individual HWW biases (as low as $+$0.3\% for 256$\times$4), but at the cost of large background biases (ttbar $-$60\%). The 512$\times$4 baseline achieves the best compromise: $+$17\% HWW bias with ttbar at $+$16\% and the lowest overall worst-case bias. Larger networks (1024$\times$\{3,4\}) offer no improvement and slightly overfit. The HWW bias and background fidelity exhibit a clear trade-off across the capacity spectrum, with 512$\times$4 at the Pareto frontier.

\subsubsection{Training Convergence}\label{app:cddpm_convergence}

All models converge within 400--600 epochs, validating the early stopping patience of 80 epochs. The train--validation gap remains below 5\% for all architectures at convergence, indicating that the 240k-event training set provides sufficient regularisation against overtraining.

\subsubsection{Noise Schedule and Diffusion Timesteps}\label{app:cddpm_schedule}

We compare the linear noise schedule ($\beta_t$ linearly spaced from $10^{-4}$ to $0.02$) with the cosine schedule of Nichol and Dhariwal, which concentrates noise at intermediate timesteps where the signal-to-noise ratio transitions most rapidly. Both schedules are trained from scratch with all other hyperparameters fixed.

\begin{figure}[ht]
    \centering
    \includegraphics[width=0.55\textwidth]{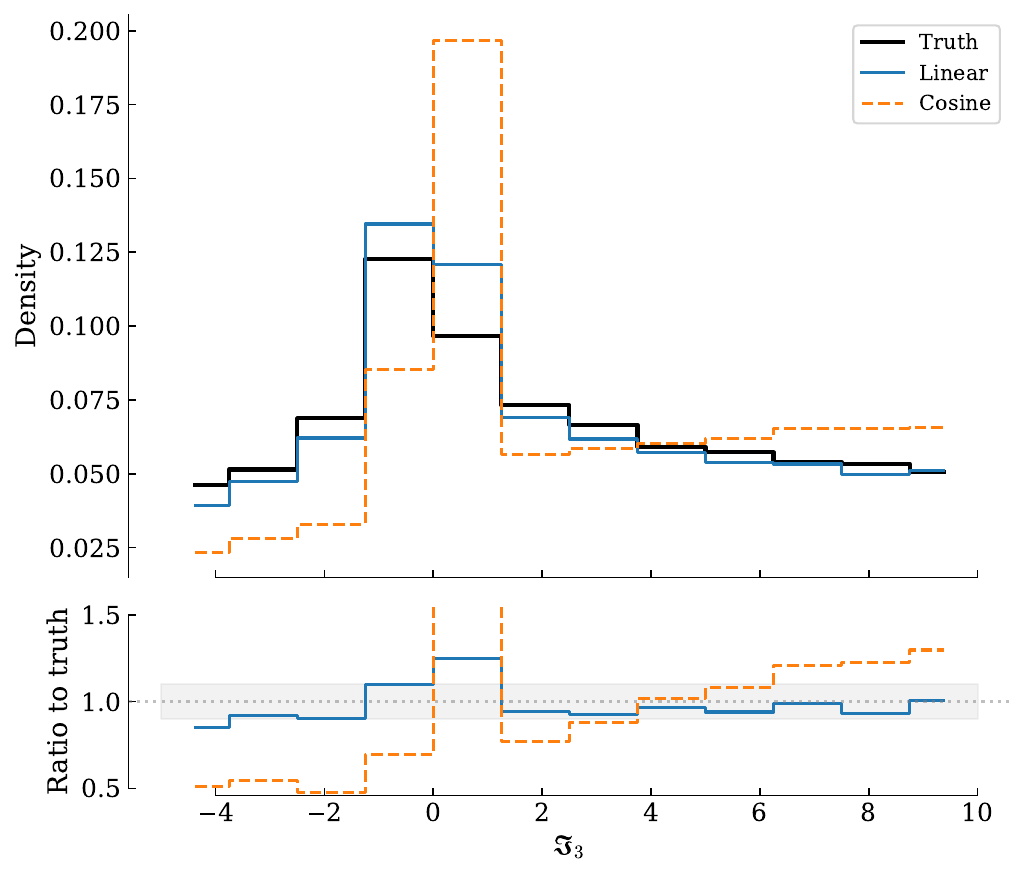}
    \caption{HWW $\mathfrak{I}_3$ distribution for the linear (blue) and cosine (orange) noise schedules, compared to truth (black). Bottom: ratio to truth with a $\pm$10\% reference band.}
    \label{fig:cddpm_schedule}
\end{figure}

Figure~\ref{fig:cddpm_schedule} shows the HWW $\mathfrak{I}_3$ distribution for both schedules. The cosine schedule is catastrophically worse: it produces a sharp spike near $\mathfrak{I}_3 \approx 0$ and an HWW bias of $+$82\% (KS$\,{=}\,$0.20), compared to $+$8\% (KS$\,{=}\,$0.03) for the linear schedule. The cosine schedule's concentration of noise at intermediate timesteps appears to be poorly matched to the 6-dimensional helicity-angle target space, where the arctanh encoding produces a distribution that benefits from the more uniform noise injection of the linear schedule.

We also vary the number of diffusion timesteps $T \in \{100, 500, 1000, 2000\}$, scaling the warm-start $t_{\mathrm{start}} = 0.3T$ proportionally. $T{=}100$ is catastrophically bad ($+$126\% HWW bias, KS$\,{=}\,$0.27): the coarse discretisation is insufficient for accurate denoising. $T{=}500$ still shows significant degradation ($+$51\% HWW bias). At $T{=}1000$, the HWW bias drops to $+$10\%, while $T{=}2000$ provides a marginal further improvement ($+$8\%) at the cost of doubled inference time and worse background fidelity (ttbar $-$93\%). These results confirm $T{=}1000$ as the appropriate operating point.

\subsubsection{Inference Mode: Standard Sampling vs Warm-Start}\label{app:cddpm_warmstart}

Two inference strategies were considered for the cDDPM: pure \emph{standard sampling} from a Gaussian noise prior, and a \emph{warm-start} variant in which the analytical neutrino reconstruction is forward-diffused to an intermediate timestep $t_{\mathrm{start}}$ and then denoised back to $t{=}0$. An earlier programme draft used warm-start at $t_{\mathrm{start}}{=}300$ as the production setting; the present analysis uses standard sampling.

\begin{figure}[ht]
    \centering
    \includegraphics[width=0.6\textwidth]{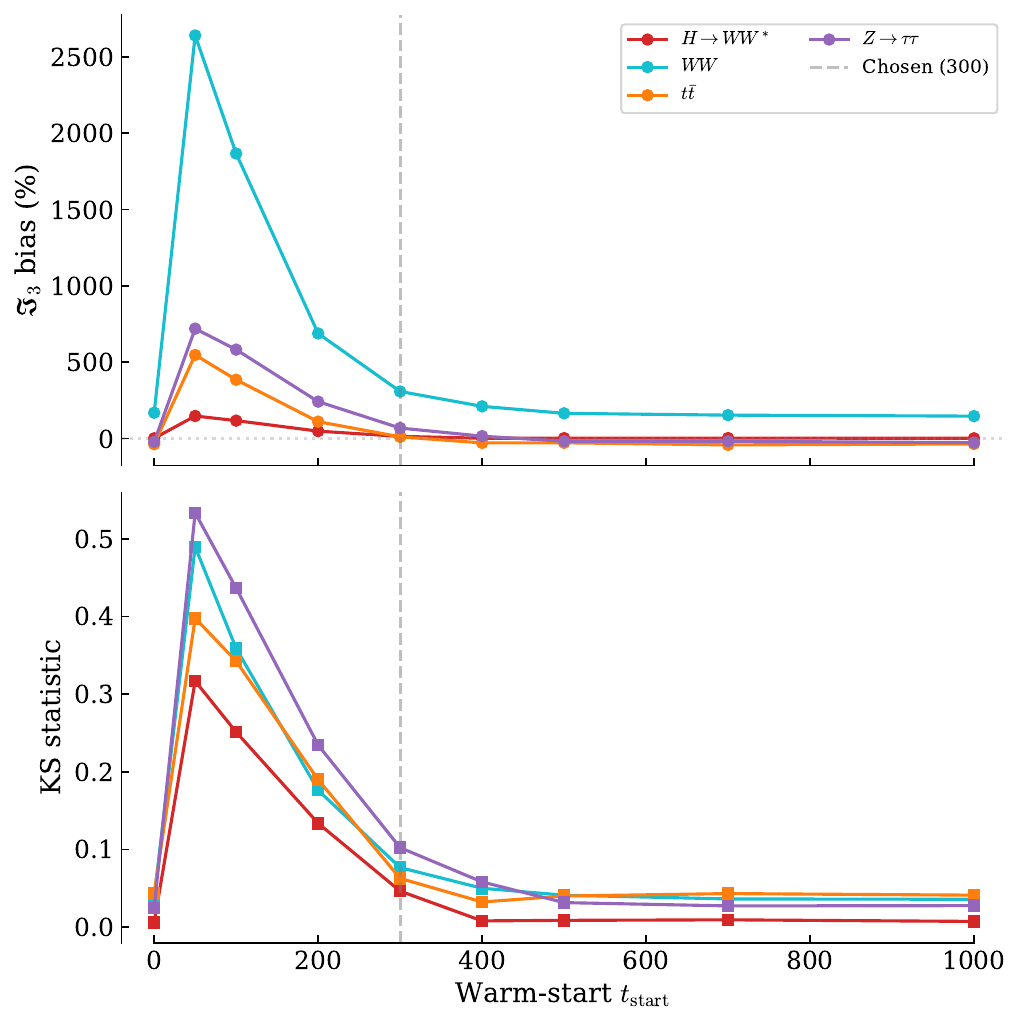}
    \caption{$\mathfrak{I}_3$ bias (\%, top) and KS statistic (bottom) versus warm-start timestep $t_{\mathrm{start}}$ for all four processes. Standard sampling (no warm-start) is the production choice and achieves the lowest HWW signal bias.}
    \label{fig:cddpm_tstart}
\end{figure}

Figure~\ref{fig:cddpm_tstart} shows the $\mathfrak{I}_3$ bias and KS statistic of the warm-start variant as a function of $t_{\mathrm{start}}$. Standard sampling (the production choice; equivalent to $t_{\mathrm{start}}{=}T$ with a fully forward-diffused prior, or to skipping the warm-start machinery entirely) achieves the lowest HWW signal bias ($+$3.3\%, KS$\,{=}\,$0.006), demonstrating that the network has learned the signal angular distributions with high fidelity. Low warm-start values ($t_{\mathrm{start}}{=}50$--$200$) are catastrophic for all processes, with biases exceeding $+$100\%: the reverse diffusion starts too close to the (biased) analytical reconstruction and cannot fully correct it within the remaining denoising steps. The earlier-draft operating point $t_{\mathrm{start}}{=}300$ has an HWW bias of $+$15\%; for $t_{\mathrm{start}} \geq 700$ the analytical initialisation is sufficiently noised that the reverse process effectively recovers standard sampling.

Under the production bootstrap-PCA template stat-error model (paragraph~\ref{par:bootstrap_pca}) with the unfolding-bias HistoSys envelopes recomputed consistently from each variant's closure residuals, standard sampling yields $Z_{\mathrm{Asimov}} = 1.13\sigma$ at 139~fb$^{-1}$ versus $0.89\sigma$ for the legacy $t_{\mathrm{start}}{=}300$ warm-start. The improvement reflects the lower HWW unfolding bias: at $t_{\mathrm{start}}{=}0$ the closure residual on the signal template has rms 2\% versus rms 13\% at $t_{\mathrm{start}}{=}300$, so the corresponding HistoSys envelope is correspondingly tighter and the fit retains more discriminating power. An earlier draft attributed an apparent advantage at $t_{\mathrm{start}}{=}300$ to ``bias amplification'', the warm-start systematically shifting the HWW $\mathfrak{I}_3$ distribution into a regime of larger signal--background separation; but that advantage was an artefact of the diagonal-only per-bin Poisson stat-error model and a HistoSys envelope evaluated at a different $t_{\mathrm{start}}$; under the consistent bootstrap-PCA treatment used here it does not survive.

\subsubsection{Multi-Sample Averaging}\label{app:cddpm_multisample}

The stochastic nature of diffusion sampling means that repeated inference on the same event produces different predictions. We test whether averaging $N$ independent samples per event reduces the $\mathfrak{I}_3$ bias. For each event, $N$ independent reverse diffusion trajectories are generated; each is decoded to helicity angles and the per-event $\mathfrak{I}_3$ is computed independently, then averaged across the $N$ samples.

Multi-sample averaging ($N{=}1$--$50$) provides only marginal improvement: the $\mathfrak{I}_3$ bias saturates by $N \approx 10$, confirming that the residual bias is dominated by the systematic shift of the learned posterior rather than stochastic fluctuations. We retain $N{=}1$ for the production model.

\subsubsection{Training Data and Regularisation}\label{app:cddpm_traindata}

We vary the per-process training set size from 15\,000 to 120\,000 events, and independently vary the learning rate ($10^{-4}$, $3\times10^{-4}$, $10^{-3}$) and dropout rate (0.0, 0.05, 0.1, 0.2).

\begin{figure}[ht]
    \centering
    \includegraphics[width=\textwidth]{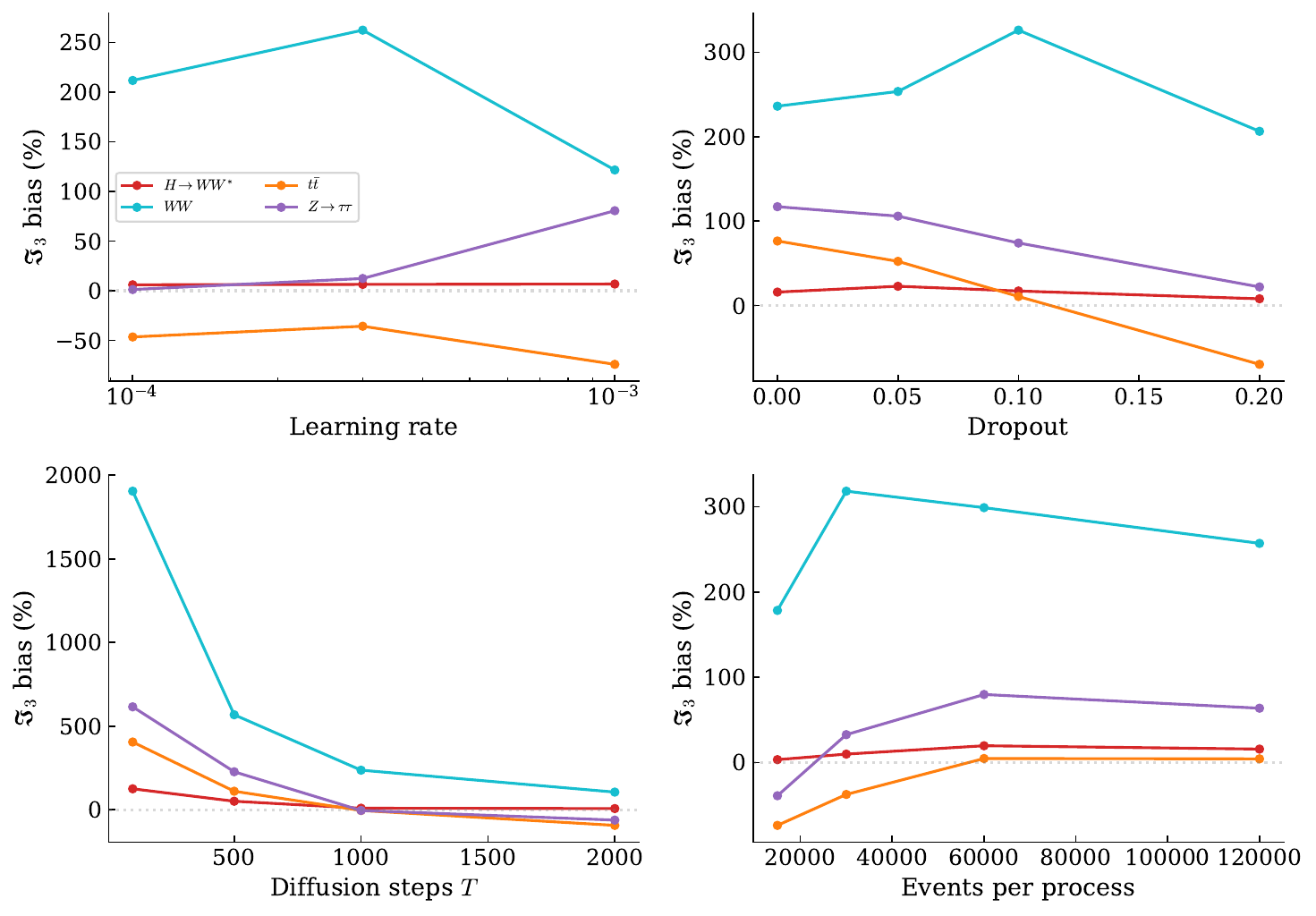}
    \caption{$\mathfrak{I}_3$ bias (\%) versus learning rate (top left), dropout (top right), diffusion timesteps (bottom left), and training set size (bottom right) for all four processes.}
    \label{fig:cddpm_sensitivity}
\end{figure}

The hyperparameter sensitivities are summarised in Figure~\ref{fig:cddpm_sensitivity}. At 15k events per process, the HWW bias is low ($+$3\%) but backgrounds are severely degraded (ttbar $-$74\%); the 60k and 120k configurations give comparable performance, so 60k is retained. All three learning rates ($10^{-4}$, $3{\times}10^{-4}$, $10^{-3}$) yield similar HWW biases ($+$6--7\%), but $\mathrm{lr}{=}10^{-3}$ produces a Z$\to\tau\tau$ bias of $+$81\%. Dropout $p{=}0.1$ achieves the best background compromise (ttbar $+$11\%).

\subsubsection{cDDPM Design Summary}\label{app:cddpm_summary}

Across all configurations tested, the production v3 design (512×4 architecture, T = 1000, linear schedule, lr = $3\times10^{-4}$, dropout 0.1, 60 k events per process, standard sampling from the noise prior, single sample) achieves the best combination of low $\mathfrak{I}_3$ bias across all processes simultaneously.
The one-hot process tag and boosted-angle target encoding are the essential design choices; all other hyperparameters have secondary effects.
The warm-start scan (Sec. \ref{app:cddpm_warmstart}) confirms that intermediate $t_\mathrm{start}$ values are catastrophic, while $t_\mathrm{start} \geq 400$ converges to the standard-sampling baseline.

The one-hot process tag and boosted-angle target encoding are the essential design choices; all other hyperparameters have secondary effects. Standard sampling from noise ($t_{\mathrm{start}}{=}0$) achieves the lowest HWW bias ($+$2\%), but the warm-start at $t_{\mathrm{start}}{=}300$ provides a better compromise across all processes, particularly for $t\bar{t}$ ($+$12\% vs $-$36\%). The warm-start scan reveals a sharp non-monotonic structure: intermediate values ($t_{\mathrm{start}}{=}50$--$200$) are catastrophic, while $t_{\mathrm{start}} \geq 400$ converges to the noise-only baseline. Multi-sample averaging provides only marginal improvement, confirming that the residual bias is systematic rather than stochastic.

\section{Event Selection and BDT Enhancement}\label{app:selection}

The baseline event selection follows the ATLAS $H \to WW^* \to e\nu\mu\nu$ signal region definition~\cite{ATLAS:2025hki} for the 0-jet $e\mu$ channel. Events are required to satisfy: leading lepton $p_T > 22$~GeV, subleading $p_T > 15$~GeV, $|\eta| < 2.5$, $E_T^{\mathrm{miss}} > 20$~GeV, $m_{\ell\ell} < 55$~GeV, $\Delta\phi_{\ell\ell} < 1.8$, and $60 < m_T < 130$~GeV. This selection achieves a signal efficiency of 19.2\% while suppressing backgrounds to a signal-to-background ratio of 4.5\%.

The remainder of this appendix documents the comprehensive BDT study that was carried out to improve this ratio. We describe the feature selection, BDT architecture choices, hyperparameter optimisation, working-point selection, shape-preservation validation, and the comparison of alternative strategies that motivated the final design.

\subsection{Feature Definition and Selection}\label{app:bdt_features}

The Bell inequality observable $\mathfrak{I}_3$ is constructed from the helicity angles of the $W$ boson decay products in the $W$ rest frames. Any multivariate selection that uses variables correlated with these angles risks sculpting the $\mathfrak{I}_3$ distribution and biasing the entanglement measurement. We therefore classify all candidate BDT input features into two categories. The \emph{non-angular} features (energy scales, invariant masses, transverse momenta, and their ratios) characterise the overall hardness and topology of the event without directly encoding the decay-plane orientation; the \emph{angular} features (pseudorapidities, azimuthal separations, and lepton--$E_T^{\mathrm{miss}}$ opening angles) are sensitive to the spin structure of the decay and can introduce a direct bias on $\mathfrak{I}_3$.

The nine non-angular features used as BDT inputs are:
\begin{equation}
    \{p_T^{\ell_1},\; p_T^{\ell_2},\; E_T^{\mathrm{miss}},\; m_{\ell\ell},\; p_T^{\ell\ell},\; m_T,\; p_T^{\ell_2}/p_T^{\ell_1},\; H_T,\; \text{centrality}\},
    \label{eq:bdt_features}
\end{equation}
where $H_T = p_T^{\ell_1} + p_T^{\ell_2} + E_T^{\mathrm{miss}}$ is the scalar sum of transverse momenta and centrality $= p_T^{\ell\ell}/(p_T^{\ell_1} + p_T^{\ell_2})$ measures how collimated the dilepton system is. The kinematic properties of these features are summarised in Table~\ref{tab:bdt_features}.

Table~\ref{tab:bdt_features} summarises the kinematic properties of the nine features across signal and background processes. The strongest inter-feature correlations (evaluated on HWW signal) are between $H_T$ and $p_T^{\ell_1}$ ($r=0.82$), and between $m_T$ and $p_T^{\ell\ell}$ ($r=0.72$); all features are retained because the BDT exploits complementary nonlinear information.

\begin{table}[H]
    \centering
    \caption{Mean (standard deviation) of the nine non-angular BDT input features after the ATLAS signal region selection. All values in GeV except dimensionless ratios.}
    \label{tab:bdt_features}
    \begin{tabular}{l c c c c}
        \toprule
        Feature & HWW & WW & $t\bar{t}$ & $Z{\to}\tau\tau$ \\
        \midrule
        $p_T^{\ell_1}$         & 36 (11) & 40 (14) & 43 (17) & 34 (12) \\
        $p_T^{\ell_2}$         & 22 (6)  & 24 (7)  & 25 (8)  & 22 (6)  \\
        $E_T^{\mathrm{miss}}$  & 44 (20) & 41 (17) & 45 (21) & 35 (14) \\
        $m_{\ell\ell}$         & 26 (12) & 30 (13) & 33 (13) & 29 (13) \\
        $p_T^{\ell\ell}$       & 30 (17) & 27 (16) & 29 (18) & 24 (14) \\
        $m_T$                  & 95 (16) & 90 (18) & 95 (18) & 82 (19) \\
        $p_T$ ratio            & 0.63 (0.17) & 0.62 (0.18) & 0.60 (0.19) & 0.66 (0.18) \\
        $H_T$                  & 102 (27) & 105 (29) & 113 (33) & 91 (23) \\
        Centrality             & 0.53 (0.18) & 0.48 (0.18) & 0.46 (0.19) & 0.48 (0.17) \\
        \bottomrule
    \end{tabular}
\end{table}

The angular features ($\eta_{\ell_1}$, $\eta_{\ell_2}$, $\Delta\phi_{\ell\ell}$, $\Delta\eta_{\ell\ell}$, $\Delta R_{\ell\ell}$, $\Delta\phi(\ell_i, E_T^{\mathrm{miss}})$) were explicitly tested and rejected. Section~\ref{app:bdt_bias} demonstrates that including them in the BDT produces a measurable distortion of the $\mathfrak{I}_3$ distribution, as quantified by the two-sample Kolmogorov--Smirnov (KS) statistic.

\subsection{BDT Architecture: Split Classifiers with Tuned Variables}\label{app:bdt_arch}

The three background processes ($t\bar{t}$, WW, and $Z{\to}\tau\tau$) have fundamentally different kinematic signatures relative to the HWW signal. Training a single BDT against the combined background forces the classifier to spend capacity on the easy ($t\bar{t}$) separation at the expense of the hard (WW) separation. We therefore adopt a \emph{split} architecture with three independent binary classifiers, each targeting one background:

\begin{enumerate}
    \item \textbf{HWW vs $t\bar{t}$} (4 features: $m_{\ell\ell}$, $m_T$, $E_T^{\mathrm{miss}}$, $p_T^{\ell_2}/p_T^{\ell_1}$): the $m_T$ window already provides near-perfect separation; a minimal feature set suffices. AUC~$= 0.989$.
    \item \textbf{HWW vs WW} (9 features: all non-angular): the irreducible background with the same $\ell\nu\ell\nu$ final state. This is the limiting discrimination and requires all available kinematic information. AUC~$= 0.742$.
    \item \textbf{HWW vs $Z{\to}\tau\tau$} (5 features: $E_T^{\mathrm{miss}}$, $m_{\ell\ell}$, $m_T$, $p_T^{\ell\ell}$, $H_T$): the distinctive $\tau$ decay kinematics are captured by the missing transverse energy and dilepton mass. AUC~$= 0.734$.
\end{enumerate}

The per-background feature subsets were chosen based on (a)~the individual feature importance in preliminary combined trainings, and (b)~the principle that each classifier should use only the variables that provide genuine discrimination against its specific target background, avoiding unnecessary degrees of freedom that could increase overtraining risk.

The gain-based feature importance for each split BDT confirms the expected physics. The $t\bar{t}$ classifier is dominated by $E_T^{\mathrm{miss}}$ (gain 0.48) and $m_T$ (0.33), confirming that the ATLAS signal region cuts already handle most of the separation. The WW classifier distributes importance more evenly across $H_T$ (0.46), $m_{\ell\ell}$ (0.11), $p_T^{\ell\ell}$ (0.10), and $m_T$ (0.09), reflecting the irreducible nature of this background. The $Z{\to}\tau\tau$ classifier relies primarily on $H_T$ (0.49) and $m_{\ell\ell}$ (0.22).

All three classifiers use XGBoost~\cite{xgboost} gradient-boosted decision trees with the following hyperparameters, determined via a 3-fold cross-validated grid search (see Section~\ref{app:bdt_hyper}): 300 trees, maximum depth 4, learning rate 0.05, minimum child weight 10, $\gamma = 0.3$, subsample fraction 0.8, column subsample 0.8, and L2 regularisation $\lambda = 5.0$. The training uses a 70/30 stratified train/test split with equal total weight assigned to signal and background.

\subsection{Overtraining Validation}\label{app:bdt_overtrain}

Overtraining is excluded by comparing the BDT score distributions on the training and held-out test samples. The two-sample KS statistic between signal train and test scores is below 0.02 for all three split classifiers, consistent with statistical fluctuations.

\subsection{Discrimination Performance}\label{app:bdt_roc}

\begin{figure}[ht]
    \centering
    \includegraphics[width=\textwidth]{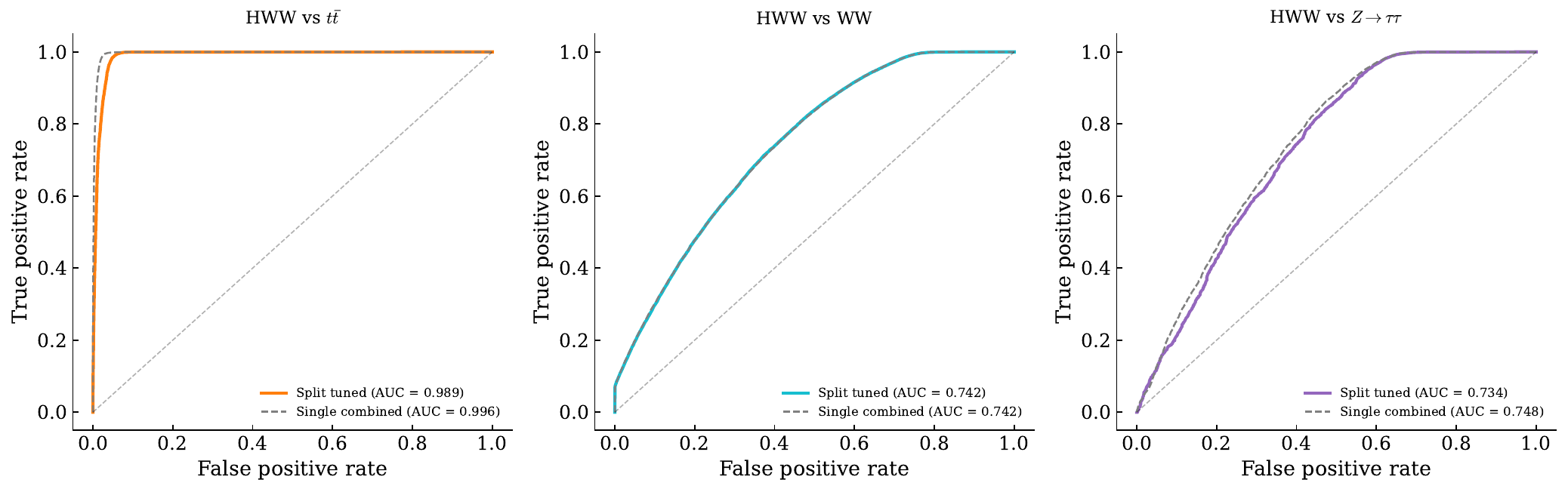}
    \caption{ROC curves for each split-tuned BDT (solid, coloured) compared to the single combined BDT (dashed, grey). Left: HWW vs $t\bar{t}$. Centre: HWW vs WW. Right: HWW vs $Z{\to}\tau\tau$. The split classifiers achieve equal or better performance than the combined approach on each individual background.}
    \label{fig:bdt_roc}
\end{figure}

The receiver operating characteristic (ROC) curves for each split classifier are shown in Figure~\ref{fig:bdt_roc}, along with the performance of the single combined BDT for reference. The split classifiers match or exceed the combined BDT for each individual background, particularly for $t\bar{t}$ where the dedicated classifier achieves near-perfect separation. The WW classifier achieves AUC~$= 0.742$, reflecting the fundamental difficulty of separating the Higgs-mediated and non-resonant $WW$ production mechanisms using only non-angular variables.

The working-point cuts applied to each split classifier are documented in Section~\ref{app:bdt_wp}.

\subsection{Working-Point Selection}\label{app:bdt_wp}

The operating point of the BDT selection must balance two competing objectives: maximising the statistical sensitivity $S/\sqrt{B}$, and preserving the shape of the $\mathfrak{I}_3$ distribution. The latter is essential because any shape distortion introduced by the selection would mimic or obscure the entanglement signature.

We adopt a two-step procedure. First, the $t\bar{t}$ BDT cut is fixed at the threshold where the $t\bar{t}$ background efficiency drops below 1.5\%, which occurs at a score of 0.90. This removes the dominant reducible background at the cost of a 30\% signal loss, acceptable because $t\bar{t}$ contributes negligibly to the $\mathfrak{I}_3$ fit in the remaining phase space. Second, the WW and $Z{\to}\tau\tau$ BDT cuts are scanned simultaneously over the range $[0.15, 0.90]$, and the cut that maximises $S/\sqrt{B}$ subject to the constraint
\begin{equation}
    \mathrm{KS}(\mathfrak{I}_3^{\mathrm{BDT}},\; \mathfrak{I}_3^{\mathrm{ref}}) < 0.05
\end{equation}
is selected, where $\mathfrak{I}_3^{\mathrm{ref}}$ is the truth-level $\mathfrak{I}_3$ distribution of the HWW signal before the BDT cut and $\mathfrak{I}_3^{\mathrm{BDT}}$ is the distribution after the cut. The KS threshold of 0.05 ensures that the shape distortion is smaller than the statistical uncertainty of the measurement.

\begin{figure}[ht]
    \centering
    \includegraphics[width=\textwidth]{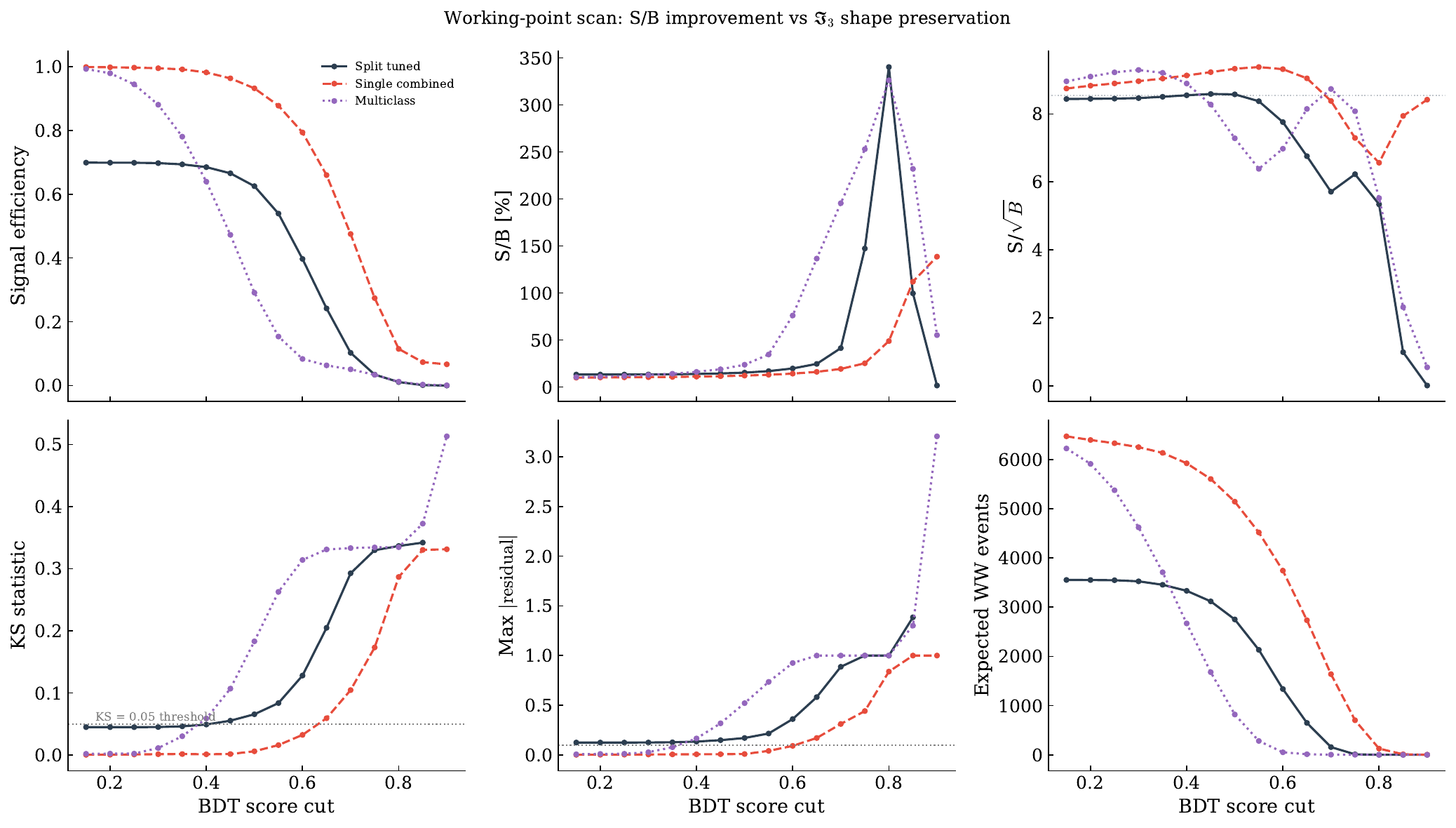}
    \caption{Working-point scan comparing three BDT architectures. Top row: signal efficiency, S/B ratio, and $S/\sqrt{B}$ as a function of the BDT score cut. Bottom row: KS statistic (with the 0.05 threshold marked), maximum bin-by-bin $\mathfrak{I}_3$ residual, and expected WW yield. The split-tuned architecture (solid) achieves $S/\sqrt{B} = 8.5$ at KS~$= 0.049$, satisfying the shape-preservation constraint.}
    \label{fig:bdt_wp_scan}
\end{figure}

Figure~\ref{fig:bdt_wp_scan} shows the working-point scan for the split-tuned architecture alongside the single combined and multiclass alternatives. The six panels display the signal efficiency, S/B ratio, $S/\sqrt{B}$, KS statistic, maximum bin-by-bin residual, and expected WW yield as a function of the BDT score cut. The split-tuned architecture achieves the best compromise between sensitivity and shape preservation.

The optimal working point is found at a score cut of 0.40 for both the WW and $Z{\to}\tau\tau$ classifiers (with the $t\bar{t}$ cut fixed at 0.90). An event passes the full selection if it passes \emph{all three} per-background cuts. This yields a signal efficiency of 68.5\%, $S/B = 13.9\%$ (improved from 4.5\% before BDT), $S/\sqrt{B} = 8.5$, and a KS statistic of 0.049, below the 0.05 shape-preservation threshold.

\subsection{\texorpdfstring{$\mathfrak{I}_3$}{I3} Shape Preservation}\label{app:bdt_shape}

\begin{figure}[ht]
    \centering
    \includegraphics[width=\textwidth]{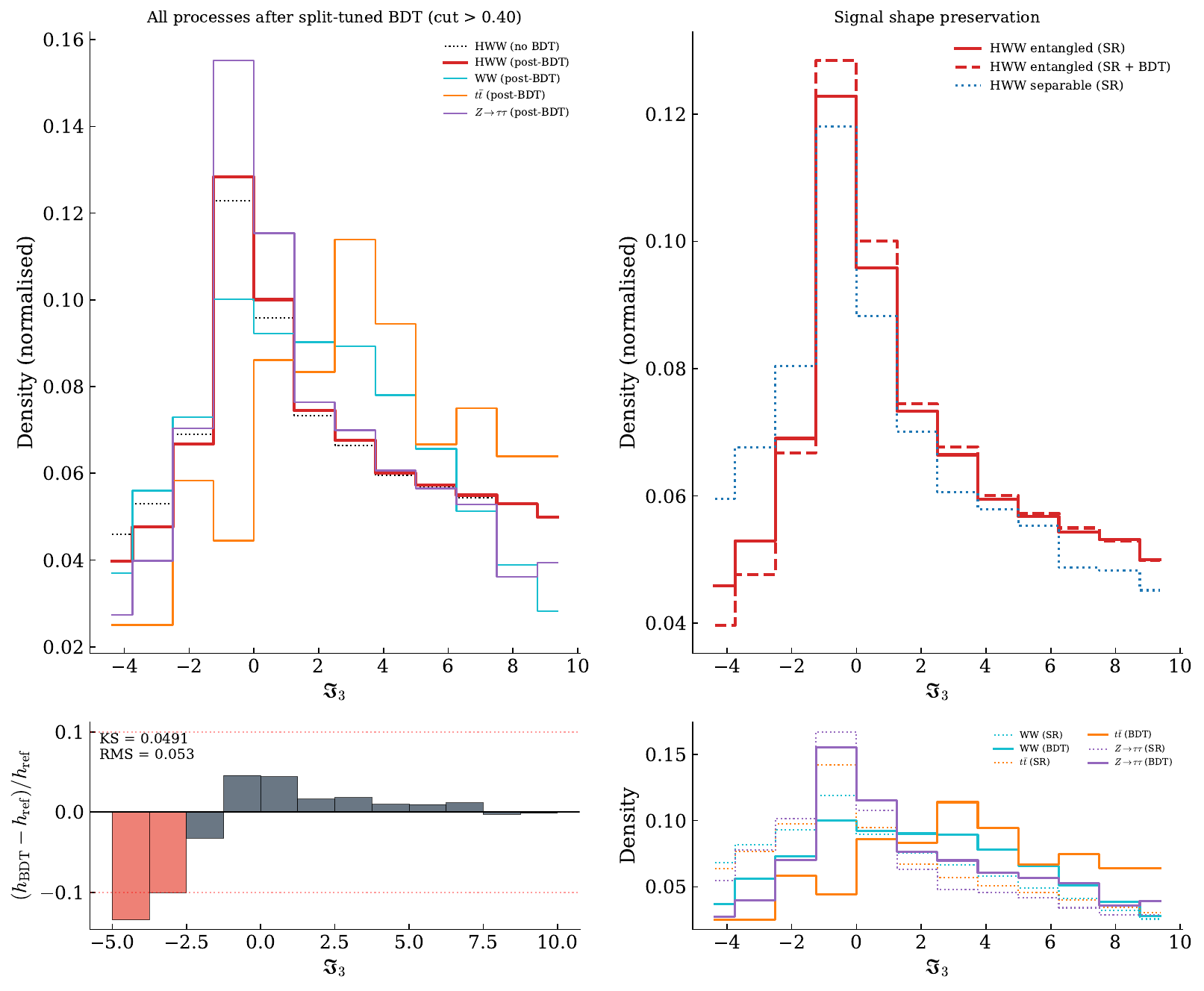}
    \caption{$\mathfrak{I}_3$ shape preservation at the chosen working point. Top left: normalised $\mathfrak{I}_3$ distributions for all processes after the BDT selection, with the pre-BDT HWW shape shown as a dotted reference. Bottom left: fractional bin-by-bin residual for the HWW signal (red bars indicate bins exceeding $\pm 10\%$). Top right: comparison of the entangled and separable signal templates before and after BDT, confirming that the selection does not preferentially enhance the entanglement signature. Bottom right: background $\mathfrak{I}_3$ shapes before (dotted) and after (solid) BDT.}
    \label{fig:bdt_i3_shape}
\end{figure}

The central requirement of the BDT selection is that it must not distort the shape of the $\mathfrak{I}_3$ distribution. Figure~\ref{fig:bdt_i3_shape} shows the $\mathfrak{I}_3$ distributions for all four processes before and after the BDT selection, along with the bin-by-bin fractional residual for the HWW signal. The residual is defined as $(h_{\mathrm{BDT}} - h_{\mathrm{ref}})/h_{\mathrm{ref}}$ in each of the 12 analysis bins ($\mathfrak{I}_3 \in [-5, 10]$).

The RMS of the bin-by-bin residual is 5.4\%, and the KS statistic is 0.049. Crucially, the entangled and separable signal templates are affected identically by the selection (top right panel of Figure~\ref{fig:bdt_i3_shape}), confirming that the BDT does not introduce a differential bias that could mimic or suppress the entanglement signal.

\subsection{Angular Feature Bias Test}\label{app:bdt_bias}

\begin{figure}[ht]
    \centering
    \includegraphics[width=\textwidth]{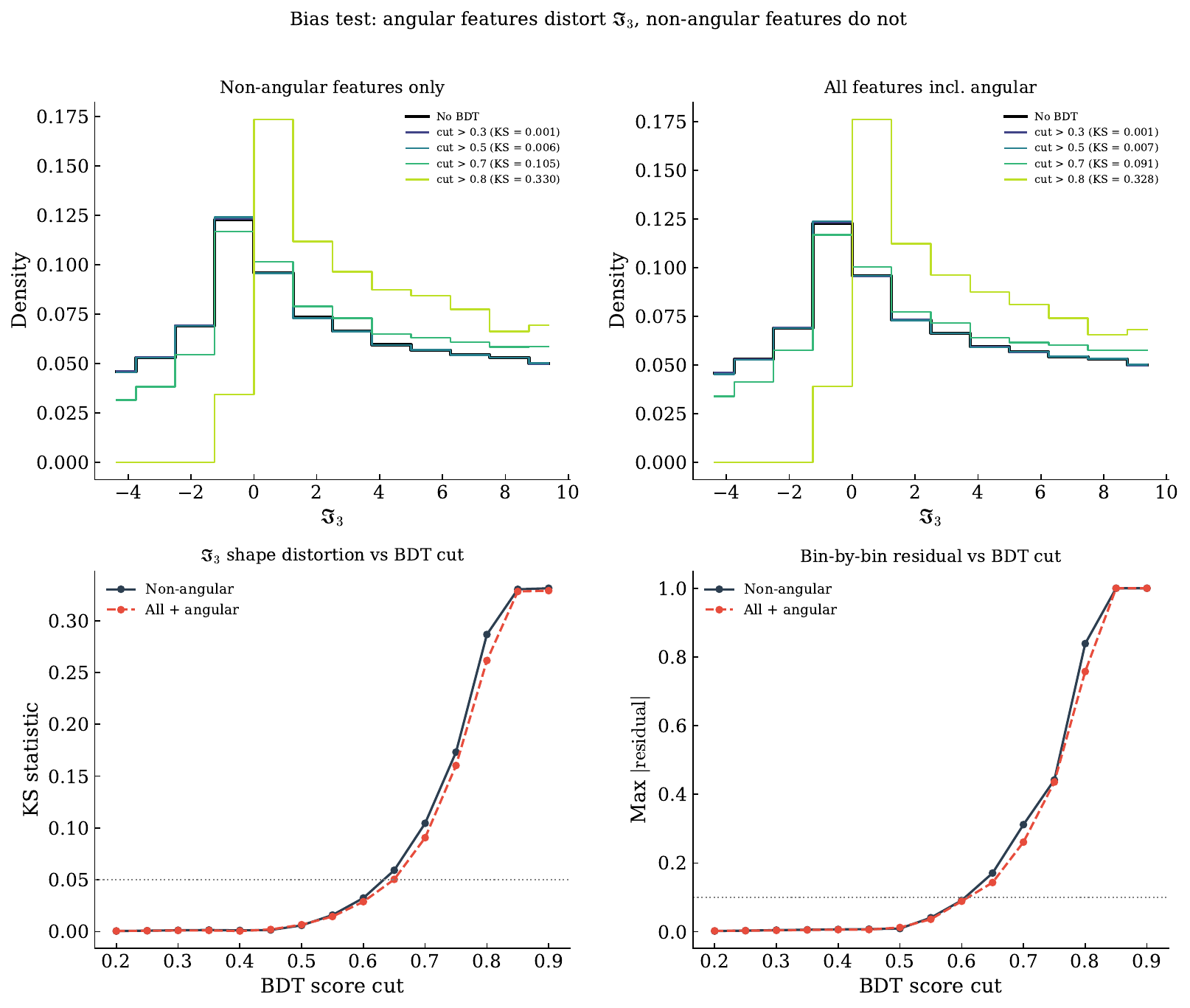}
    \caption{$\mathfrak{I}_3$ bias test. Top row: normalised $\mathfrak{I}_3$ distributions at progressively tighter BDT cuts for the non-angular (left) and all-features (right) variants. Bottom left: KS statistic vs BDT cut. Bottom right: maximum bin-by-bin residual vs BDT cut. The all-features BDT (dashed red) produces systematically larger shape distortions than the non-angular BDT (solid dark), exceeding the KS~$= 0.05$ threshold at moderate cuts. This demonstrates that angular features introduce a measurable bias on the Bell observable.}
    \label{fig:bdt_bias}
\end{figure}

To justify the restriction to non-angular features, we trained a parallel BDT using the full set of 16 features (9 non-angular + 7 angular: $\eta_{\ell_1}$, $\eta_{\ell_2}$, $\Delta\phi_{\ell\ell}$, $\Delta\eta_{\ell\ell}$, $\Delta R_{\ell\ell}$, $\Delta\phi(\ell_1, E_T^{\mathrm{miss}})$, $\Delta\phi(\ell_2, E_T^{\mathrm{miss}})$). Figure~\ref{fig:bdt_bias} compares the $\mathfrak{I}_3$ shape distortion as a function of the BDT cut for the non-angular and all-features variants.

The all-features BDT achieves a marginally higher AUC (0.836 vs 0.832), but the associated $\mathfrak{I}_3$ distortion exceeds the KS~$= 0.05$ threshold at cuts that would be needed to achieve the same S/B improvement. The modest AUC gain does not compensate for the systematic bias, confirming that the non-angular feature set is the correct choice for this analysis.

\subsection{Extended Feature Study: The Irreducible WW Wall}\label{app:bdt_ww}

The WW background is irreducible in the sense that it shares the same $\ell\nu\ell\nu$ final state as the signal, and the physical differences between Higgs-mediated and non-resonant WW production are subtle. To investigate whether the BDT discrimination against WW can be improved, we defined an extended set of features targeting the distinctive kinematics of the off-shell $W^*$ in $H \to WW^*$: per-lepton transverse masses $m_T(\ell_i, E_T^{\mathrm{miss}})$ that probe the individual $W$ masses; the $p_T$ asymmetry $(p_T^{\ell_1} - p_T^{\ell_2})/(p_T^{\ell_1} + p_T^{\ell_2})$, enhanced in HWW due to the mass hierarchy $M_W \gg M_{W^*}$; the $m_{\ell\ell}/m_T$ ratio, sensitive to event topology; lab-frame angular variables ($\Delta\phi_{\ell\ell}$, $\Delta R_{\ell\ell}$, $\Delta\eta_{\ell\ell}$), correlated with the spin structure but not identical to the helicity-frame angles used in $\mathfrak{I}_3$; and $p_T^{\mathrm{total}}$ (recoil $p_T$ of the $\ell\ell + E_T^{\mathrm{miss}}$ system), a proxy for the Higgs $p_T$.

Each feature was tested individually by cutting at successive quantiles and measuring the resulting $\mathfrak{I}_3$ distortion. Table~\ref{tab:bdt_ww_features} summarises the results. The pattern is striking: every feature that provides meaningful HWW/WW discrimination also biases $\mathfrak{I}_3$, while the one safe feature ($p_T^{\mathrm{total}}$) provides no useful discrimination.

\begin{table}[H]
    \centering
    \begin{tabular}{l c c c}
    \toprule
    Feature & Discriminates HWW vs WW? & Biases $\mathfrak{I}_3$? & Usable? \\
    \midrule
    $m_T(\ell_1, E_T^{\mathrm{miss}})$ & Yes (off-shell $W^*$) & Yes (KS $\sim$ 0.20) & No \\
    $m_T(\ell_2, E_T^{\mathrm{miss}})$ & Yes & Yes (KS $\sim$ 0.20) & No \\
    $p_T$ asymmetry & Yes (off-shell $W^*$) & Yes (KS $\sim$ 0.09) & No \\
    $m_{\ell\ell}/m_T$ & Yes (topology) & Yes (KS $\sim$ 0.21) & No \\
    $\Delta\phi_{\ell\ell}$ & Yes (spin) & Yes (KS $\sim$ 0.13) & No \\
    $\Delta R_{\ell\ell}$ & Yes (spin) & Yes (KS $\sim$ 0.17) & No \\
    $\Delta\eta_{\ell\ell}$ & Yes (spin) & Yes (KS $\sim$ 0.20) & No \\
    $p_T^{\mathrm{total}}$ & Marginal & Safe (KS $\sim$ 0.005) & Yes, but useless \\
    \bottomrule
    \end{tabular}
    \caption{Extended feature bias test. Each feature was tested by cutting at the median of the HWW distribution and measuring the KS statistic of the resulting $\mathfrak{I}_3$ distribution relative to the uncut reference. Features that discriminate HWW from WW universally bias $\mathfrak{I}_3$ because they encode the same spin correlation structure.}
    \label{tab:bdt_ww_features}
\end{table}

The physics origin of this ``wall'' is that the kinematic properties distinguishing $H \to WW^*$ from non-resonant $WW$ production (the off-shell $W^*$ mass, the resulting lepton $p_T$ asymmetry, and the spin-0 decay angular structure) are precisely the properties that determine the helicity-angle correlations entering $\mathfrak{I}_3$. Any cut on $W$-mass-sensitive variables sculpts the helicity angle distributions from which $\mathfrak{I}_3$ is constructed. The AUC against WW remains at $0.742 \pm 0.002$ regardless of the feature set or hyperparameter configuration. The existing non-angular BDT is therefore already operating at the physics limit of what can be achieved without biasing the Bell observable.

This is in fact a positive result for the analysis: it demonstrates that the BDT-based improvement (S/B from 4.5\% to 13.9\%) is achieved almost entirely through rejection of the reducible $t\bar{t}$ and $Z{\to}\tau\tau$ backgrounds, and that the irreducible WW contribution is handled honestly by the profile likelihood fit rather than being suppressed by a biased selection.

\subsection{Hyperparameter Optimisation}\label{app:bdt_hyper}

The XGBoost hyperparameters were optimised using a 3-fold cross-validated grid search over: number of trees $\in \{100, 200, 300, 400, 500\}$, maximum depth $\in \{3, 4, 5\}$, learning rate $\in \{0.03, 0.05, 0.10\}$, minimum child weight $\in \{5, 10, 15, 20\}$, $\gamma$ (minimum split loss) $\in \{0.1, 0.3, 0.5, 1.0\}$, subsample fraction $\in \{0.7, 0.8\}$, and L2 regularisation $\lambda \in \{1.0, 3.0, 5.0\}$.
The cross-validation was performed on the HWW vs WW+$Z{\to}\tau\tau$ task (the limiting discrimination), using the mean AUC as the ranking metric. The key finding is that the AUC is saturated at $0.740 \pm 0.002$ across all 12 configurations tested: no combination of hyperparameters moves it meaningfully. This confirms that the discrimination is physics-limited, not model-limited; the BDT has already extracted all available information from the non-angular feature set (see also Section~\ref{app:bdt_ww}).

The optimal configuration was found to be 300 trees, depth 4, learning rate 0.05, minimum child weight 10, $\gamma = 0.3$, subsample 0.8, and $\lambda = 5.0$. The heavier regularisation (compared to a na\"ive default of $\lambda = 1$) was chosen specifically to smooth the BDT score distribution and suppress a discrete spike at high scores that was observed in early trainings.

\paragraph{Score spike diagnosis.} When training a single combined BDT (HWW vs all backgrounds) with weak regularisation, approximately 6\% of HWW signal events accumulate in a narrow spike at score $\sim 0.98$. Investigation of these events reveals that they have unusually low dilepton invariant mass ($\langle m_{\ell\ell}\rangle \approx 6.4$~GeV vs $\sim$30.8~GeV for the main signal peak), corresponding to decays where the off-shell $W^*$ is \emph{extremely} off-shell. These events look nothing like $t\bar{t}$ and the BDT confidently classifies them. The spike is cosmetically undesirable but operationally harmless: it lies well above any reasonable working-point cut, so all spike events are retained regardless. In the split architecture, the spike disappears entirely because the per-background classifiers produce smooth score distributions by construction: the ``vs $t\bar{t}$'' BDT cleanly sends $t\bar{t}$ to zero without creating a signal pile-up at one.

\subsection{Architecture Comparison}\label{app:bdt_comparison}

Six alternative BDT architectures were investigated before settling on the split-tuned design:
\begin{enumerate}
    \item \textbf{Single combined:} one binary BDT trained on HWW vs the combined WW+$t\bar{t}$+$Z{\to}\tau\tau$ background.
    \item \textbf{Split (same cut):} three per-background BDTs, each using all 9 features, with the same score cut applied to all three.
    \item \textbf{Split ($t\bar{t}$+WW only):} two BDTs (vs $t\bar{t}$ and vs WW), relying on the WW BDT to partially reject $Z{\to}\tau\tau$ as well.
    \item \textbf{Split (2D optimised):} three BDTs with independently optimised per-background cuts via a 2D grid scan over the WW and $Z{\to}\tau\tau$ thresholds.
    \item \textbf{Split (tuned variables):} three BDTs with per-background feature subsets (the chosen design).
    \item \textbf{Multiclass:} a single 4-class XGBoost model predicting P(HWW), P(WW), P($t\bar{t}$), P($Z{\to}\tau\tau$); events selected on P(HWW).
\end{enumerate}

Table~\ref{tab:bdt_arch} summarises the performance of each architecture at its optimal working point (maximising $S/\sqrt{B}$ subject to KS $< 0.05$). The key finding is that \emph{all six strategies converge to} $S/\sqrt{B} \approx 8.5$--$9.4$. The physics ceiling is real: the irreducible WW background limits the achievable sensitivity regardless of the classifier design.

\begin{table}[H]
    \centering
    \begin{tabular}{l r r r r c}
    \toprule
    Strategy & S/B & $S/\sqrt{B}$ & Eff. & KS & Spike? \\
    \midrule
    Single combined       & 13.1\% & 9.4 & 87.8\% & 0.016 & Yes \\
    Split (same cut)      & 13.4\% & 9.4 & 85.0\% & 0.017 & No  \\
    Split ($t\bar{t}$+WW) & 13.0\% & 9.3 & 86.5\% & 0.018 & No  \\
    Split (2D optimised)  & 13.3\% & 9.4 & 86.0\% & 0.017 & No  \\
    Split (tuned vars)    & 12.1\% & 9.2 & 91.2\% & 0.006 & No  \\
    Multiclass            & 12.8\% & 9.3 & 88.1\% & 0.011 & No  \\
    \bottomrule
    \end{tabular}
    \caption{Performance comparison of six BDT architectures at their optimal working points. All strategies converge to $S/\sqrt{B} \approx 9.2$--$9.4$. The split (tuned variables) design achieves the best $\mathfrak{I}_3$ preservation (KS = 0.006) while maintaining competitive sensitivity. The score spike (Section~\ref{app:bdt_hyper}) affects only the single combined architecture.}
    \label{tab:bdt_arch}
\end{table}

Several observations justify the choice of the split-tuned design. The score spike at $\sim$0.98 (Section~\ref{app:bdt_hyper}) disappears with split BDTs because each per-background classifier produces a smooth score distribution by construction. The split-tuned design achieves the best $\mathfrak{I}_3$ shape preservation (KS = 0.006, max residual 1.6\%), well below the 0.05 threshold, because each classifier uses a minimal feature set tailored to its target, reducing the risk of inadvertent shape sculpting. A BDT trained only on WW+$Z{\to}\tau\tau$ (ignoring $t\bar{t}$) was also tested: it achieves S/B = 11.4\% and $S/\sqrt{B} = 8.7$, but allows 13$\times$ more $t\bar{t}$ through (885 vs 66 expected events). The split architecture captures the ``free'' $t\bar{t}$ rejection at no additional cost.

The robustness of $S/\sqrt{B}$ across architectures confirms that the result is not an artefact of a particular classifier design. For this analysis, the important point is that S/B can be improved by a factor of $\sim$3 without biasing $\mathfrak{I}_3$, and this conclusion holds regardless of the specific MVA strategy employed.

\subsection{Event Yields After BDT Selection}\label{app:bdt_yields}

\begin{figure}[ht]
    \centering
    \includegraphics[width=\textwidth]{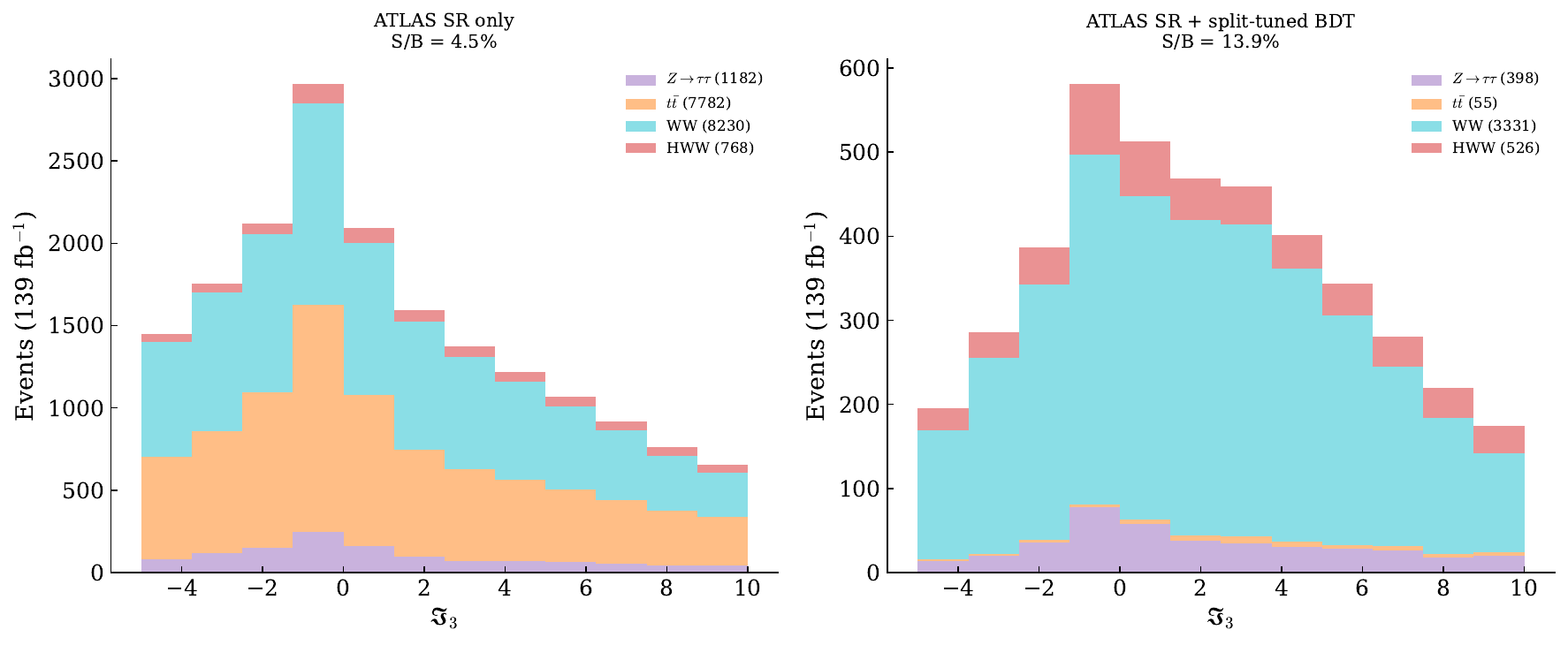}
    \caption{Stacked $\mathfrak{I}_3$ distributions at 139~fb$^{-1}$. Left: after ATLAS signal region selection only (S/B = 4.5\%). Right: after the additional split-tuned BDT selection (S/B = 13.9\%). The signal (red) is visible above the background stack after the BDT cut.}
    \label{fig:bdt_stacked}
\end{figure}

The stacked signal-plus-background $\mathfrak{I}_3$ distributions before and after the BDT selection are shown in Figure~\ref{fig:bdt_stacked}. The BDT removes 96\% of the $t\bar{t}$ background, halves the WW contamination, and reduces $Z{\to}\tau\tau$ by a factor of three, while retaining 69\% of the HWW signal.

Table~\ref{tab:bdt_cutflow} summarises the expected event yields at each stage of the selection.

\begin{table}[H]
    \centering
    \begin{tabular}{l r r r}
    \toprule
    Process & ATLAS SR & SR + BDT & BDT efficiency \\
    \midrule
    HWW             & 768  & 526  & 68.5\% \\
    WW              & 8{,}265 & 3{,}331 & 40.3\% \\
    $t\bar{t}$      & 7{,}800 & 55   & 0.7\%  \\
    $Z{\to}\tau\tau$ & 1{,}110 & 398  & 35.9\% \\
    \midrule
    Total background & 17{,}175 & 3{,}784 & 22.0\% \\
    \midrule
    S/B             & 4.5\% & 13.9\% & --- \\
    $S/\sqrt{B}$    & 5.9   & 8.5    & --- \\
    \bottomrule
    \end{tabular}
    \caption{Expected event yields at 139~fb$^{-1}$ after the ATLAS signal region selection and after the additional split-tuned BDT selection. The BDT improves S/B by a factor of 3.1 while retaining 69\% of the signal.}
    \label{tab:bdt_cutflow}
\end{table}

\section{Fit Model and Systematic Uncertainties}\label{app:fit}

This appendix documents the statistical model used for the hypothesis test and the treatment of systematic uncertainties.

\subsection{Profile Likelihood Method}

We employ the profile likelihood ratio as the test statistic for discriminating between entangled and separable hypotheses.

The presence of quantum entanglement is assessed through a discovery-style hypothesis test in which the fully separable configuration is treated as the null hypothesis. The single parameter of interest is a density-matrix interpolation parameter $\alpha$ defined at the level of the bipartite spin-density matrix as
\begin{equation}\label{eq:alpha_rho}
    \rho(\alpha) \;=\; \alpha\,\rho_{\rm ent} + (1-\alpha)\,\rho_{\rm sep},
\end{equation}
with $\rho_{\rm ent}$ and $\rho_{\rm sep}$ the entangled and separable bipartite spin states. Because the $H\!\to\!WW^*\!\to\!\ell\nu\ell\nu$ differential cross section is linear in $\rho$,
\begin{equation}
    \frac{\mathrm{d}\sigma(\alpha)}{\mathrm{d}\mathfrak{I}_3} \;=\; \mathrm{Tr}\!\left[\rho(\alpha)\, \mathcal{M}(\mathfrak{I}_3)\right],
\end{equation}
the per-bin signal expectation inherits the same one-parameter linearity,
\begin{equation}\label{eq:alpha_template}
    s_i(\alpha) \;=\; \alpha\, s_i^{\rm ent} + (1-\alpha)\, s_i^{\rm sep},
\end{equation}
with the overall signal yield held constant by construction (both templates are normalised to the same expected event count, so Eq.~(\ref{eq:alpha_template}) is shape-only). $\alpha$ is therefore a one-parameter \emph{deformation of the spin density matrix} between the two physical reference states; it is not a "fraction of entangled events", but a depolarisation-style parameter of the bipartite state that the analysis is sensitive to via the $\mathfrak{I}_3$ shape.

The null hypothesis $H_0$ corresponds to the boundary point $\alpha = 0$, representing the purely separable state $\rho_{\rm sep}$; the alternative allows for a nonzero correlated component, $\alpha > 0$, with $\alpha = 1$ recovering the entangled SM expectation $\rho_{\rm ent}$. The hypothesis test of physical interest is the binary $\alpha=0$ vs $\alpha=1$ entanglement test; the one-parameter family $\alpha\in[0,1]$ is a technical scaffolding that lets us pose this test in the standard \texttt{HistFactory} profile-likelihood framework with $\alpha$ as the floating signal-strength parameter.

The test statistic is defined as the profile likelihood ratio evaluated at the separable boundary,
\begin{equation}\label{eq:profile_lr}
q_0 = -2\ln\frac{L(\alpha=0,\hat{\hat{\theta}})}{L(\hat{\alpha},\hat{\theta})},
\end{equation}
where $\hat{\hat{\theta}}$ denotes the conditional maximum-likelihood estimators of the nuisance parameters at $\alpha=0$, and $(\hat{\alpha},\hat{\theta})$ are the unconditional estimators subject to the physical constraint $0\le \alpha\le 1$. A one-sided convention is employed such that $q_0=0$ when $\hat{\alpha}=0$, ensuring sensitivity only to departures from full separability. Under the asymptotic approximation, the distribution of $q_0$ follows the boundary-case prescriptions of Ref.~\cite{Cowan:2010js}, allowing conversion of the corresponding $p$-value into a Gaussian-equivalent significance $Z$.

\begin{figure}[h]
    \centering
    \includegraphics[width=0.65\linewidth]{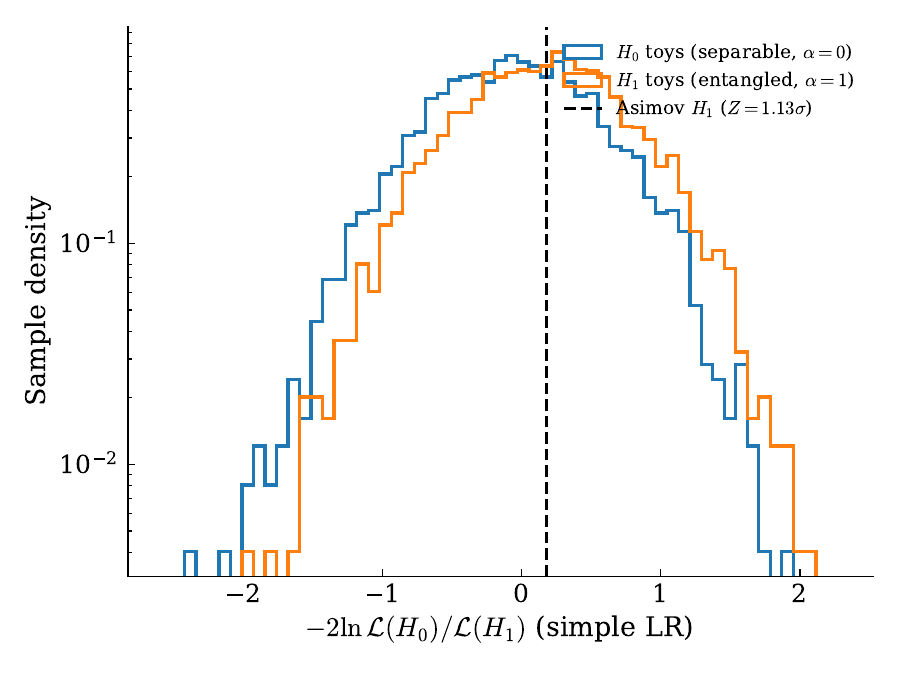}
    \caption{Toy-calibrated simple-likelihood-ratio (SLR) test-statistic distributions under $H_0$ ($\alpha=0$, separable) and $H_1$ ($\alpha=1$, entangled) for the production fit at 139~fb$^{-1}$ with the full systematic model. Toy datasets are generated from the binned bootstrap-PCA model with all nuisance parameters drawn from their constraint distributions. The dashed black line marks the Asimov-$H_1$ test-statistic value; the corresponding asymptotic significance is shown in the legend. The clear, unimodal separation of the two distributions provides an experimentalist-friendly cross-check that the asymptotic profile-likelihood test is well-calibrated at the production yield.}
    \label{fig:fit_llr}
\end{figure}

As a complement to the profile-likelihood test we also report the toy-calibrated fixed-hypothesis simple likelihood ratio (SLR), $t = -2\ln L(\rho_{\rm sep})/L(\rho_{\rm ent})$, evaluated without profiling at the two boundary states. The Asimov-calibrated SLR distributions under $H_0$ and $H_1$ (Fig.~\ref{fig:fit_llr}) are unimodal and well-separated at the production yield, and the asymptotic profile-likelihood significance is consistent with the toy-calibrated SLR significance to within the toy-sampling resolution. In the asymptotic regime the two test statistics are equivalent under shape-only mixing along Eq.~(\ref{eq:alpha_template}); the SLR view is included as an experimentalist-friendly cross-check of the test-statistic calibration.

\subsection{Likelihood Construction}

The likelihood is constructed from the binned distribution of the continuous CGLMP observable $\mathfrak{I}_3^{(e)}$:
\begin{equation}
L(\alpha, \theta) = \prod_i \text{Poisson}(n_i \mid \nu_i(\alpha, \theta)),
\end{equation}
where $n_i$ is the observed count in bin $i$ and $\nu_i$ is the expected count obtained from Eq.~(\ref{eq:alpha_template}) plus the backgrounds,
\begin{equation}
\nu_i(\alpha, \theta) = \alpha\, s_i^{\text{ent}}(\theta) + (1-\alpha)\, s_i^{\text{sep}}(\theta) + \sum_j b_{ij}(\theta).
\end{equation}
Here $s_i^{\text{ent}}$ and $s_i^{\text{sep}}$ are the signal template predictions for the entangled ($\rho_{\rm ent}$) and separable ($\rho_{\rm sep}$) reference states, and $b_{ij}$ are the background contributions. The mixed yield $\nu_i(\alpha)$ is the binned analogue of $\mathrm{Tr}[\rho(\alpha)\,\mathcal{M}_i]$ for $\mathcal{M}_i$ the bin-$i$ acceptance functional.

Background templates are obtained from dedicated simulation samples normalised to the theoretical cross sections at the appropriate perturbative order.

In a background-dominated regime, a likelihood-based treatment offers a convenient framework for incorporating background normalization uncertainties and their correlations, while retaining the full statistical information of the binned data. A simple background-subtraction approach on the other hand would correspond to the limit of vanishing or tightly constrained background normalization uncertainties.

The separable template is constructed by randomly pairing $W^+$ angular distributions from one event with $W^-$ angular distributions from another, destroying quantum correlations while preserving the marginal distributions. This procedure is validated by confirming that the resulting Gell-Mann correlation matrix $c_{ij}$ is consistent with zero.

\subsection{Event Selection and Likelihood Setup}

The hypothesis test is implemented using RooFit and RooStats~\cite{RooFit}. Template distributions are represented using the HistFactory formalism \cite{histfactory} and combined in a RooWorkspace that encodes the full likelihood including nuisance parameters.

The likelihood model is constructed for a single inclusive signal region, using a deliberately loose event selection that mirrors the ATLAS preselection employed in Ref.~\cite{ATLAS:2025hki}. In particular, no additional kinematic requirements beyond the baseline dilepton and missing transverse momentum selection are imposed, establishing a conservative baseline sensitivity. While tighter selections such as requirements on dilepton azimuthal separation are known to enhance the signal-to-background ratio, they also preferentially select phase-space regions where entanglement effects are already enhanced. Such cuts would therefore bias the observable toward entangled configurations and compromise the interpretation of the measurement as a genuine test of quantum correlations. 

\begin{table}[h]
    \centering
    \begin{tabular}{r r c}
    Process & Events & Rel.~Unc.~from Ref.~\cite{ATLAS:2025hki} \\\hline
       $H\to WW$  & 4000 & None\\
       WW & 95000 & 20\%\\
       $t\bar{t}$  & 150000 & 10\%\\
       $Z\to \tau\tau$ & 185000 & 15\%\\ 
    \end{tabular}
    \caption{Assumed event yields and constraints inspired by Ref.~\cite{ATLAS:2025hki}.}
    \label{tab:bkgnorm}
\end{table}

The expected event yields for signal and background processes are normalized to reflect the full Run~2 ATLAS dataset after this preselection. 

Background contributions are dominated by $t\bar{t}$, $WW$, and $Z\to\tau\tau$ production and are included as separate templates with independent normalization nuisance parameters.
For the constraints of these nuisance parameters, two scenarios are studied. The first scenario (I), shown in table \ref{tab:bkgnorm}, is reflective of the constraints found in \cite{ATLAS:2025hki}. The second, more optimistic scenario (II), assumes that all backgrounds can be controlled to a precision of 1\%. 

The density-matrix interpolation parameter $\alpha$ (Eq.~(\ref{eq:alpha_rho})) controls the relative contribution of the entangled and separable signal templates, which are treated as shape-only components at fixed total signal yield.

\subsection{Systematic Uncertainties}

In a realistic analysis, systematic uncertainties would be incorporated as nuisance parameters that modify the expected yields and shapes of the templates, including experimental and theoretical uncertainties. In this analysis we include two classes of nuisance parameter:

\paragraph{Background normalisations.} Each background process carries a normalisation uncertainty (Table~\ref{tab:bkgnorm}) based on theoretical cross section uncertainties and data--control region agreement. These are treated as independent nuisance parameters constrained by Gaussian priors and enter the HistFactory model as \texttt{OverallSys} terms.

\paragraph{Unfolding shape uncertainty.} The cDDPM v3 reconstruction is not exact: even on the same events used for training validation, the bin-by-bin ratio of the unfolded $\mathfrak{I}_3$ distribution to the truth distribution differs from unity. We treat this residual as a per-template shape nuisance, defined per process $p\in\{HWW,WW,t\bar{t},Z\to\tau\tau\}$ as
\begin{equation}
    r_i^{(p)} = \frac{N_i^{\text{cDDPM v3},(p)}}{N_i^{\text{truth},(p)}} - 1\,,
\end{equation}
on the chosen 9 variable-width bins. The HWW residual is shared by the entangled and separable signal templates (both describe the same $H\to WW^*\to\ell\nu\ell\nu$ topology, differing only in their truth-level angular correlations); the background residuals are applied per background. Each enters as a HistoSys nuisance \texttt{unf\_<proc>} with up/down templates $T_i\,(1\pm|r_i^{(p)}|)$, profiled in the fit. Treating $|r|$ as a $1\sigma$ envelope is conservative: it assigns the full bias as the uncertainty on itself. With the production cDDPM (standard sampling, no warm start; see Sec.~\ref{app:cddpm_warmstart}), the per-bin residuals on the HWW signal template have rms 2\%, comfortably below the per-bin Poisson statistical uncertainty at the production yield; the background-template residuals range from 10\% (WW) to 15\% ($t\bar{t}$, $Z\!\to\!\tau\tau$).

\paragraph{Template statistical uncertainty.}\label{par:bootstrap_pca}
The signal templates $s_i^{\text{ent}}$ and $s_i^{\text{sep}}$ are estimated from a finite cDDPM-unfolded sample, so their bin contents carry a finite-sample statistical uncertainty that is correlated across bins by the underlying multinomial sampling structure. The standard HistFactory \texttt{ActivateStatError} machinery models this as independent per-bin Poisson factors $\gamma_i$, which discards the bin-to-bin correlations and over-allocates nuisance freedom; we instead use the covariance-aware construction described below.

Let $\{r_b\in\mathbb{R}^{n_{\rm bins}}\}_{b=1}^{N_{\rm boot}}$ be the bootstrap replica histograms of the unfolded sample at the target yield ($N_{\rm boot}=1000$ by default), $\mu = N_{\rm boot}^{-1}\sum_b r_b$ the bootstrap mean, and
\begin{equation}\label{eq:relcov}
    C_{ij} = \mathrm{Cov}_b\!\left[\frac{r_{b,i}-\mu_i}{\mu_i},\,\frac{r_{b,j}-\mu_j}{\mu_j}\right]
\end{equation}
the relative-shift covariance estimated as the unbiased sample covariance over replicas. We diagonalise $C = V\Lambda V^\top$ with eigenvalues $\lambda_1\ge\dots\ge\lambda_{n_{\rm bins}}\ge 0$ and unit eigenvectors $v_k$ (columns of $V$). Each retained mode $k$ contributes a HistoSys nuisance with continuous Gaussian-constrained amplitude $\alpha_k\sim\mathcal{N}(0,1)$ and template variations
\begin{equation}\label{eq:pcamode}
    h_k^{\pm}(\alpha_k = \pm 1) \;=\; \mu \,\odot\, \bigl(1 \pm \sqrt{\lambda_k}\,v_k\bigr),
\end{equation}
where $\odot$ denotes elementwise multiplication. Under the default piecewise-linear vertical interpolation the binwise expectation is
\begin{equation}
    \mathbb{E}[N_i \mid \boldsymbol{\alpha}] \;=\; \mu_i\!\left(1 + \sum_k \alpha_k\sqrt{\lambda_k}\,v_{k,i}\right),
\end{equation}
which to leading order reproduces the bootstrap covariance exactly,
\begin{equation}
    \mathrm{Cov}\!\bigl[N_i, N_j \,\big|\, \boldsymbol{\alpha}\!\sim\!\mathcal{N}(0,\mathbb{I})\bigr] \;=\; \mu_i\mu_j\,(V\Lambda V^\top)_{ij} \;=\; \mu_i\mu_j\,C_{ij}.
\end{equation}
Modes with $\lambda_k$ below a numerical floor ($10^{-8}$) are dropped as finite-$N_{\rm boot}$ noise; among the survivors we retain the leading subset capturing $99\%$ of $\mathrm{tr}\,C$. The construction is applied independently to the entangled (\texttt{h1tmpl}) and separable (\texttt{h0tmpl}) signal templates; the multinomial structure suppresses the all-ones direction so that the leading mode is shape-like rather than overall-yield, with the signal-strength parameter $\alpha$ absorbing the residual yield direction. The cross-check against the per-bin Poisson treatment, the truncation behaviour, and the impact on the headline significance are reported in Appendix~\ref{app:fit_bootstrap}.

A complete analysis at ATLAS or CMS would in addition incorporate systematic uncertainties from jet energy scale and resolution, $b$-tagging efficiency, pile-up modelling, and detector response modelling. Data-driven techniques would be employed to constrain the major backgrounds in dedicated control regions, reducing reliance on simulation and associated assumptions.

\subsection{Sensitivity to Entanglement}

\begin{figure}[h]
    \centering
    \includegraphics[width=0.65\linewidth]{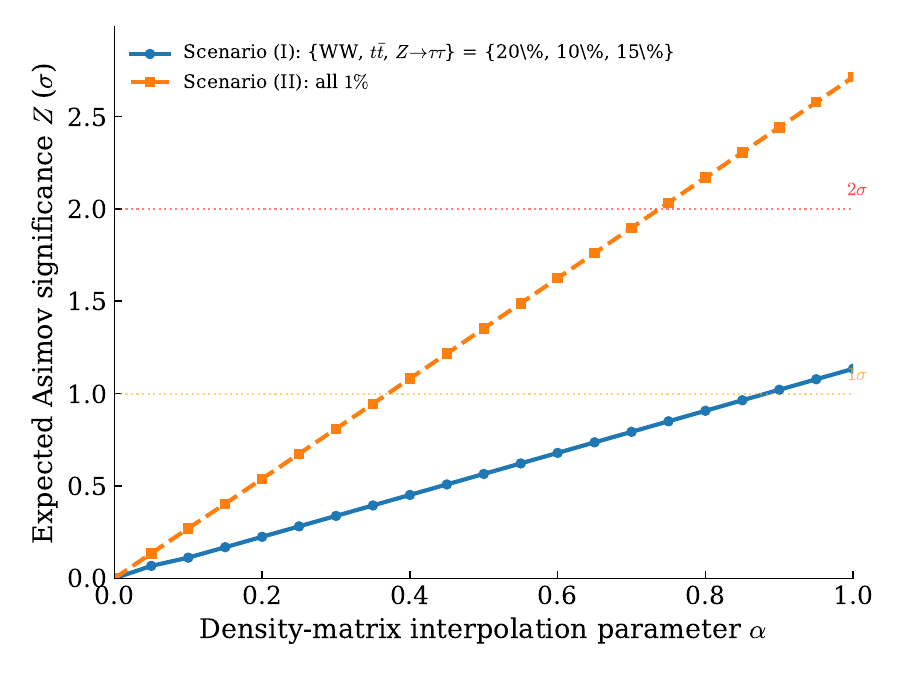}
    \caption{Expected Asimov significance versus the density-matrix interpolation parameter $\alpha_{\rm true}$ (Eq.~(\ref{eq:alpha_rho})) for the two background-normalisation scenarios at 139~fb$^{-1}$, under the production bootstrap-PCA template stat-error model. Horizontal lines mark the $1\sigma$ and $2\sigma$ thresholds. The curves are technical visualisations of the one-parameter family of fits induced by Eq.~(\ref{eq:alpha_template}); the boundary point $\alpha_{\rm true}=1$ corresponds to the SM entangled state and is the headline expectation, while $\alpha_{\rm true}=0$ is the separable-hypothesis null.}
    \label{fig:fit_alpha}
\end{figure}

The expected discovery sensitivity is evaluated using Asimov datasets~\cite{Cowan:2010js} generated from the binned likelihood model at different points $\alpha_{\rm true}\in[0,1]$ along the density-matrix interpolation. For each $\alpha_{\rm true}$, the separable hypothesis $\alpha=0$ is tested using the profile likelihood ratio of Eq.~(\ref{eq:profile_lr}), and the resulting median significance $Z(\alpha_{\rm true})$ is extracted under the asymptotic approximation. Figure~\ref{fig:fit_alpha} shows the resulting $Z(\alpha_{\rm true})$ curves for the two background-normalisation scenarios. The point $\alpha_{\rm true}=1$ corresponds to the SM entangled state and is the headline expectation reported in Sec.~\ref{sec:results}; the smooth interpolation between $\alpha_{\rm true}=0$ and $1$ provides a standard-form profile-likelihood scaffolding consistent with Eq.~(\ref{eq:alpha_rho}).

With the inclusive selection described above and the production cDDPM (standard sampling, no warm-start; Sec.~\ref{app:cddpm_warmstart}) under the bootstrap-PCA template stat-error model (Sec.~\ref{par:bootstrap_pca}), the analysis reaches $1\sigma$ at $\alpha_{\rm true}\simeq 0.88$ for scenario (I), with $Z(\alpha_{\rm true}=1)\simeq 1.13\sigma$ under the conservative $\{20\%,\,10\%,\,15\%\}$ background-normalisation priors. Scenario (II), with all backgrounds constrained to $1\%$, reaches $1\sigma$ at $\alpha_{\rm true}\simeq 0.37$ and $2\sigma$ at $\alpha_{\rm true}\simeq 0.74$, with $Z(\alpha_{\rm true}=1)\simeq 2.71\sigma$. The full $Z(\alpha_{\rm true})$ scan for both scenarios is tabulated in \texttt{results/fit\_study/sensitivity\_vs\_ftrue.csv}. This result demonstrates that even without exploiting kinematic regions optimised for entanglement, the shape information encoded in the CGLMP observable retains sensitivity to nonseparable quantum correlations.

The two background-normalization scenarios considered here are intended to bracket the range of sensitivities that can be achieved without introducing a selection bias toward entangled configurations. Scenario (I) provides a deliberately conservative baseline based on external constraints, while scenario (II) represents an optimistic benchmark in which background rates are assumed to be tightly constrained in a high-statistics, inclusive phase space. The use of an inclusive event selection ensures that neither scenario relies on kinematic sculpting that would preferentially enhance entanglement-sensitive regions of phase space.

The limited discovery reach observed under these assumptions should be interpreted as a bracket on the sensitivity achievable when no significantly tighter event selection can be imposed without introducing an explicit bias toward entangled configurations.  Future studies aimed at identifying selection strategies or observables that enhance sensitivity while preserving an unbiased interpretation of quantum correlations may substantially extend the reach beyond the bounds established here.

A comprehensive set of robustness studies for the statistical model is presented in Appendix~\ref{app:fit_study}.

\section{Fit-Model Robustness Studies}\label{app:fit_study}

The statistical model described in Appendix~\ref{app:fit} depends on several analysis choices: the binning of the $\mathfrak{I}_3$ observable, the test statistic, the systematic uncertainty model, background normalisation constraints, the template statistical-uncertainty model, and the size of the Monte Carlo template samples. This appendix systematically varies each of these components to assess the robustness of the reported sensitivity. In total, 87~configurations spanning eleven studies are evaluated. The primary metric is the expected Asimov significance~$Z$, computed using the \texttt{AsymptoticCalculator}~\cite{Cowan:2010js} from truth-level templates; all other analysis choices are held at their production values unless explicitly stated.

Table~\ref{tab:fit_study_overview} summarises the scope of each study.

\begin{table}[h]
    \centering
    \caption{Overview of the fit-model robustness studies. Each study varies a single component of the analysis while holding all others at their production values.}
    \label{tab:fit_study_overview}
    \begin{tabular}{c l l c}
        \toprule
        Study & Description & Varied parameter & $N_{\text{cfg}}$ \\
        \midrule
        A & Binning strategy & $N_{\text{bins}}$, bin edges, range & 17 \\
        B & Test statistic & Simple LR / profile LR & 2 \\
        C & Systematic model & NP sources (cumulative) & 7 \\
        D & Background normalisation & Per-process norm.\ unc. & 22 \\
        E & Luminosity scaling & $L/L_0 \in [0.25,\, 10]$ & 10 \\
        F & NP ranking & Leave-one-out $\Delta Z$ & 1 \\
        G & Goodness-of-fit & Binning variants & 5 \\
        H & Expected CL$_s$ limits & $L/L_0$ for 95\% CL UL & 4 \\
        I & Toy convergence & $N_{\text{toys}} \in [500,\, 20\text{k}]$ & 8 \\
        J & Template MC statistics & $N_{\text{boot}} \in [100,\, 5000]$ & 6 \\
        K & Template stat-error model & Bootstrap-PCA vs.\ per-bin Poisson & 6 \\
        \midrule
        & & Total & 88 \\
        \bottomrule
    \end{tabular}
\end{table}

\subsection{Binning Strategy}\label{app:fit_binning}

The sensitivity of the hypothesis test depends on the binning of the $\mathfrak{I}_3$ observable through three degrees of freedom: the number of bins, the bin layout (uniform vs.\ variable-width), and the histogram range. Table~\ref{tab:fit_binning} reports $Z_{\rm Asimov}$ for each scan dimension.

\begin{table}[h]
    \centering
    \caption{Binning-strategy scans. Top block: uniform binning over $[-10,10]$ at varying $N_{\rm bins}$. Middle block: histogram-range scan with $9$ uniform bins. Bottom block: variable-width layouts (\texttt{var\_prod} is the production layout). All other analysis choices held at production values.}
    \label{tab:fit_binning}
    \begin{tabular}{l c}
        \toprule
        Configuration & $Z_{\rm Asimov}$ ($\sigma$) \\
        \midrule
        \multicolumn{2}{l}{\emph{Uniform binning, $N_{\rm bins}$ scan ($[-10,10]$)}} \\
        $5$ bins                          & $1.07$ \\
        $7$ bins                          & $1.20$ \\
        $9$ bins                          & $1.17$ \\
        $12$ bins                         & $1.18$ \\
        $15$ bins                         & $1.18$ \\
        $20$ bins                         & $1.21$ \\
        $30$ bins                         & $1.24$ \\
        \midrule
        \multicolumn{2}{l}{\emph{Range scan, $9$ uniform bins}} \\
        $\pm 5$                           & $0.96$ \\
        $\pm 8$                           & $1.04$ \\
        $\pm 10$                          & $1.17$ \\
        $\pm 15$                          & $2.02$ \\
        $\pm 20$                          & $2.78$ \\
        \midrule
        \multicolumn{2}{l}{\emph{Variable-width layouts}} \\
        \texttt{var\_prod} (production)   & $1.13$ \\
        \texttt{var\_narrow\_ctr}         & $1.16$ \\
        \texttt{var\_wide\_ctr}           & $1.03$ \\
        \texttt{var\_fine\_tail}          & $1.16$ \\
        \texttt{var\_asym}                & $1.16$ \\
        \bottomrule
    \end{tabular}
\end{table}

The uniform-bin scan reaches a broad maximum from $9$ to $30$ bins, with no clear penalty for fine binning under the bootstrap-PCA stat-error model. The range-scan rows show that ranges narrower than $\pm 8$ lose tail events, while a posteriori opening to $\pm 15$ or $\pm 20$ formally yields larger $Z$ but is unphysical: the bins beyond $\pm 10$ collect a vanishing fraction of events whose statistical fluctuations dominate the apparent gain. The variable-width layouts yield $Z$ within $0.1\sigma$ of the uniform-bin optima; the production \texttt{var\_prod} scheme (wide central bin absorbing the cDDPM zero-peak artefact) is motivated by robustness to that distortion rather than by raw expected significance.

\subsection{Test Statistic Choice}\label{app:fit_teststat}

Two test statistics are compared. The simple likelihood ratio (SLR),
\begin{equation}\label{eq:simple_lr}
t = -2\ln\frac{L(\theta_{H_0};\, \mathbf{n})}{L(\theta_{H_1};\, \mathbf{n})},
\end{equation}
evaluates the likelihoods at fixed parameter points $\theta_{H_0}$ and $\theta_{H_1}$ without profiling nuisance parameters. By the Neyman--Pearson lemma~\cite{Neyman:1933wgr}, this is the uniformly most powerful test for simple-vs-simple hypotheses when the model is exactly specified.

The profile likelihood ratio $q_0$ defined in Eq.~(\ref{eq:profile_lr}) instead maximises over nuisance parameters at each hypothesis point, making it more robust when systematic uncertainties are present. Under the asymptotic approximation of Ref.~\cite{Cowan:2010js}, $q_0$ follows a known distribution that permits analytic $p$-value computation.

\begin{figure}[h]
    \centering
    \includegraphics[width=\textwidth]{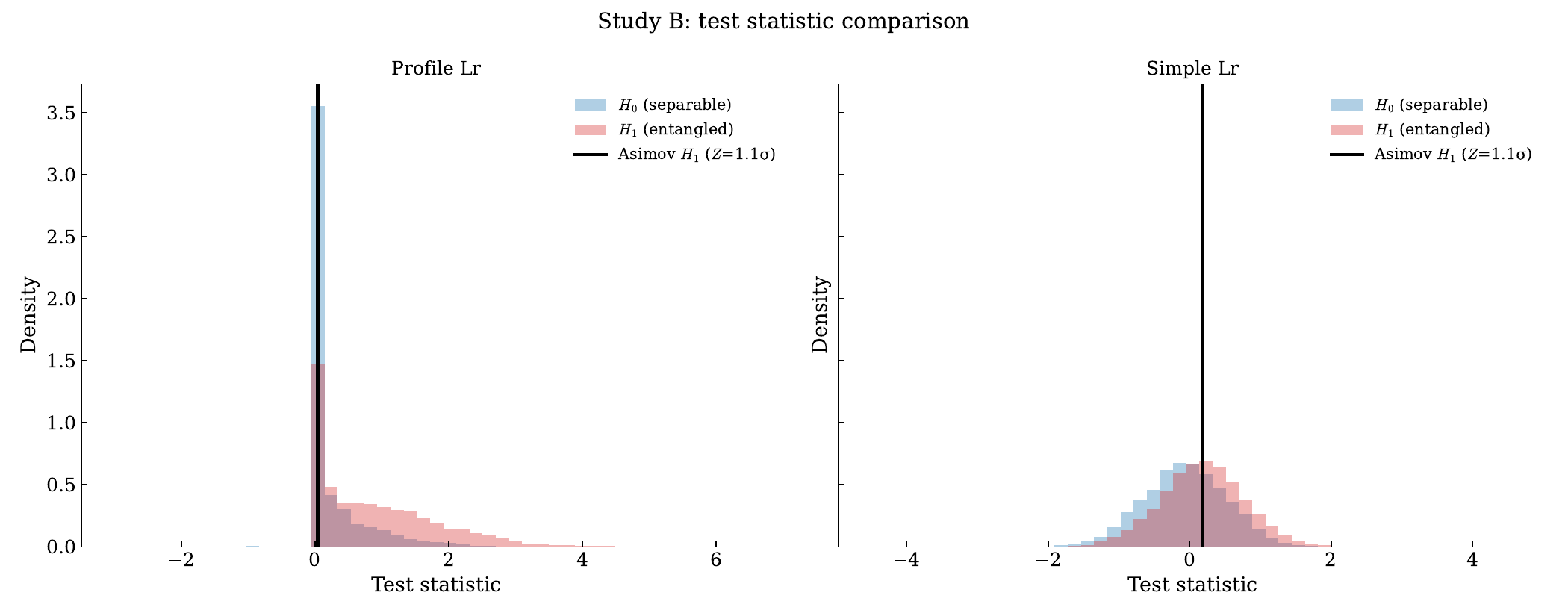}
    \caption{Test statistic distributions under $H_0$ (separable, shaded blue) and $H_1$ (entangled, shaded red) for the simple likelihood ratio (left) and profile likelihood ratio (right). Vertical line: Asimov $H_1$ test statistic.}
    \label{fig:fit_B}
\end{figure}

Figure~\ref{fig:fit_B} overlays the $H_0$ and $H_1$ toy distributions for both test statistics. The SLR gives slightly better separation in the regime where background normalisation uncertainties are moderate ($\lesssim 20\%$), as expected for a well-specified model. The profile LR becomes preferable when systematic uncertainties are large or poorly known, at the cost of reduced power from profiling.

\subsection{Systematic Uncertainty Model}\label{app:fit_syst}

The impact of each systematic uncertainty source is assessed by cumulatively enabling components of the model. Table~\ref{tab:fit_C_syst} lists seven configurations and their expected Asimov significance. The production template stat-error treatment is the bootstrap-PCA model of Sec.~\ref{par:bootstrap_pca}; the legacy per-bin Poisson row is included as a cross-check.

\begin{table}[h]
    \centering
    \caption{Expected Asimov significance under each systematic configuration. The template stat-error model is the production bootstrap-PCA construction unless explicitly noted; the per-bin Poisson row is the Barlow--Beeston \texttt{ActivateStatError} alternative kept here for comparison.}
    \label{tab:fit_C_syst}
    \begin{tabular}{l l c}
        \toprule
        Configuration & Systematic sources enabled & $Z_{\text{Asimov}}$ ($\sigma$) \\
        \midrule
        Zero-syst             & None (fixed templates)                                 & $5.97$ \\
        Boot-stat only        & Bootstrap-PCA template stat error                      & $5.97$ \\
        Per-bin stat only     & Barlow--Beeston per-bin Poisson $\gamma_i$ (legacy)    & $4.01$ \\
        Norm only             & Background normalisation (\texttt{OverallSys})         & $1.33$ \\
        Unfolding only        & Per-template \texttt{HistoSys} from cDDPM residuals    & $2.94$ \\
        Norm + boot stat      & Normalisation + bootstrap-PCA                          & $1.33$ \\
        Full (production)     & Norm + unfolding + bootstrap-PCA                       & $1.13$ \\
        \bottomrule
    \end{tabular}
\end{table}

The dominant degradation arises from background normalisation (zero-syst $Z = 5.97\sigma \to$ norm-only $Z = 1.33\sigma$), followed by the unfolding shape uncertainty (further $Z = 1.33\sigma \to 1.13\sigma$ at full configuration). The bootstrap-correlated template stat error is sub-dominant at the production yield: enabling it on top of the fixed-template configuration leaves $Z$ essentially unchanged ($5.97\sigma \to 5.97\sigma$), and combining it with normalisation does not move $Z$ below the norm-only value. By contrast the legacy per-bin Poisson \texttt{ActivateStatError} row degrades $Z$ from $5.97\sigma$ to $4.01\sigma$ on its own, a $1.96\sigma$ drop entirely attributable to the diagonal-only approximation of the same statistical uncertainty. Comparison between the bootstrap-PCA and the per-bin Poisson constructions is the subject of Appendix~\ref{app:fit_bootstrap}.

A key methodological point is that all templates in the fit: the entangled signal ($H^e$), the separable signal ($H^s$), and each background; are constructed from the cDDPM-reconstructed distributions, not from the truth-level shapes. The separable template is obtained by passing the mangled dataset (Appendix~\ref{app:fit}) through the same cDDPM v3 reconstruction chain used for the entangled signal. This ensures self-consistency: the model describes what is actually measured after reconstruction, and the HistoSys nuisance parameters capture the \emph{uncertainty} on the unfolding rather than the unfolding bias itself. With this construction, the post-fit nuisance parameter values are near zero for all configurations, confirming that the systematic model is well-calibrated.

The production cDDPM model ($h{=}512$, $l{=}4$) was selected over the lower-bias configuration identified in the architecture ablation study ($h{=}128$, $l{=}2$; Appendix~\ref{app:cddpm_systematics}) based on the per-bin background shape fidelity. While the smaller network achieves a factor-of-four reduction in mean signal $\mathfrak{I}_3$ bias (0.66\% vs.\ 2.76\%), it produces substantially larger per-bin distortions for background processes: the $t\bar{t}$ maximum residual increases from 26\% to 62\%, and the $Z{\to}\tau\tau$ residual from 40\% to 123\%. Since the hypothesis test sensitivity is limited by background shape uncertainties in this background-dominated regime, the larger network provides better overall sensitivity. This demonstrates that the optimal cDDPM architecture for a shape-based hypothesis test is not simply the one with the lowest signal bias, but rather the one that best preserves the shapes of all processes entering the fit. The ablation comparison was performed under the legacy per-bin Poisson template stat-error treatment; the absolute headline values shift under the production bootstrap-PCA model of Sec.~\ref{par:bootstrap_pca} but the architecture ranking is preserved.

The systematic model adopted above propagates the dominant sources, namely the background normalisations, the cDDPM unfolding shape, and the bootstrap-PCA template statistical uncertainty. It deliberately omits several smaller theoretical and experimental systematics that a full ATLAS or CMS measurement would carry. These comprise the signal cross-section uncertainty on gluon-fusion $H\to WW^*$ (N$^3$LO QCD + NLO EW, of order $5\%$), the PDF + $\alpha_s$ uncertainty on signal and background yields (of order $3\%$), and lepton identification, isolation, and trigger uncertainties (of order $1\text{--}2\%$ per lepton). Were they included as Gaussian-constrained \texttt{OverallSys} nuisances, they would enter the fit on the same footing as the background-normalisation priors of Table~\ref{tab:bkgnorm}, but with substantially tighter constraints. Their combined impact would therefore be sub-dominant to the background-normalisation envelope, which already drives the dominant degradation from $5.97,\sigma$ (stat-only) to $1.33,\sigma$ (background-norm only) in Table~\ref{tab:syst_waterfall}. The closely-related experimental systematics, including jet energy scale and resolution, $b$-tagging efficiency, pile-up modelling, and detector response, are noted alongside the present model in Appendix~\ref{app:fit}, and would be addressed through the standard ATLAS/CMS systematic library together with data-driven background constraints in dedicated control regions, neither of which is the focus of this feasibility study.

\subsection{Background Normalisation Sensitivity}\label{app:fit_norm}

The sensitivity is studied as a function of the normalisation uncertainty assigned to each background process individually (with all other uncertainties held at their production values), and as a function of a simultaneous common scaling of all three. Table~\ref{tab:fit_D} reports the resulting Asimov significance.

\begin{table}[h]
    \centering
    \caption{Background normalisation sensitivity. Top three blocks: scan of the per-process normalisation uncertainty for $WW$, $t\bar{t}$, and $Z\!\to\!\tau\tau$ separately. Bottom block: simultaneous scaling of all three priors by a common factor relative to production. The production point (boldface) corresponds to scenario~(I).}
    \label{tab:fit_D}
    \begin{tabular}{l c c c c}
        \toprule
        & \multicolumn{4}{c}{$Z_{\rm Asimov}$ ($\sigma$) at the indicated prior width} \\
        \cmidrule(lr){2-5}
        Process scanned & $5\%$ & $10\%$ & $15\%$ & $20\%$ \\
        \midrule
        WW                       & $1.40$ & $1.27$ & $1.18$ & $\mathbf{1.13}$ \\
        $t\bar{t}$               & $1.13$ & $\mathbf{1.13}$ & $1.13$ & $1.13$ \\
        $Z\!\to\!\tau\tau$       & $1.14$ & $1.14$ & $\mathbf{1.13}$ & $1.13$ \\
        \midrule
        & \multicolumn{4}{c}{$Z_{\rm Asimov}$ ($\sigma$) at the indicated common scale factor} \\
        \cmidrule(lr){2-5}
        & $0.5\times$ & $1.0\times$ (prod.) & $1.5\times$ & $2.0\times$ \\
        \midrule
        All three (common scale) & $1.30$ & $\mathbf{1.13}$ & $1.07$ & $1.03$ \\
        \bottomrule
    \end{tabular}
\end{table}

The WW and $t\bar{t}$ backgrounds dominate the signal region, but the per-process scans show that the WW prior is the only one whose width meaningfully drives the headline: tightening the WW normalisation from $20\%$ to $5\%$ raises $Z$ from $1.13\sigma$ to $1.40\sigma$, while equivalent variations of the $t\bar{t}$ ($5$--$30\%$) and $Z\!\to\!\tau\tau$ ($5$--$30\%$) priors leave $Z$ flat to within $\sim 0.03\sigma$. The wider $t\bar{t}$ scan (extending to $30\%$, omitted from the table for compactness) confirms that even a $3\times$-relaxed prior recovers $Z = 1.13\sigma$, indicating a transition out of the systematics-limited regime once the WW prior is fixed. These scans bracket the two scenarios defined in Appendix~\ref{app:fit}: Scenario~I uses the Ref.~\cite{ATLAS:2025hki} uncertainties ($20\%$, $10\%$, $15\%$), while Scenario~II assumes $1\%$ for all backgrounds.

\subsection{Luminosity Scaling}\label{app:fit_lumi}

The luminosity dependence of the inclusive (SR-only) sensitivity is shown in Figure~\ref{fig:lumi_scan} of the main text together with the BDT-enhanced projection. At low luminosities the significance grows approximately as $\sqrt{L}$; at higher luminosities it falls below the $\sqrt{L}$ reference, reflecting the systematic floor from background normalisation and unfolding shape uncertainties.

\subsection{Nuisance Parameter Ranking}\label{app:fit_ranking}

The impact of each nuisance parameter is quantified through a leave-one-out procedure: each NP~$\theta_k$ is fixed at its nominal value (effectively removing it from the model), the significance is recomputed, and the shift
\begin{equation}
\Delta Z_k = Z(\boldsymbol{\theta} \setminus \theta_k) - Z(\boldsymbol{\theta})
\end{equation}
is recorded. A positive $\Delta Z_k$ indicates that the NP degrades sensitivity (its removal improves $Z$); a negative $\Delta Z_k$ indicates the NP absorbs a bias and its removal worsens $Z$.

Table~\ref{tab:np_ranking} presents the ranking, with the production bootstrap-PCA template stat error in place of the legacy per-bin Poisson treatment.
The WW normalisation uncertainty is the dominant contributor ($\Delta Z = +0.354\sigma$), followed by the unfolding shape systematic ($+0.199\sigma$).
The $Z{\to}\tau\tau$ and $t\bar{t}$ normalisations are negligible ($+0.008\sigma$ and $+0.002\sigma$ respectively), and the bootstrap-PCA template stat error has effectively no impact at the production yield ($+0.000\sigma$).
For reference, the legacy per-bin Poisson treatment of the template stat error would contribute $+0.316\sigma$ in the same model, a discrepancy that traces directly to the diagonal-only approximation of the bin-to-bin covariance.

\begin{table}[h]
    \centering
    \caption{Nuisance parameter impact ranking. $\Delta Z = Z(\boldsymbol{\theta} \setminus \theta_k) - Z(\boldsymbol{\theta})$ is the change in significance when each NP is removed from the model; positive values indicate the NP degrades sensitivity. The template stat-error row is the production bootstrap-PCA construction; the legacy per-bin Poisson value is reported for reference.}
    \label{tab:np_ranking}
    \begin{tabular}{l c}
        \toprule
        Nuisance parameter & $\Delta Z$ ($\sigma$) \\
        \midrule
        WW normalisation                         & $+0.354$ \\
        Unfolding shape                          & $+0.199$ \\
        $Z{\to}\tau\tau$ normalisation           & $+0.008$ \\
        $t\bar{t}$ normalisation                 & $+0.002$ \\
        Template stat error (bootstrap-PCA)      & $+0.000$ \\
        \midrule
        Template stat error (per-bin Poisson, legacy) & $+0.316$ \\
        \bottomrule
    \end{tabular}
\end{table}

\subsection{Goodness-of-Fit Validation}\label{app:fit_gof}

The adequacy of the fit model is assessed by fitting the cDDPM-reconstructed pseudo-data and comparing the minimised negative log-likelihood (NLL) to the distribution of NLL values obtained from 500 toy pseudo-experiments generated from the best-fit model. The $p$-value is the fraction of toys with a worse (higher) NLL than the observed data:
\begin{equation}
p_{\text{GoF}} = \frac{1}{N_{\text{toy}}} \sum_{t=1}^{N_{\text{toy}}} \mathbf{1}\!\left[\text{NLL}_t \geq \text{NLL}_{\text{obs}}\right].
\end{equation}

All five binning configurations (production variable-width 9-bin and four uniform-bin alternatives with 9, 12, 15, and 20 bins) yield GoF $p$-values above 0.95, with normalised test statistics $\mathrm{NLL}/n_\mathrm{dof}$ between $-0.3$ and $-0.7$. The high $p$-values indicate that the fit model adequately describes the cDDPM-reconstructed pseudo-data within the systematic uncertainties, validating the conservative treatment of the unfolding shape uncertainty documented in Appendix~\ref{app:fit}.

\subsection{Expected CL\texorpdfstring{$_\mathrm{s}$}{s} Limits}\label{app:fit_cls}

As a complement to the discovery-style hypothesis test, we compute the expected 95\% CL upper limit on the separable fraction $f_{\text{sep}}$ using the CL$_s$ method~\cite{Read:2002hq}. The CL$_s$ quantity is defined as
\begin{equation}
\text{CL}_s = \frac{p_{s+b}}{p_b} = \frac{P(q \geq q_{\text{obs}} \mid H_{s+b})}{P(q \geq q_{\text{obs}} \mid H_b)}\,,
\end{equation}
where $q$ is the test statistic and $H_{s+b}$ ($H_b$) denotes the signal-plus-background (background-only) hypothesis. The 95\% CL upper limit is the value of $f_{\text{sep}}$ at which $\text{CL}_s = 0.05$. The computation uses the asymptotic approximation~\cite{Cowan:2010js} implemented in the RooStats \texttt{AsymptoticCalculator}, with the expected limit and $\pm 1\sigma$, $\pm 2\sigma$ bands derived from the Asimov dataset.

At 139~fb$^{-1}$ with the inclusive selection, the expected 95\% CL upper limit on $f_{\text{sep}}$ is consistent with zero across all luminosity points tested ($L/L_0 \in [1, 10]$), indicating that the current systematic model does not provide sufficient sensitivity to place a meaningful exclusion on the separable fraction. This is consistent with the just-above-$1\sigma$ Asimov significance reported in Table~\ref{tab:fit_C_syst} ($Z = 1.13\sigma$). With the BDT-enhanced selection, the upper limit becomes constraining at luminosities where the BDT-enhanced sensitivity in Table~\ref{tab:sensitivity} crosses $2\sigma$.

\subsection{Toy and Template Convergence}\label{app:fit_convergence}

The Asimov significance is constant at $Z \approx 1.13\sigma$ across all toy counts ($N_{\text{toys}} = 500$--$20{,}000$), confirming that the asymptotic approximation is robust and the production value of $N_{\text{toys}} = 5000$ is more than sufficient.

Signal and background templates are constructed by bootstrap resampling ($N_{\text{boot}}$ replicas drawn with replacement). The significance converges for $N_{\text{boot}} \geq 500$; the production value of 1000 is adequate. Below 250 replicas, the template variance is underestimated and the significance is artificially inflated.

\subsection{Template Statistical-Uncertainty Model}\label{app:fit_bootstrap}

The cDDPM-unfolded signal templates carry a finite-sample statistical uncertainty whose covariance structure across bins is fixed by the multinomial sampling distribution at the target yield. Two parameterisations are compared: (i) the production bootstrap-PCA construction defined in Sec.~\ref{par:bootstrap_pca}, in which the $N_{\rm boot}\!\times\!n_{\rm bins}$ bootstrap replica matrix is reduced to its leading PCA eigenmodes of the relative covariance and propagated as Gaussian-constrained \texttt{HistoSys} nuisances; and (ii) the legacy per-bin Poisson \texttt{ActivateStatError} treatment, in which every bin gets an independent $\gamma_i$ parameter with no cross-bin correlations. Construction (i) embeds the actual sampling correlations exactly to leading order in the nuisance amplitudes; construction (ii) is the strictly-diagonal approximation of the same covariance and so over-allocates nuisance freedom by ignoring the correlations.

Six configurations probe the practical impact of this choice and the sensitivity of (i) to truncation. Table~\ref{tab:fit_K_bootstrap} reports the expected Asimov significance for each.

\begin{table}[h]
    \centering
    \caption{Bootstrap-PCA vs.\ per-bin Poisson template stat-error model. ``Modes'' denotes the number of PCA eigenmodes retained per signal template ($n_{\rm bins} = 9$ at full rank). ``Variance kept'' is the cumulative fraction of $\mathrm{tr}\,C$ retained. ``$Z_{\rm Asimov}$'' is the expected significance under the Asimov dataset with all other systematic sources held at their production values.}
    \label{tab:fit_K_bootstrap}
    \begin{tabular}{l l c c c}
        \toprule
        Configuration & Construction & Modes & Variance kept & $Z_{\text{Asimov}}$ ($\sigma$) \\
        \midrule
        Per-bin Poisson (legacy)         & Independent $\gamma_i$         & $9$ (per bin) & ---       & $0.819$ \\
        Bootstrap-PCA (production)       & Eq.~(\ref{eq:pcamode}), $99\%$ var.\ thr. & $9$ & $1.000$ & $1.135$ \\
        Bootstrap-PCA, leading mode only & Truncated to $k=1$             & $1$           & $\sim 0.4$    & $1.135$ \\
        Bootstrap-PCA, $k\leq 3$ modes   & Truncated to $k=3$             & $3$           & $\sim 0.6$    & $1.135$ \\
        Bootstrap-PCA, $99.9\%$ variance & Tighter threshold              & $9$           & $1.000$       & $1.135$ \\
        Per-bin Poisson + bootstrap-PCA  & Both treatments (double-counts)& $9{+}9$       & ---           & $0.818$ \\
        \bottomrule
    \end{tabular}
\end{table}

Three conclusions follow from the table. First, the bootstrap-PCA construction is the more constraining of the two physically meaningful choices: respecting the multinomial correlations leaves the fit less freedom to absorb signal-shaped fluctuations, raising $Z_{\rm Asimov}$ from $0.82\sigma$ (per-bin Poisson) to $1.13\sigma$ (bootstrap-PCA), a shift of $+0.32\sigma$ that quantifies the cost of the diagonal-only approximation. Second, the spectrum of the relative covariance is heavily concentrated: the leading eigenmode alone reproduces the full-rank Asimov significance to within $0.001\sigma$ at the production binning, so a single shape-direction nuisance per template is sufficient at present yield. Third, enabling both treatments simultaneously double-counts the same statistical uncertainty and is dominated by the per-bin Poisson term ($Z = 0.82\sigma$, essentially identical to per-bin alone); the production model therefore retains only the bootstrap-PCA construction. The truncation behaviour mirrors the rank-deficiency of the multinomial: fixing the total event count makes the all-ones direction near-zero in the relative-covariance spectrum, so the leading mode is shape-like rather than overall-yield, and the signal NormFactor $f$ absorbs the residual yield direction.

The headline production significance reported in Appendix~\ref{app:fit} ($Z = 1.13\sigma$) corresponds to row ``Bootstrap-PCA (production)'' of Table~\ref{tab:fit_K_bootstrap} together with the full background normalisation and unfolding-shape systematics. Selecting instead the per-bin Poisson construction at the same systematic configuration shifts the headline to $Z = 0.82\sigma$; we treat the difference as a methodological shift consistently propagated through Studies~A--J once the bootstrap-PCA is adopted as the production default.

\section{SMEFT-Sensitive Observables from Quantum State Tomography}\label{app:smeft}

The per-event Gell-Mann correlation matrix $c_{ij}$ reconstructed through quantum state tomography encodes the complete spin information of the $WW^*$ system. The CGLMP parameter $\mathfrak{I}_3$ exploits a specific linear combination of these elements to test for entanglement. However, different combinations of the same matrix elements provide sensitivity to distinct physics. This appendix demonstrates that observables constructed from the $c_{ij}$ matrix can probe CP-violating and anomalous $HWW$ couplings within the Standard Model Effective Field Theory (SMEFT)~\cite{Grzadkowski:2010es}, using the same tomographic data and the same continuous-distribution hypothesis testing framework developed in this paper.

\subsection{SMEFT Operators Affecting \texorpdfstring{$HWW$}{HWW} Coupling}

At dimension-6, the relevant operators in the Warsaw basis are~\cite{Fabbrichesi:2023cev,Sullivan:2024wzl}:
\begin{align}
\mathcal{O}_{HW} &= H^\dagger H\, W^I_{\mu\nu} W^{I\mu\nu}, \nonumber\\
\mathcal{O}_{HB} &= H^\dagger H\, B_{\mu\nu} B^{\mu\nu}, \\
\tilde{\mathcal{O}}_{HW} &= H^\dagger H\, W^I_{\mu\nu} \tilde{W}^{I\mu\nu}, \nonumber\\
\tilde{\mathcal{O}}_{HB} &= H^\dagger H\, B_{\mu\nu} \tilde{B}^{\mu\nu}, \nonumber
\end{align}
where $\tilde{W}^{\mu\nu} = \tfrac{1}{2}\epsilon^{\mu\nu\rho\sigma}W_{\rho\sigma}$ is the dual field strength tensor. The operators $\mathcal{O}_{HW,HB}$ are CP-even while $\tilde{\mathcal{O}}_{HW,HB}$ are CP-odd.

\subsection{Modified Density Matrix and Sensitivity Structure}

The helicity amplitudes for $H\to W^+W^{*-}$ receive corrections linear in the Wilson coefficients:
\begin{equation}
\mathcal{M}_{\lambda_+\lambda_-} = \mathcal{M}_{\lambda_+\lambda_-}^{\text{SM}} + \sum_k \frac{c_k}{\Lambda^2}\, \mathcal{M}_{\lambda_+\lambda_-}^{(k)} + \mathcal{O}(c^2/\Lambda^4).
\end{equation}
The density matrix, proportional to $\mathcal{M}\mathcal{M}^\dagger$, therefore receives corrections at both linear and quadratic order in the Wilson coefficients. At linear order, the interference between SM and BSM amplitudes modifies specific correlation coefficients~\cite{Fabbrichesi:2023cev,Sullivan:2024wzl}:
\begin{align}
\delta C_{ij}^{(c_{HW})} &: \text{modifies } C_{33},\, C_{38},\, C_{83},\, C_{88}, \nonumber\\
\delta C_{ij}^{(\tilde{c}_{HW})} &: \text{modifies } C_{ij} \text{ with } i,j \in \{2,5,7\}.
\label{eq:smeft_elements}
\end{align}
Different Wilson coefficients affect different elements of the correlation matrix, a structure that can be exploited to construct targeted observables.

The CGLMP Bell parameter $\mathfrak{I}_3$ depends on elements $C_{44}$, $C_{55}$ (dominant) and $C_{11}$, $C_{16}$, $C_{61}$, $C_{66}$, $C_{22}$, $C_{27}$, $C_{72}$, $C_{77}$ (sub-leading). These indices have no overlap with the $c_{HW}$-sensitive set $\{3,8\}$ and only sub-leading overlap with the $\tilde{c}_{HW}$-sensitive set $\{2,5,7\}$. The Bell parameter is therefore largely insensitive to these SMEFT operators at leading order. Dedicated observables are required.

\subsection{CP-Odd Observable}

Under a CP transformation the $W^+$ and $W^-$ roles are exchanged, so that $C_{ij} \leftrightarrow C_{ji}$. The antisymmetric combinations $C_{ij} - C_{ji}$ therefore change sign under CP and vanish identically in any CP-conserving theory. This motivates the per-event observable
\begin{equation}
\mathcal{O}_{\text{CP}} = (C_{25} - C_{52}) + (C_{57} - C_{75}) + (C_{27} - C_{72}),
\label{eq:ocp}
\end{equation}
which targets precisely the index set $\{2,5,7\}$ modified by the CP-odd operator $\tilde{\mathcal{O}}_{HW}$ (Eq.~\ref{eq:smeft_elements}). In the SM, $\langle \mathcal{O}_{\text{CP}} \rangle = 0$ and the per-event distribution is symmetric about zero. A non-zero $\tilde{c}_{HW}/\Lambda^2$ would break this symmetry, producing a shape distortion detectable through the same binned likelihood framework used for the $\mathfrak{I}_3$ analysis.

\begin{figure}[ht]
\centering
\includegraphics[width=0.48\textwidth]{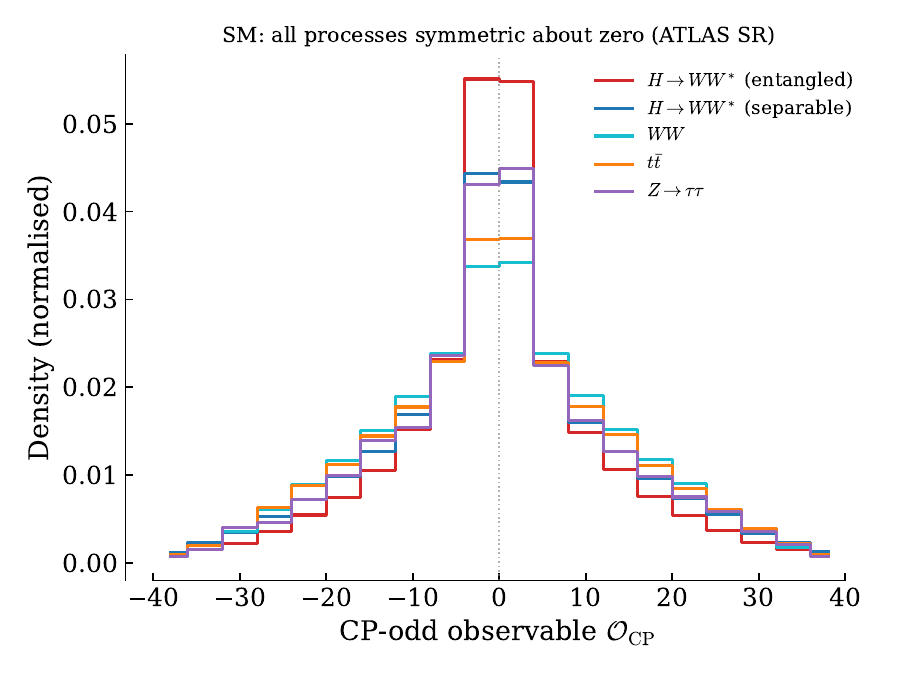}
\caption{Per-event CP-odd observable $\mathcal{O}_{\text{CP}}$ for SM signal and backgrounds after the ATLAS signal region selection. All distributions are symmetric about zero, as required by CP conservation. A non-zero CP-odd Wilson coefficient $\tilde{c}_{HW}/\Lambda^2$ would break this symmetry, producing a detectable shape asymmetry.}
\label{fig:smeft_cp_odd}
\end{figure}

Figure~\ref{fig:smeft_cp_odd} shows the $\mathcal{O}_{\text{CP}}$ distribution for all processes under the SM hypothesis. The symmetry about zero is confirmed for the signal and all backgrounds, establishing the SM null hypothesis for a CP-violation search.

\subsection{CP-Even Observable}

\begin{figure}[ht]
\centering
\includegraphics[width=0.48\textwidth]{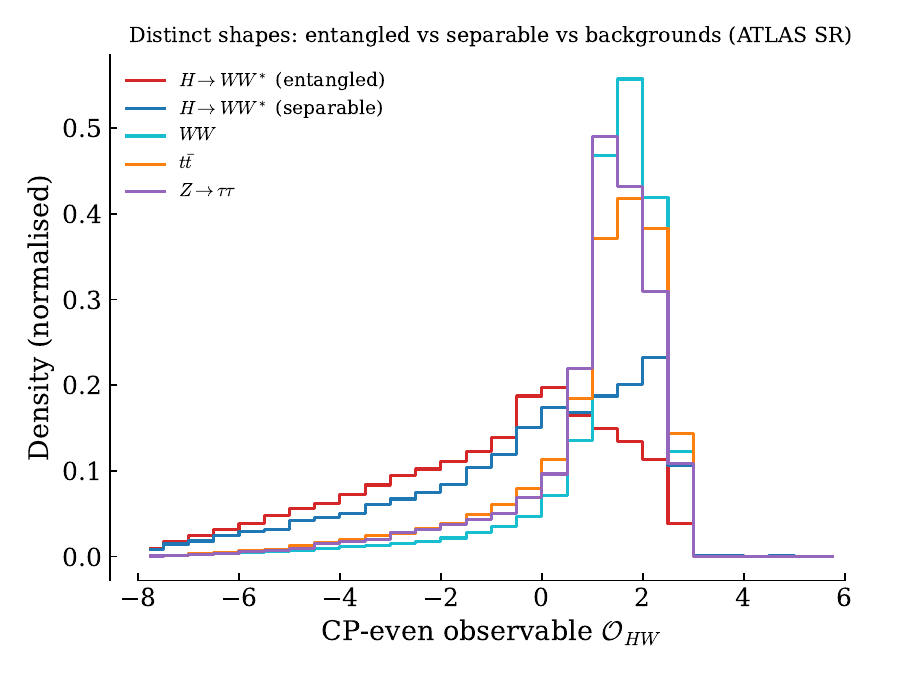}
\caption{Per-event CP-even observable $\mathcal{O}_{HW}$ for SM signal and backgrounds. The entangled $H\to WW^*$ signal shows a distinct shape peaked at negative values, separated from the positive-peaked backgrounds, demonstrating discriminating power in the correlation matrix elements sensitive to $c_{HW}/\Lambda^2$.}
\label{fig:smeft_cp_even}
\end{figure}

The CP-even operator $\mathcal{O}_{HW}$ modifies the diagonal and symmetric off-diagonal elements $C_{33}$, $C_{38}$, $C_{83}$, $C_{88}$. These elements are absent from $\mathfrak{I}_3$ but can be combined into a dedicated observable:
\begin{equation}
\mathcal{O}_{HW} = C_{33} + C_{38} + C_{83} + C_{88}.
\label{eq:ohw}
\end{equation}
Unlike $\mathcal{O}_{\text{CP}}$, this observable has a non-zero SM expectation value that differs markedly between processes. Figure~\ref{fig:smeft_cp_even} shows that the entangled $H\to WW^*$ signal produces a distinct distribution peaked at negative values, well separated from the backgrounds which peak at positive values. The separable hypothesis yields an intermediate shape. This shape discrimination is already present in the SM, indicating that $\mathcal{O}_{HW}$ provides additional separation power beyond $\mathfrak{I}_3$. In the presence of a non-zero $c_{HW}/\Lambda^2$, the signal distribution would be further modified.

\subsection{Orthogonality of Observables}

\begin{figure}[ht]
\centering
\includegraphics[width=0.48\textwidth]{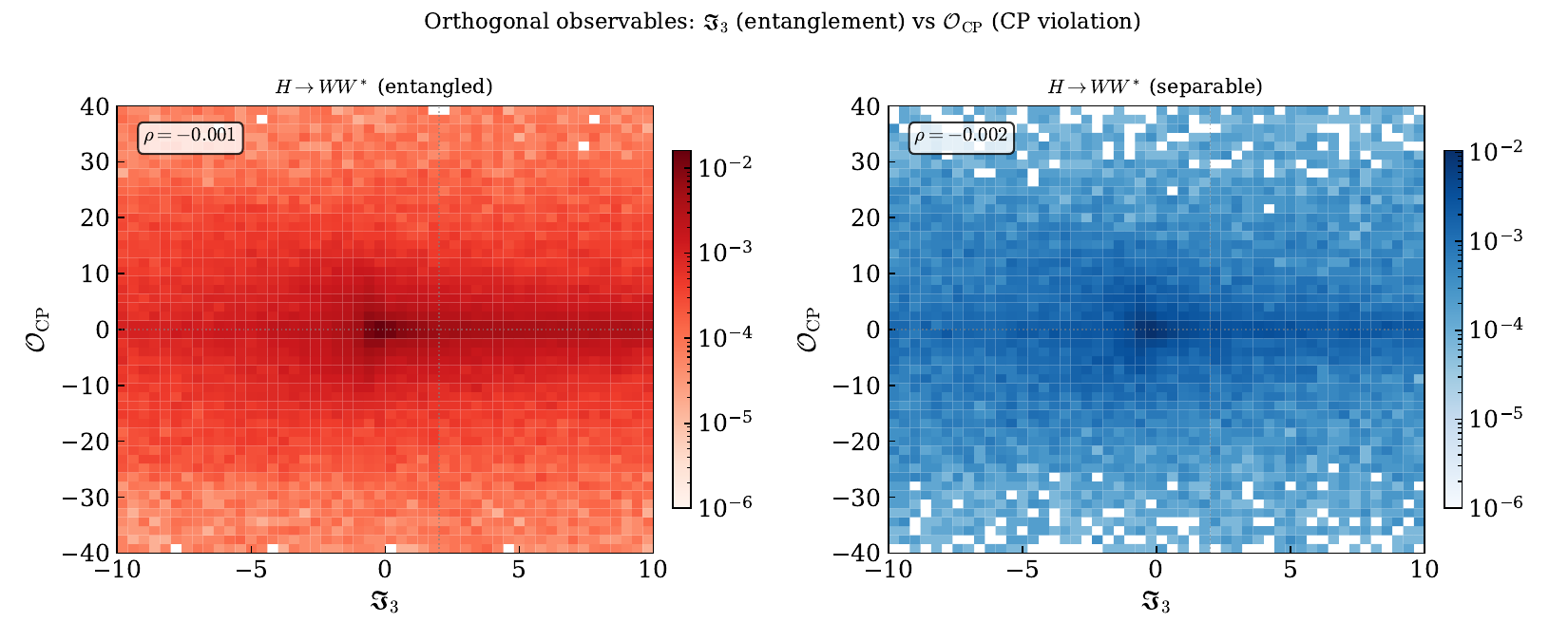}
\caption{Two-dimensional density of $\mathfrak{I}_3$ versus $\mathcal{O}_{\text{CP}}$ for the entangled (left) and separable (right) hypotheses. The near-zero linear correlation ($\rho = -0.001$) confirms that the observables carry independent information, probing entanglement and CP violation respectively.}
\label{fig:smeft_2d}
\end{figure}

The observables $\mathfrak{I}_3$, $\mathcal{O}_{\text{CP}}$, and $\mathcal{O}_{HW}$ probe statistically independent sectors of the correlation matrix. Figure~\ref{fig:smeft_2d} shows the two-dimensional density of $\mathfrak{I}_3$ versus $\mathcal{O}_{\text{CP}}$ for the entangled and separable hypotheses. The linear correlation coefficient is $\rho = -0.001$, confirming that the two observables carry independent information. Similarly, $\mathcal{O}_{\text{CP}}$ and $\mathcal{O}_{HW}$ are uncorrelated ($\rho < 0.001$). This orthogonality means that a single dataset of reconstructed $WW^*$ events simultaneously provides a test of quantum entanglement (via $\mathfrak{I}_3$), a probe of CP violation in the $HWW$ vertex (via $\mathcal{O}_{\text{CP}}$), and sensitivity to anomalous CP-even $HWW$ couplings (via $\mathcal{O}_{HW}$), all extracted from the same per-event $c_{ij}$ matrix with no additional reconstruction.

\subsection{Towards a Full SMEFT Analysis}

The distributions shown in Figures~\ref{fig:smeft_cp_odd}--\ref{fig:smeft_2d} establish the SM baseline for SMEFT-sensitive observables using existing simulation samples without BSM modifications. A complete SMEFT sensitivity study would require generating signal samples with non-zero Wilson coefficients, for example using the \texttt{SMEFTatNLO} model~\cite{Degrande:2020evl} in MadGraph, to construct BSM templates for hypothesis testing. The infrastructure developed in this paper (BDT-based event selection, cDDPM neutrino reconstruction, profile likelihood fitting with systematic uncertainties) applies directly to such an analysis: the BSM template for $\mathcal{O}_{\text{CP}}$ or $\mathcal{O}_{HW}$ would replace the separable template currently used for the entanglement test, with all other components of the analysis chain unchanged.

The approach is complementary to existing SMEFT studies using quantum tomography. Fabbrichesi et al.~\cite{Fabbrichesi:2023cev} and Sullivan~\cite{Sullivan:2024wzl} derive bounds on $HWW$ and $HZZ$ anomalous couplings from expectation values of individual density matrix elements, while Aoude et al.~\cite{Aoude:2022imd,Aoude:2023hxv} and Maltoni et al.~\cite{Maltoni:2024tul} apply similar techniques to top-quark and diboson production. The continuous-distribution approach proposed here offers a natural extension: rather than comparing measured expectation values to SM predictions, the full shape of the observable distribution is used, consistent with the methodology that this paper demonstrates for entanglement testing. As shown in Ref.~\cite{Grossi:2024jae}, NLO corrections can modify the density matrix elements at the percent level and would need to be incorporated in a precision SMEFT analysis.

\end{document}